\documentclass[sigconf, nonacm]{acmart}

\usepackage{booktabs}
\usepackage{drawings}
\usepackage{enumitem}
\usepackage{linearizability}
\usepackage{lipsum}
\usepackage{pervasives}
\usepackage{subcaption}
\usepackage{tikz}
\usetikzlibrary{backgrounds}
\usetikzlibrary{calc}
\usetikzlibrary{positioning}

\ifdefined\turnontechreport
  \toggletrue{techreportenabled}
\else
  \togglefalse{techreportenabled}
\fi
\toggletrue{techreportenabled}

\newcommand{\msgfont}[1]{\textsc{#1}}
\newcommand{\msg}[2]{\msgfont{#1}$\langle #2 \rangle$}

\newcommand\vldbdoi{XX.XX/XXX.XX}
\newcommand\vldbpages{XXX-XXX}
\newcommand\vldbvolume{14}
\newcommand\vldbissue{1}
\newcommand\vldbyear{2020}
\newcommand\vldbauthors{\authors}
\newcommand\vldbtitle{\shorttitle}
\newcommand\vldbavailabilityurl{http://vldb.org/pvldb/format_vol14.html}
\newcommand\vldbpagestyle{plain}

\newcommand{\bds}[3]{
  \begin{tcolorbox}[colback=flatblue!25,
                    colframe=white,
                    before skip=0pt,
                    after skip=6pt]
    \textbf{Bottleneck:} \emph{#1}

    \textbf{Decouple:} \emph{#2}

    \textbf{Scale:} \emph{#3}
  \end{tcolorbox}
}

\begin{document}
\title{Scaling Replicated State Machines with Compartmentalization}
\iftoggle{techreportenabled}{\subtitle{Technical Report. \today.}}{}

\author{Michael Whittaker}
\affiliation{\institution{UC Berkeley}}
\email{mjwhittaker@berkeley.edu}

\author{Ailidani Ailijiang}
\affiliation{\institution{Microsoft}}
\email{aiailiji@microsoft.com}

\author{Aleksey Charapko}
\affiliation{\institution{University of New Hampshire}}
\email{aleksey.charapko@unh.edu}

\author{Murat Demirbas}
\affiliation{\institution{University at Buffalo}}
\email{demirbas@buffalo.edu}

\author{Neil Giridharan}
\affiliation{\institution{UC Berkeley}}
\email{giridhn@berkeley.edu }

\author{Joseph M. Hellerstein}
\affiliation{\institution{UC Berkeley}}
\email{hellerstein@berkeley.edu}

\author{Heidi Howard}
\affiliation{\institution{University of Cambridge}}
\email{hh360@cst.cam.ac.uk}

\author{Ion Stoica}
\affiliation{\institution{UC Berkeley}}
\email{istoica@berkeley.edu}

\author{Adriana Szekeres}
\affiliation{\institution{VMWare}}
\email{aszekeres@vmware.com}

{\begin{abstract}
  State machine replication protocols, like MultiPaxos and Raft, are a critical
  component of many distributed systems and databases. However, these protocols
  offer relatively low throughput due to several bottlenecked components.
  Numerous existing protocols fix different bottlenecks in isolation but fall
  short of a complete solution. When you fix one bottleneck, another arises. In
  this paper, we introduce compartmentalization, the first comprehensive
  technique to eliminate state machine replication bottlenecks.
  Compartmentalization involves decoupling individual bottlenecks into distinct
  components and scaling these components independently. Compartmentalization
  has two key strengths. First, compartmentalization leads to strong
  performance. In this paper, we demonstrate how to compartmentalize MultiPaxos
  to increase its throughput by $6\times$ on a write-only workload and
  $16\times$ on a mixed read-write workload. Unlike other approaches, we
  achieve this performance without the need for specialized hardware. Second,
  compartmentalization is a technique, not a protocol.  Industry practitioners
  can apply compartmentalization to their protocols incrementally without
  having to adopt a completely new protocol.
\end{abstract}
}

\maketitle

\pagestyle{\vldbpagestyle}
\begingroup\small\noindent\raggedright\textbf{PVLDB Reference Format:}\\
\vldbauthors. \vldbtitle. PVLDB, \vldbvolume(\vldbissue): \vldbpages, \vldbyear.\\
\href{https://doi.org/\vldbdoi}{doi:\vldbdoi}
\endgroup
\begingroup
\renewcommand\thefootnote{}\footnote{\noindent
This work is licensed under the Creative Commons BY-NC-ND 4.0 International License. Visit \url{https://creativecommons.org/licenses/by-nc-nd/4.0/} to view a copy of this license. For any use beyond those covered by this license, obtain permission by emailing \href{mailto:info@vldb.org}{info@vldb.org}. Copyright is held by the owner/author(s). Publication rights licensed to the VLDB Endowment. \\
\raggedright Proceedings of the VLDB Endowment, Vol. \vldbvolume, No. \vldbissue\ %
ISSN 2150-8097. \\
\href{https://doi.org/\vldbdoi}{doi:\vldbdoi} \\
}\addtocounter{footnote}{-1}\endgroup

\ifdefempty{\vldbavailabilityurl}{}{
\vspace{.3cm}
\begingroup\small\noindent\raggedright\textbf{PVLDB Artifact Availability:}\\
The source code, data, and/or other artifacts have been made available at \url{\vldbavailabilityurl}.
\endgroup
}

{\section{Introduction}
State machine replication protocols are a crucial component of many distributed
systems and databases~\cite{corbett2013spanner, thomson2012calvin,
burrows2006chubby, taft2020cockroachdb, cosmos2019website, tidb2019website,
yugabyte2019website, cassandra2019website}. In many state machine replication
protocols, a single node has multiple responsibilities. For example, a Raft
leader acts as a batcher, a sequencer, a broadcaster, \emph{and} a state
machine replica. These overloaded nodes are often a throughput bottleneck,
which can be disastrous for systems that rely on state machine replication.


Many databases, for example, rely on state machine replication to replicate
large data partitions of tens of
gigabytes~\cite{schultz2019tunable,cosmos2019website}. These databases require
high-throughput state machine replication to handle all the requests in a
partition. However, in such systems, it is not uncommon to exceed the
throughput budget of a partition. For example, Cosmos DB will split a partition
if it experiences high throughput despite being under the storage limit. The
split, aside from costing resources, may have additional adverse effects on
applications, as Cosmos DB provides strongly consistent transactions only
within the partition. Eliminating state machine replication bottlenecks can
help avoid such unnecessary partition splits and improve performance,
consistency, and resource utilization.

Researchers have studied how to eliminate throughput bottlenecks, often
by inventing new state machine replication protocols that eliminate a
\emph{single} throughput bottleneck~\cite{moraru2013there, arun2017speeding,
mao2008mencius, ailijiang2019wpaxos, lamport2005generalized, lamport2006fast,
charapko2019linearizable, howard2016flexible, zhu2019harmonia,
terrace2009object, biely2012s}. However, eliminating a \emph{single} bottleneck
is not enough to achieve the best possible throughput. When you eliminate one
bottleneck, another arises. To achieve the best possible throughput, we have to
eliminate \emph{all} of the bottlenecks.


The key to eliminating these throughput bottlenecks is scaling, but it is
widely believed that state machine replication protocols don't
scale~\markrevisions{\cite{kapritsos2010scalable, zhang2018building,
mao2008mencius, moraru2013there, arun2017speeding}}. In this paper, we show
that this is not true. State machine replication protocols can indeed scale. As
a concrete illustration, we analyze the throughput bottlenecks of MultiPaxos
and systematically eliminate them using a combination of decoupling and
scaling, a technique we call \defword{compartmentalization}. For example,
consider the MultiPaxos leader, a notorious throughput bottleneck. The leader
has two distinct responsibilities. First, it sequences state machine commands
into a log. It puts the first command it receives into the first log entry, the
next command into the second log entry, and so on. Second, it broadcasts the
commands to the set of MultiPaxos acceptors, receives their responses, and then
broadcasts the commands again to a set of state machine replicas. To
compartmentalize the MultiPaxos leader, we first \defword{decouple} these two
responsibilities. There's no fundamental reason that the leader has to sequence
commands \emph{and} broadcast them. Instead, we have the leader sequence
commands and introduce a new set of nodes, called proxy leaders, to broadcast
the commands. Second, we \defword{scale} up the number of proxy leaders. We
note that broadcasting commands is embarrassingly parallel, so we can increase
the number of proxy leaders to avoid them becoming a bottleneck.  Note that
this scaling wasn't possible when sequencing and broadcasting were coupled on
the leader since sequencing is not scalable. Compartmentalization has three key
strengths.

\textbf{(1) Strong Performance Without Strong Assumptions.}
\markrevisions{%
  We compartmentalize MultiPaxos and increase its throughput by a factor of $6
  \times$ on a write-only workload using $6.66\times$ the number of machines
  and $16 \times$ on a mixed read-write workload using $4.33\times$ the number
  of machines.
}
%
Moreover, we achieve our strong performance without the strong assumptions made
by other state machine replication protocols with comparable
performance~\cite{terrace2009object, zhu2019harmonia, van2004chain,
takruri2020flair, jin2018netchain}.  For example, we do not assume a perfect
failure detector, we do not assume the availability of specialized hardware, we
do not assume uniform data access patterns, we do not assume clock synchrony,
and we do not assume key-partitioned state machines.

\textbf{(2) General and Incrementally Adoptable.}
%
Compartmentalization is \emph{not} a protocol. Rather, it's a technique that
can be systematically applied to existing protocols. Industry practitioners can
incrementally apply compartmentalization to their current protocols without
having to throw out their battle-tested implementations for something new and
untested.
We demonstrate the generality of compartmentalization by applying it three
other protocols~\cite{mao2008mencius, biely2012s, ding2020scalog} in addition
to MultiPaxos.

\markrevisions{%
  \textbf{(3) Easy to Understand.}
  Researchers have invented new state machine replication protocols to
  eliminate throughput bottlenecks, but these new protocols are often subtle
  and complicated. As a result, these sophisticated protocols have been largely
  ignored in industry due to their high barriers to adoption.
  Compartmentalization is based on the simple principles of decoupling and
  scaling and is designed to be easily understood.
}

In summary, we present the following contributions
\begin{itemize}
  \item
    We characterize all of MultiPaxos' throughput bottlenecks and explain why,
    historically, it was believed that they could not be scaled.

  \item
    We introduce the concept of compartmentalization: a technique to decouple
    and scale throughput bottlenecks.

  \item
    We apply compartmentalization to systematically eliminate MultiPaxos'
    throughput bottlenecks. In doing so, we debunk the widely held belief that
    MultiPaxos and similar state machine replication protocols do not scale.

\end{itemize}
}
{\section{Background}

\subsection{System Model}
Throughout the paper, we assume an asynchronous network model in which messages
can be arbitrarily dropped, delayed, and reordered. We assume machines can fail
by crashing but do not act maliciously; i.e., we do not consider Byzantine
failures. We assume that machines operate at arbitrary speeds, and we do not
assume clock synchronization. Every protocol discussed in this paper assumes
that at most $f$ machines will fail for some configurable $f$.

\subsection{Paxos}
\defword{Consensus} is the act of choosing a single value among a set of
proposed values, and \defword{Paxos}~\cite{lamport1998part} is the de facto
standard consensus protocol. We assume the reader is familiar with Paxos, but
we pause to review the parts of the protocol that are most important to
understand for the rest of this paper.

A Paxos deployment that tolerates $f$ faults consists of an arbitrary number of
clients, at least $f+1$ \defword{proposers}, and $2f+1$ \defword{acceptors}, as
illustrated in \figref{PaxosBackgroundDiagram}. When a client wants to propose
a value, it sends the value to a proposer $p$. The proposer then initiates a
two-phase protocol. In Phase 1, the proposer contacts the acceptors and learns
of any values that may have already been chosen. In Phase 2, the proposer
proposes a value to the acceptors, and the acceptors vote on whether or not to
choose the value. If a value receives votes from a majority of the acceptors,
the value is considered chosen.

More concretely, in Phase 1, $p$ sends \msgfont{Phase1a} messages to at least a
majority of the $2f+1$ acceptors. When an acceptor receives a \msgfont{Phase1a}
message, it replies with a \msgfont{Phase1b} message. When the leader receives
\msgfont{Phase1b} messages from a majority of the acceptors, it begins Phase 2.
In Phase 2, the proposer sends \msg{Phase2a}{x} messages to the acceptors with
some value $x$. Upon receiving a \msg{Phase2a}{x} message, an acceptor can
either ignore the message, or vote for the value $x$ and return a
\msg{Phase2b}{x} message to the proposer. Upon receiving \msg{Phase2b}{x}
messages from a majority of the acceptors, the proposed value $x$ is considered
chosen.

{
\tikzstyle{proc}=[draw, circle, thick, inner sep=2pt]
\tikzstyle{client}=[proc, fill=clientcolor!25]
\tikzstyle{proposer}=[proc, fill=proposercolor!25]
\tikzstyle{acceptor}=[proc, fill=acceptorcolor!25]

\tikzstyle{proclabel}=[inner sep=0pt, align=center, font=\small]

\tikzstyle{component}=[draw, thick, flatgray, rounded corners]

\tikzstyle{comm}=[-latex, thick]
\tikzstyle{commnum}=[fill=white, inner sep=0pt]

\begin{figure}[ht]
  \centering
  \begin{subfigure}[b]{0.45\columnwidth}
    \centering
    \begin{tikzpicture}[xscale=1.25]
      \node[client] (c1) at (0, 2) {$c_1$};
      \node[client] (c2) at (0, 1) {$c_2$};
      \node[client] (c3) at (0, 0) {$c_3$};
      \node[proposer] (p1) at (1, 1.5) {$p_1$};
      \node[proposer] (p2) at (1, 0.5) {$p_2$};
      \node[acceptor] (a1) at (2, 2) {$a_1$};
      \node[acceptor] (a2) at (2, 1) {$a_2$};
      \node[acceptor] (a3) at (2, 0) {$a_3$};

      \node[proclabel] (clients) at (0, 3) {Clients};
      \node[proclabel] (proposers) at (1, 3) {$f+1$\\Proposers};
      \node[proclabel] (acceptors) at (2, 3) {$2f+1$\\Acceptors};
      \halffill{clients}{clientcolor!25}
      \quarterfill{proposers}{proposercolor!25}
      \quarterfill{acceptors}{acceptorcolor!25}

      \draw[comm] (c1) to node[commnum]{1} (p1);
      \draw[comm, bend left=15] (p1) to node[commnum]{2} (a1);
      \draw[comm, bend left=15] (p1) to node[commnum]{2} (a2);
      \draw[comm, bend left=15] (a1) to node[commnum]{3} (p1);
      \draw[comm, bend left=15] (a2) to node[commnum]{3} (p1);
    \end{tikzpicture}%
    \caption{Phase 1}\figlabel{PaxosBackgroundDiagramPhase1}
  \end{subfigure}\hspace{0.09\columnwidth}
  \begin{subfigure}[b]{0.45\columnwidth}
    \centering
    \begin{tikzpicture}[xscale=1.25]
      \node[client] (c1) at (0, 2) {$c_1$};
      \node[client] (c2) at (0, 1) {$c_2$};
      \node[client] (c3) at (0, 0) {$c_3$};
      \node[proposer] (p1) at (1, 1.5) {$p_1$};
      \node[proposer] (p2) at (1, 0.5) {$p_2$};
      \node[acceptor] (a1) at (2, 2) {$a_1$};
      \node[acceptor] (a2) at (2, 1) {$a_2$};
      \node[acceptor] (a3) at (2, 0) {$a_3$};

      \node[proclabel] (clients) at (0, 3) {Clients};
      \node[proclabel] (proposers) at (1, 3) {$f+1$\\Proposers};
      \node[proclabel] (acceptors) at (2, 3) {$2f+1$\\Acceptors};
      \halffill{clients}{clientcolor!25}
      \quarterfill{proposers}{proposercolor!25}
      \quarterfill{acceptors}{acceptorcolor!25}

      \draw[comm, bend left=15] (p1) to node[commnum]{4} (a1);
      \draw[comm, bend left=15] (p1) to node[commnum]{4} (a2);
      \draw[comm, bend left=15] (a1) to node[commnum]{5} (p1);
      \draw[comm, bend left=15] (a2) to node[commnum]{5} (p1);
      \draw[comm] (p1) to node[commnum]{6} (c1);
    \end{tikzpicture}
    \caption{Phase 2}\figlabel{PaxosBackgroundDiagramPhase2}
  \end{subfigure}
  \caption{An example execution of Paxos ($f=1$).}%
  \figlabel{PaxosBackgroundDiagram}
\end{figure}
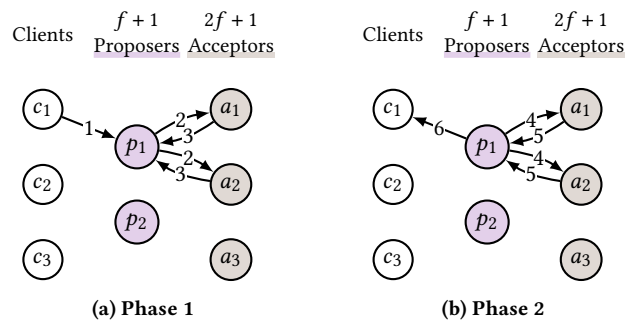}


\subsection{MultiPaxos}
While consensus is the act of choosing a single value, \defword{state machine
replication} is the act of choosing a sequence (a.k.a.\ log) of values. A state
machine replication protocol manages a number of \defword{replicas} of a
deterministic state machine. Over time, the protocol constructs a growing log
of state machine commands, and replicas execute the commands in log order. By
beginning in the same initial state, and by executing commands in the same
order, all state machine replicas are kept in sync. This is illustrated in
\figref{StateMachineReplicationExample}.

{\newlength{\logentryinnersep}
\setlength{\logentryinnersep}{4pt}
\newlength{\logentrylinewidth}
\setlength{\logentrylinewidth}{1pt}
\newlength{\logentrywidth}
\setlength{\logentrywidth}{\widthof{$X$}+2\logentryinnersep}
\newcommand{\cmdi}{$x$}
\newcommand{\cmdii}{$y$}
\newcommand{\cmdiii}{$z$}

\tikzstyle{logentry}=[%
  draw,
  inner sep=\logentryinnersep,
  line width=\logentrylinewidth,
  minimum height=\logentrywidth,
  minimum width=\logentrywidth]
\tikzstyle{executed}=[fill=flatgreen, opacity=0.2, draw opacity=1, text opacity=1]
\tikzstyle{logindex}=[gray, font=\small]

\newcommand{\rightof}[1]{-\logentrylinewidth of #1}

\newcommand{\multipaxoslog}[6]{%
  \node[logentry, label={[logindex]90:0}, #2] (0) {#1};
  \node[logentry, label={[logindex]90:1}, right=\rightof{0}, #4] (1) {#3};
  \node[logentry, label={[logindex]90:2}, right=\rightof{1}, #6] (2) {#5};
}

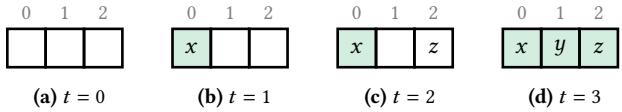
\begin{figure}[ht]
  \centering
  \begin{subfigure}[b]{0.2\columnwidth}
    \begin{tikzpicture}
      \multipaxoslog{}{}%
                    {}{}%
                    {}{}
    \end{tikzpicture}
    \caption{$t=0$}\figlabel{StateMachineReplicationExampleA}
  \end{subfigure}\hspace{12pt}
  \begin{subfigure}[b]{0.2\columnwidth}
    \begin{tikzpicture}
      \multipaxoslog{\cmdi}{executed}%
                    {}{}%
                    {}{}
    \end{tikzpicture}
    \caption{$t=1$}\figlabel{StateMachineReplicationExampleB}
  \end{subfigure}\hspace{12pt}
  \begin{subfigure}[b]{0.2\columnwidth}
    \begin{tikzpicture}
      \multipaxoslog{\cmdi}{executed}%
                    {}{}%
                    {\cmdiii}{}
    \end{tikzpicture}
    \caption{$t=2$}\figlabel{StateMachineReplicationExampleC}
  \end{subfigure}\hspace{12pt}
  \begin{subfigure}[b]{0.2\columnwidth}
    \begin{tikzpicture}
      \multipaxoslog{\cmdi}{executed}%
                    {\cmdii}{executed}%
                    {\cmdiii}{executed}
    \end{tikzpicture}
    \caption{$t=3$}\figlabel{StateMachineReplicationExampleD}
  \end{subfigure}
  \caption{%
    At time $t=0$, no state machine commands are chosen. At time $t=1$ command
    $x$ is chosen in slot $0$. At times $t=2$ and $t=3$, commands $z$ and $y$
    are chosen in slots $2$ and $1$. Executed commands are shaded green. Note
    that all state machines execute the commands $x$, $y$, $z$ in log order.
  }\figlabel{StateMachineReplicationExample}
\end{figure}
}

\defword{MultiPaxos} is one of the most widely used state machine replication
protocols. Again, we assume the reader is familiar with MultiPaxos, but we
review the most salient bits.
MultiPaxos uses one instance of Paxos for every log entry, choosing the command
in the $i$th log entry using the $i$th instance of Paxos.
A MultiPaxos deployment that tolerates $f$ faults consists of an arbitrary
number of clients, at least $f+1$ proposers, and $2f+1$ acceptors (like Paxos),
as well as at least $f+1$ replicas, as illustrated in
\figref{MultiPaxosBackgroundDiagram}.

{
\tikzstyle{proc}=[draw, circle, thick, inner sep=2pt]
\tikzstyle{client}=[proc, fill=clientcolor!25]
\tikzstyle{proposer}=[proc, fill=proposercolor!25]
\tikzstyle{acceptor}=[proc, fill=acceptorcolor!25]
\tikzstyle{replica}=[proc, fill=replicacolor!25]

\tikzstyle{proclabel}=[inner sep=0pt, align=center]

\tikzstyle{component}=[draw, thick, flatgray, rounded corners]

\tikzstyle{comm}=[-latex, thick]
\tikzstyle{commnum}=[fill=white, inner sep=0pt]

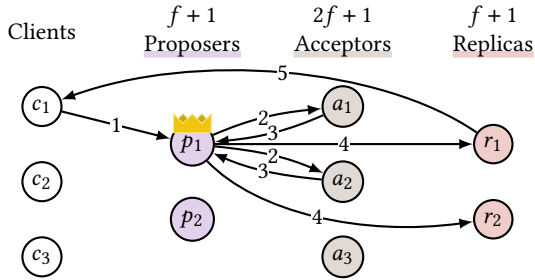
\begin{figure}[h]
  \centering
  \begin{tikzpicture}[xscale=2]
    \node[client] (c1) at (0, 2) {$c_1$};
    \node[client] (c2) at (0, 1) {$c_2$};
    \node[client] (c3) at (0, 0) {$c_3$};
    \node[proposer] (p1) at (1, 1.5) {$p_1$};
    \node[proposer] (p2) at (1, 0.5) {$p_2$};
    \node[acceptor] (a1) at (2, 2) {$a_1$};
    \node[acceptor] (a2) at (2, 1) {$a_2$};
    \node[acceptor] (a3) at (2, 0) {$a_3$};
    \node[replica] (r1) at (3, 1.5) {$r_1$};
    \node[replica] (r2) at (3, 0.5) {$r_2$};

    \crown{(p1.north)++(0,-0.15)}{0.25}{0.25}
    \node[proclabel] (clients) at (0, 3) {Clients};
    \node[proclabel] (proposers) at (1, 3) {$f+1$\\Proposers};
    \node[proclabel] (acceptors) at (2, 3) {$2f+1$\\Acceptors};
    \node[proclabel] (replicas) at (3, 3) {$f+1$\\Replicas};
    \halffill{clients}{clientcolor!25}
    \quarterfill{proposers}{proposercolor!25}
    \quarterfill{acceptors}{acceptorcolor!25}
    \quarterfill{replicas}{replicacolor!25}

    \draw[comm] (c1) to node[commnum]{1} (p1);
    \draw[comm, bend left=15] (p1) to node[commnum]{2} (a1);
    \draw[comm, bend left=15] (p1) to node[commnum]{2} (a2);
    \draw[comm, bend left=15] (a1) to node[commnum]{3} (p1);
    \draw[comm, bend left=15] (a2) to node[commnum]{3} (p1);
    \draw[comm] (p1) to node[commnum]{4} (r1);
    \draw[comm, bend right=40] (p1) to node[commnum]{4} (r2);
    \draw[comm, bend right=45] (r1) to node[commnum]{5} (c1);
  \end{tikzpicture}
  \caption{%
    An example execution of MultiPaxos ($f=1$). The leader is adorned with a
    crown.
  }%
  \figlabel{MultiPaxosBackgroundDiagram}
\end{figure}}

Initially, one of the proposers is elected leader and runs Phase 1 of Paxos for
every log entry. When a client wants to propose a state machine command
$x$, it sends the command to the leader (1). The leader assigns the command a
log entry $i$ and then runs Phase 2 of the $i$th Paxos instance to get the
value $x$ chosen in entry $i$. That is, the leader sends \msg{Phase2a}{i, x}
messages to the acceptors to vote for value $x$ in slot $i$ (2). In the normal
case, the acceptors all vote for $x$ in slot $i$ and respond with
\msg{Phase2b}{i, x} messages (3). Once the leader learns that a command has
been chosen in a given log entry (i.e.\ once the leader receives
\msg{Phase2b}{i, x} messages from a majority of the acceptors), it informs the
replicas (4). Replicas insert commands into their logs and execute the logs in
prefix order.

Note that the leader assigns log entries to commands in increasing order. The
first received command is put in entry $0$, the next command in entry $1$, the
next command in entry $2$, and so on.
Also note that even though every replica executes every command, for any given
state machine command $x$, only one replica needs to send the result of
executing $x$ back to the client (5). For example, log entries can be
round-robin partitioned across the replicas.
%

\subsection{MultiPaxos Doesn't Scale?}\seclabel{MultiPaxosDoesntScale}
It is widely believed that MultiPaxos does not
scale~\markrevisions{\cite{kapritsos2010scalable, zhang2018building,
mao2008mencius, moraru2013there, arun2017speeding}}. Throughout the paper, we
will explain that this is not true, but it first helps to understand why trying
to scale MultiPaxos in the straightforward and obvious way does not work.
MultiPaxos consists of proposers, acceptors, and replicas. We discuss each.

First, increasing the number of proposers \emph{does not improve performance}
because every client must send its requests to the leader regardless of the
number proposers. The non-leader replicas are idle and do not contribute to the
protocol during normal operation.

Second, increasing the number of acceptors \emph{hurts performance}. To get a
value chosen, the leader must contact a majority of the acceptors. When we
increase the number of acceptors, we increase the number of acceptors that the
leader has to contact. This decreases throughput because the leader---which is
the throughput bottleneck---has to send and receive more messages per command.
Moreover, every acceptor processes at least half of all commands regardless of
the number of acceptors.

Third, increasing the number of replicas \emph{hurts performance}. The leader
broadcasts chosen commands to all of the replicas, so when we increase the
number of replicas, we increase the load on the leader and decrease MultiPaxos'
throughput. Moreover, every replica must execute every state machine command,
so increasing the number of replicas does not decrease the replicas' load.
}
{\section{Compartmentalizing MultiPaxos}\seclabel{MultiPaxos}
We now compartmentalize MultiPaxos. Throughout the paper, we introduce six
compartmentalizations, summarized in \tabref{CompartmentalizationSummary}. For
every compartmentalization, we identify a throughput bottleneck and then
explain how to decouple and scale it.

\begin{table*}[ht]
  \centering
  \caption{A summary of the compartmentalizations presented in this paper.}%
  \tablabel{CompartmentalizationSummary}
  \begin{tabular}{llll}
    \toprule
    Compartmentalization                                          & Bottleneck          & Decouple                                    & Scale \\\midrule
    1 \textcolor{black!80}{(\secref{MultiPaxos/ProxyLeaders})}    & leader              & command sequencing and command broadcasting & the number of proxy leaders \\
    2 \textcolor{black!80}{(\secref{MultiPaxos/AcceptorGrids})}   & acceptors           & read quorums and write quorums              & the number of write quorums \\
    3 \textcolor{black!80}{(\secref{MultiPaxos/MoreReplicas})}    & replicas            & command sequencing and command broadcasting & the number of replicas \\
    4 \textcolor{black!80}{(\secref{MultiPaxos/LeaderlessReads})} & leader and replicas & read path and write path                    & the number of read quorums \\
    5 \textcolor{black!80}{(\secref{Batchers})}                   & leader              & batch formation and batch sequencing        & the number of batchers \\
    6 \textcolor{black!80}{(\secref{Unbatchers})}                 & replicas            & batch processing and batch replying         & the number of unbatchers \\
    \bottomrule
  \end{tabular}
\end{table*}

\subsection{Compartmentalization 1: Proxy Leaders}\seclabel{MultiPaxos/ProxyLeaders}
\bds{leader}
    {command sequencing and broadcasting}
    {the number of command broadcasters}

\paragraph{Bottleneck}
The MultiPaxos leader is a well known throughput bottleneck for the following
reason. Refer again to \figref{MultiPaxosBackgroundDiagram}. To process a
single state machine command from a client, the leader must receive a message
from the client, send at least $f+1$ \msgfont{Phase2a} messages to the
acceptors, receive at least $f+1$ \msgfont{Phase2b} messages from the
acceptors, and send at least $f+1$ messages to the replicas. In total, the
leader sends and receives at least $3f+4$ messages per command. Every acceptor
on the other hand processes only $2$ messages, and every replica processes
either $1$ or $2$. Because every state machine command goes through the leader,
and because the leader has to perform disproportionately more work than every
other component, the leader is the throughput bottleneck.

\paragraph{Decouple}
To alleviate this bottleneck, we first decouple the leader. To do so, we note
that a MultiPaxos leader has two jobs.
The first is \defword{sequencing}. The leader sequences commands by assigning
each command a log entry. Log entry $0$, then $1$, then $2$, and so on.
The second is \defword{broadcasting}. The leader sends \msgfont{Phase2a}
messages, collects \msgfont{Phase2b} responses, and broadcasts chosen values to
the replicas.
Historically, these two responsibilities have both fallen on the leader, but
this is not fundamental. We instead decouple the two responsibilities. We
introduce a set of at least $f+1$ \defword{proxy leaders}, as shown in
\figref{ProxyLeaders}. The leader is responsible for sequencing commands, while
the proxy leaders are responsible for getting commands chosen and broadcasting
the commands to the replicas.

{
\tikzstyle{proc}=[draw, circle, thick, inner sep=2pt]
\tikzstyle{newproc}=[proc, draw=flatgreen]
\tikzstyle{client}=[proc, fill=clientcolor!25]
\tikzstyle{proposer}=[proc, fill=proposercolor!25]
\tikzstyle{proxyleader}=[proc, fill=proxyleadercolor!25]
\tikzstyle{acceptor}=[proc, fill=acceptorcolor!25]
\tikzstyle{replica}=[proc, fill=replicacolor!25]

\tikzstyle{proclabel}=[inner sep=0pt, align=center, font=\small]

\tikzstyle{component}=[draw, thick, flatgray, rounded corners]

\tikzstyle{comm}=[-latex, thick]
\tikzstyle{commnum}=[fill=white, inner sep=0pt]
\tikzstyle{newcomm}=[comm, flatgreen]
\tikzstyle{newcommnum}=[commnum, flatred]
\tikzstyle{oldcomm}=[comm, black!50]
\tikzstyle{oldcommnum}=[commnum, black!50]

\begin{figure}[ht]
  \centering
  \begin{tikzpicture}[xscale=1.75]
    \node[client] (c1) at (0, 2) {$c_1$};
    \node[client] (c2) at (0, 1) {$c_2$};
    \node[client] (c3) at (0, 0) {$c_3$};
    \node[proposer] (p1) at (1, 1.5) {$p_1$};
    \node[proposer] (p2) at (1, 0.5) {$p_2$};
    \node[proxyleader, newproc] (pl1) at (2, 2) {$l_1$};
    \node[proxyleader, newproc] (pl2) at (2, 1) {$l_2$};
    \node[proxyleader, newproc] (pl3) at (2, 0) {$l_3$};
    \node[acceptor] (a1) at (3, 2) {$a_1$};
    \node[acceptor] (a2) at (3, 1) {$a_2$};
    \node[acceptor] (a3) at (3, 0) {$a_3$};
    \node[replica] (r1) at (4, 1.5) {$r_1$};
    \node[replica] (r2) at (4, 0.5) {$r_2$};

    \crown{(p1.north)++(0,-0.15)}{0.28}{0.25}
    \node[proclabel] (clients) at (0, 3) {Clients};
    \node[proclabel] (proposers) at (1, 3) {$f+1$\\Proposers};
    \node[proclabel] (proxyleaders) at (2, 3) {$\geq f+1$\\Proxy Leaders};
    \node[proclabel] (acceptors) at (3, 3) {$2f+1$\\Acceptors};
    \node[proclabel] (replicas) at (4, 3) {$f+1$\\Replicas};
    \halffill{clients}{clientcolor!25}
    \quarterfill{proposers}{proposercolor!25}
    \quarterfill{proxyleaders}{proxyleadercolor!25}
    \quarterfill{acceptors}{acceptorcolor!25}
    \quarterfill{replicas}{replicacolor!25}

    \draw[oldcomm] (c1) to node[commnum]{1} (p1);
    \draw[newcomm] (p1) to node[commnum]{2} (pl2);
    \draw[newcomm, near start, bend left=40] (pl2) to node[commnum]{3} (a1);
    \draw[newcomm, near start, bend right=40] (pl2) to node[commnum]{3} (a3);
    \draw[newcomm, near start, bend right=20] (a1) to node[commnum]{4} (pl2);
    \draw[newcomm, near start, bend left=20] (a3) to node[commnum]{4} (pl2);
    \draw[newcomm, bend left=20] (pl2) to node[commnum]{5} (r1);
    \draw[newcomm, bend right=20] (pl2) to node[commnum]{5} (r2);
    \draw[oldcomm, bend right=35] (r1) to node[commnum]{6} (c1);
  \end{tikzpicture}
  \caption{%
    An example execution of Compartmentalized MultiPaxos with three proxy
    leaders ($f=1$).  Throughout the paper, nodes and messages that were not
    present in previous iterations of the protocol are highlighted in green.
  }%
  \figlabel{ProxyLeaders}
\end{figure}
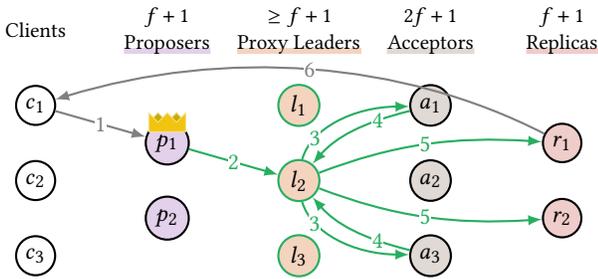}

More concretely, when a leader receives a command $x$ from a client (1), it
assigns the command $x$ a log entry $i$ and then forms a \msgfont{Phase2a}
message that includes $x$ and $i$. The leader does \emph{not} send the
\msgfont{Phase2a} message to the acceptors. Instead, it sends the
\msgfont{Phase2a} message to a randomly selected proxy leader (2). Note that
every command can be sent to a different proxy leader. The leader balances load
evenly across all of the proxy leaders.
Upon receiving a \msgfont{Phase2a} message, a proxy leader broadcasts it to the
acceptors (3), gathers a quorum of $f+1$ \msgfont{Phase2b} responses (4), and
notifies the replicas of the chosen value (5). All other aspects of the
protocol remain unchanged.

Without proxy leaders, the leader processes $3f+4$ messages per command. With
proxy leaders, the leader only processes $2$. This makes the leader
significantly less of a throughput bottleneck, or potentially eliminates it as
the bottleneck entirely.

\paragraph{Scale}
The leader now processes fewer messages per command, but every proxy leader has
to process $3f+4$ messages. Have we really eliminated the leader as a
bottleneck, or have we just moved the bottleneck into the proxy leaders? To
answer this, we note that the proxy leaders are embarrassingly parallel. They
operate independently from one another. Moreover, the leader distributes load
among the proxy leaders equally, so the load on any single proxy leader
decreases as we increase the number of proxy leaders. Thus, we can trivially
increase the number of proxy leaders until they are no longer a throughput
bottleneck.

\paragraph{Discussion}
Note that decoupling \emph{enables} scaling. As discussed in
\secref{MultiPaxosDoesntScale}, we cannot naively increase the number of
proposers. Without decoupling, the leader is both a sequencer and broadcaster,
so we cannot increase the number of leaders to increase the number of
broadcasters because doing so would lead to multiple sequencers, which is not
permitted. Only by decoupling the two responsibilities can we scale one without
scaling the other.

Also note that the protocol remains tolerant to $f$ faults regardless of the
number of machines. However, increasing the number of machines does decrease
the expected time to $f$ failures (this is true for every protocol that scales
up the number of machines, not just our protocol).  We believe that increasing
throughput at the expense of a shorter time to $f$ failures is well worth it in
practice because failed machines can be replaced with new machines using a
reconfiguration protocol~\cite{lamport2001paxos, ongaro2014search}. The time
required to perform a reconfiguration is many orders of magnitude smaller than
the mean time between failures.

\subsection{Compartmentalization 2: Acceptor Grids}\seclabel{MultiPaxos/AcceptorGrids}
\bds{acceptors}
    {read quorums and write quorums}
    {the number of write quorums}

\paragraph{Bottleneck}
After compartmentalizing the leader, it is possible that the acceptors are the
throughput bottleneck. It is widely believed that acceptors do not scale:
``using more than $2f+1$ [acceptors] for $f$ failures is possible but illogical
because it requires a larger quorum size with no additional
benefit''~\cite{zhang2018building}. As explained in
\secref{MultiPaxosDoesntScale}, there are two reasons why naively increasing
the number of acceptors is ill-advised.

First, increasing the number of acceptors increases the number of messages that
the leader has to send and receive. This increases the load on the leader, and
since the leader is the throughput bottleneck, this decreases throughput. This
argument no longer applies. With the introduction of proxy leaders, the leader
no longer communicates with the acceptors. Increasing the number of acceptors
increases the load on every individual proxy leader, but the increased load
will not make the proxy leaders a bottleneck because we can always scale them
up.

Second, every command must be processed by a majority of the acceptors. Thus,
even with a large number of acceptors, every acceptor must process at
least half of all state machine commands. This argument still holds.

\paragraph{Decouple}
We compartmentalize the acceptors by using flexible
quorums~\cite{howard2016flexible}. MultiPaxos---the vanilla version, not the
compartmentalized version---requires $2f+1$ acceptors, and the leader
communicates with $f+1$ acceptors in both Phase 1 and Phase 2 (a majority of
the acceptors). The sets of $f+1$ acceptors are called \defword{quorums}, and
MultiPaxos' correctness relies on the fact that any two quorums intersect.
While majority quorums are sufficient for correctness, they are not necessary.
MultiPaxos is correct as long as every quorum contacted in Phase 1 (called a
\defword{read quorum}) intersects every quorum contacted in Phase 2 (called a
\defword{write quorum}). Read quorums do not have to intersect other read
quorums, and write quorums do not have to intersect other write quorums.

By decoupling read quorums from write quorums, we can reduce the load on the
acceptors by eschewing majority quorums for a more efficient set of quorums.
Specifically, we arrange the acceptors into an $r \times w$ rectangular grid,
where $r, w \geq f+1$. Every row forms a read quorum, and every column forms a
write quorum ($r$ stands for row and for read). That is, a leader contacts an
arbitrary row of acceptors in Phase 1 and an arbitrary column of acceptors for
every command in Phase 2. Every row intersects every column, so this is a valid
set of quorums.

A $2 \times 3$ acceptor grid is illustrated in \figref{AcceptorGrids}. There
are two read quorums (the rows $\set{a_1, a_2, a_3}$ and $\set{a_4, a_5, a_6}$)
and three write quorums (the columns $\set{a_1, a_4}$, $\set{a_2, a_5}$,
$\set{a_3, a_6}$). Because there are three write quorums, every acceptor only
processes one third of all the commands.  This is not possible with majority
quorums because with majority quorums, every acceptor processes at least half
of all the commands, regardless of the number of acceptors.

{
\tikzstyle{proc}=[draw, circle, thick, inner sep=2pt]
\tikzstyle{newproc}=[draw=flatgreen]
\tikzstyle{client}=[proc, fill=clientcolor!25]
\tikzstyle{proposer}=[proc, fill=proposercolor!25]
\tikzstyle{proxyleader}=[proc, fill=proxyleadercolor!25]
\tikzstyle{acceptor}=[proc, fill=acceptorcolor!25]
\tikzstyle{replica}=[proc, fill=replicacolor!25]

\tikzstyle{proclabel}=[inner sep=0pt, align=center, font=\footnotesize]

\tikzstyle{component}=[draw, thick, flatgray, rounded corners]

\tikzstyle{comm}=[-latex, thick]
\tikzstyle{commnum}=[fill=white, inner sep=0pt]
\tikzstyle{newcomm}=[comm, flatgreen]
\tikzstyle{oldcomm}=[comm, black!50]

\begin{figure}[ht]
  \centering
  \begin{tikzpicture}[xscale=1.5]
    \node[client] (c1) at (0, 2) {$c_1$};
    \node[client] (c2) at (0, 1) {$c_2$};
    \node[client] (c3) at (0, 0) {$c_3$};
    \node[proposer] (p1) at (1, 1.5) {$p_1$};
    \node[proposer] (p2) at (1, 0.5) {$p_2$};
    \node[proxyleader] (pl1) at (2, 2) {$l_1$};
    \node[proxyleader] (pl2) at (2, 1) {$l_2$};
    \node[proxyleader] (pl3) at (2, 0) {$l_3$};
    \node[acceptor, newproc] (a1) at (3, 2) {$a_1$};
    \node[acceptor, newproc] (a2) at (3.5, 2) {$a_2$};
    \node[acceptor, newproc] (a3) at (4, 2) {$a_3$};
    \node[acceptor, newproc] (a4) at (3, 0) {$a_4$};
    \node[acceptor, newproc] (a5) at (3.5, 0) {$a_5$};
    \node[acceptor, newproc] (a6) at (4, 0) {$a_6$};
    \node[replica] (r1) at (5, 1.5) {$r_1$};
    \node[replica] (r2) at (5, 0.5) {$r_2$};

    \crown{(p1.north)++(0,-0.15)}{0.333}{0.25}
    \node[proclabel] (clients) at (0, 3) {Clients};
    \node[proclabel] (proposers) at (1, 3) {$f+1$\\Proposers};
    \node[proclabel] (proxyleaders) at (2, 3) {$\geq f+1$\\Proxy Leaders};
    \node[proclabel] (acceptors) at (3.5, 3) {$(\geq\!f\!+\!1) \times (\geq\!f\!+\!1)$\\Acceptors};
    \node[proclabel] (replicas) at (5, 3) {$f+1$\\Replicas};
    \halffill{clients}{clientcolor!25}
    \quarterfill{proposers}{proposercolor!25}
    \quarterfill{proxyleaders}{proxyleadercolor!25}
    \quarterfill{acceptors}{acceptorcolor!25}
    \quarterfill{replicas}{replicacolor!25}

    \tikzstyle{quorum}=[rounded corners]
    \tikzstyle{readquourum}=[quorum, draw=red]
    \tikzstyle{writequorum}=[quorum, dashed, draw=blue]
    \foreach \x/\y in {a1/a3, a4/a6}{
      \draw[readquourum] ($(\x.north west) + (-0.1, 0.1)$) rectangle
                         ($(\y.south east) + (0.1, -0.1)$);
    }
    \foreach \x/\y in {a1/a4, a2/a5, a3/a6}{
      \draw[writequorum] ($(\x.north west) + (-0.07, 0.15)$) rectangle
                         ($(\y.south east) + (0.07, -0.15)$);
    }

    \draw[oldcomm, near start] (c1) to node[commnum]{1} (p1);
    \draw[oldcomm] (p1) to node[commnum]{2} (pl2);
    \draw[newcomm, near start] (pl2) to node[commnum]{3} (a1);
    \draw[newcomm, near start] (pl2) to node[commnum]{3} (a4);
    \draw[newcomm, near start, bend left=20] (a1) to node[commnum]{4} (pl2);
    \draw[newcomm, near start, bend right=20] (a4) to node[commnum]{4} (pl2);
    \draw[oldcomm] (pl2) to node[commnum]{5} (r1);
    \draw[oldcomm] (pl2) to node[commnum]{5} (r2);
    \draw[oldcomm, bend right=30] (r1) to node[commnum]{6} (c1);
  \end{tikzpicture}
  \caption{%
    An execution of Compartmentalized MultiPaxos with a $2 \times 3$ grid of
    acceptors ($f=1$). The two read quorums---$\set{a_1, a_2, a_3}$ and
    $\set{a_4, a_5, a_6}$---are shown in solid red rectangles. The three write
    quorums---$\set{a_1, a_4}$, $\set{a_2, a_5}$, and $\set{a_3, a_6}$---are
    shown in dashed blue rectangles.
  }%
  \figlabel{AcceptorGrids}
\end{figure}
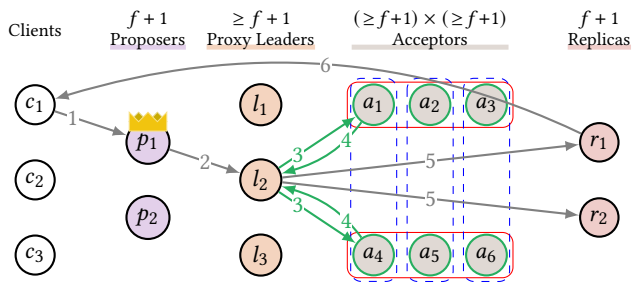}

\paragraph{Scale}
With majority quorums, every acceptor has to process at least half of all state
machines commands. With grid quorums, every acceptor only has to process
$\frac{1}{w}$ of the state machine commands. Thus, we can increase $w$ (i.e.\
increase the number of columns in the grid) to reduce the load on the acceptors
and eliminate them as a throughput bottleneck.

\paragraph{Discussion}
Note that, like with proxy leaders, decoupling \emph{enables} scaling. With
majority quorums, read and write quorums are coupled, so we cannot increase the
number of acceptors without also increasing the size of all quorums. Acceptor
grids allow us to decouple the number of acceptors from the size of write
quorums, allowing us to scale up the acceptors and decrease their load.

Also note that increasing the number of write quorums increases the size of
read quorums which increases the number of acceptors that a leader has to
contact in Phase 1. We believe this is a worthy trade-off since Phase 2 is
executed in the normal case and Phase 1 is only run in the event of a leader
failure.

\subsection{Compartmentalization 3: More Replicas}\seclabel{MultiPaxos/MoreReplicas}
\bds{replicas}
    {command sequencing and broadcasting}
    {the number of replicas}

\paragraph{Bottleneck}
After compartmentalizing the leader and the acceptors, it is possible that the
replicas are the bottleneck. Recall from \secref{MultiPaxosDoesntScale} that
naively scaling the replicas does not work for two reasons. First, every
replica must receive and execute every state machine command. This is not
actually true, but we leave that for the next compartmentalization. Second,
like with the acceptors, increasing the number of replicas increases the load
on the leader. Because we have already decoupled sequencing from broadcasting
on the leader and introduced proxy leaders, this is no longer true, so we are
free to increase the number of replicas. In \figref{MoreReplicas}, for example,
we show MultiPaxos with three replicas instead of the minimum required two.

\paragraph{Scale}
If every replica has to execute every command, does increasing the number of
replicas decrease their load?  Yes. Recall that while every replica has to
execute every state machine, only \emph{one} of the replicas has to send the
result of executing the command back to the client. Thus, with $n$ replicas,
every replica only has to send back results for $\frac{1}{n}$ of the commands.
If we scale up the number of replicas, we reduce the number of messages that
each replica has to send. This reduces the load on the replicas and helps
prevent them from becoming a throughput bottleneck. In \figref{MoreReplicas}
for example, with three replicas, every replica only has to reply to one third
of all commands. With two replicas, every replica has to reply to half of all
commands. In the next compartmentalization, we'll see another major advantage
of increasing the number of replicas.

{
\tikzstyle{proc}=[draw, circle, thick, inner sep=2pt]
\tikzstyle{newproc}=[draw=flatgreen]
\tikzstyle{client}=[proc, fill=clientcolor!25]
\tikzstyle{proposer}=[proc, fill=proposercolor!25]
\tikzstyle{proxyleader}=[proc, fill=proxyleadercolor!25]
\tikzstyle{acceptor}=[proc, fill=acceptorcolor!25]
\tikzstyle{replica}=[proc, fill=replicacolor!25]

\tikzstyle{proclabel}=[inner sep=0pt, align=center, font=\footnotesize]

\tikzstyle{component}=[draw, thick, flatgray, rounded corners]

\tikzstyle{comm}=[-latex, thick]
\tikzstyle{commnum}=[fill=white, inner sep=0pt]
\tikzstyle{newcomm}=[comm, flatgreen]
\tikzstyle{oldcomm}=[comm, black!50]

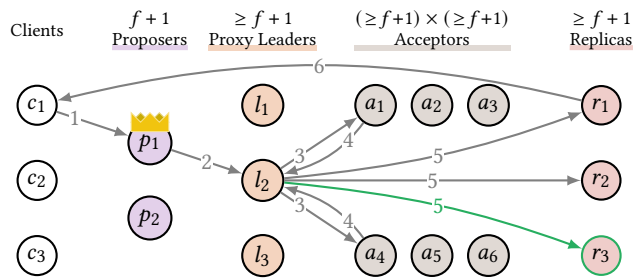
\begin{figure}[ht]
  \centering
  \begin{tikzpicture}[xscale=1.5]
    \node[client] (c1) at (0, 2) {$c_1$};
    \node[client] (c2) at (0, 1) {$c_2$};
    \node[client] (c3) at (0, 0) {$c_3$};
    \node[proposer] (p1) at (1, 1.5) {$p_1$};
    \node[proposer] (p2) at (1, 0.5) {$p_2$};
    \node[proxyleader] (pl1) at (2, 2) {$l_1$};
    \node[proxyleader] (pl2) at (2, 1) {$l_2$};
    \node[proxyleader] (pl3) at (2, 0) {$l_3$};
    \node[acceptor] (a1) at (3, 2) {$a_1$};
    \node[acceptor] (a2) at (3.5, 2) {$a_2$};
    \node[acceptor] (a3) at (4, 2) {$a_3$};
    \node[acceptor] (a4) at (3, 0) {$a_4$};
    \node[acceptor] (a5) at (3.5, 0) {$a_5$};
    \node[acceptor] (a6) at (4, 0) {$a_6$};
    \node[replica] (r1) at (5, 2) {$r_1$};
    \node[replica] (r2) at (5, 1) {$r_2$};
    \node[replica, newproc] (r3) at (5, 0) {$r_3$};

    \crown{(p1.north)++(0,-0.15)}{0.333}{0.25}
    \node[proclabel] (clients) at (0, 3) {Clients};
    \node[proclabel] (proposers) at (1, 3) {$f+1$\\Proposers};
    \node[proclabel] (proxyleaders) at (2, 3) {$\geq f+1$\\Proxy Leaders};
    \node[proclabel] (acceptors) at (3.5, 3) {$(\geq\!f\!+\!1) \times (\geq\!f\!+\!1)$\\Acceptors};
    \node[proclabel] (replicas) at (5, 3) {$\geq f+1$\\Replicas};
    \halffill{clients}{clientcolor!25}
    \quarterfill{proposers}{proposercolor!25}
    \quarterfill{proxyleaders}{proxyleadercolor!25}
    \quarterfill{acceptors}{acceptorcolor!25}
    \quarterfill{replicas}{replicacolor!25}

    \draw[oldcomm, near start] (c1) to node[commnum]{1} (p1);
    \draw[oldcomm] (p1) to node[commnum]{2} (pl2);
    \draw[oldcomm, near start] (pl2) to node[commnum]{3} (a1);
    \draw[oldcomm, near start] (pl2) to node[commnum]{3} (a4);
    \draw[oldcomm, near start, bend left=20] (a1) to node[commnum]{4} (pl2);
    \draw[oldcomm, near start, bend right=20] (a4) to node[commnum]{4} (pl2);
    \draw[oldcomm, bend right=10] (pl2) to node[commnum]{5} (r1);
    \draw[oldcomm] (pl2) to node[commnum]{5} (r2);
    \draw[newcomm, bend left=10] (pl2) to node[commnum]{5} (r3);
    \draw[oldcomm, bend right=20] (r1) to node[commnum]{6} (c1);
  \end{tikzpicture}
  \caption{%
    An example execution of Compartmentalized MultiPaxos with three replicas as
    opposed to the minimum required two ($f=1$).
  }%
  \figlabel{MoreReplicas}
\end{figure}}

\paragraph{Discussion}
Again decoupling \emph{enables} scaling. Without decoupling the leader and
introducing proxy leaders, increasing the number of replicas hurts rather than
helps performance.

\subsection{Compartmentalization 4: Leaderless Reads}\seclabel{MultiPaxos/LeaderlessReads}
\bds{leader and replicas}
    {read path and write path}
    {the number of read quorums}

\paragraph{Bottleneck}
We have now compartmentalized the leader, the acceptors, and the replicas. At
this point, the bottleneck is in one of two places. Either the leader is still
a bottleneck, or the replicas are the bottleneck. Fortunately, we can bypass
both bottlenecks with a single compartmentalization.

\paragraph{Decouple}
We call commands that modify the state of the state machine \defword{writes}
and commands that don't modify the state of the state machine \defword{reads}.
The leader must process every write because it has to linearize the writes with
respect to one another, and \emph{every} replica must process every write
because otherwise the replicas' state would diverge (imagine if one replica
performs a write but the other replicas don't). However, because reads do not
modify the state of the state machine, the leader does not have to linearize
them (reads commute), and only a single replica (as opposed to every replica)
needs to execute a read.

We take advantage of this observation by decoupling the read path from the
write path. Writes are processed as before, but we bypass the leader and
perform a read on a single replica by using the idea from Paxos Quorum Reads
(PQR)~\cite{charapko2019linearizable}.
Specifically, to perform a read, a client sends a \msg{PreRead}{} message to a
read quorum of acceptors. Upon receiving a \msg{PreRead}{} message, an acceptor
$a_i$ returns a \msg{PreReadAck}{w_i} message where $w_i$ is the index of the
largest log entry in which the acceptor has voted (i.e.\ the largest log entry
in which the acceptor has sent a \msgfont{Phase2b} message). We call this $w_i$
a vote watermark. When the client receives \msgfont{PreReadAck} messages from a
read quorum of acceptors, it computes $i$ as the maximum of all received vote
watermarks. It then sends a \msg{Read}{x, i} request to any one of the replicas
where $x$ is an arbitrary read (i.e.\ a command that does not modify the state
of the state machine).

When a replica receives a \msg{Read}{x, i} request from a client, it waits
until it has executed the command in log entry $i$. Recall that replicas
execute commands in log order, so if the replica has executed the
command in log entry $i$, then it has also executed all of the commands in log
entries less than $i$. After the replica has executed the command in log entry
$i$, it executes $x$ and returns the result to the client. Note that upon
receiving a \msg{Read}{x, i} message, a replica may have already executed the
log beyond $i$. That is, it may have already executed the commands in log
entries $i+1$, $i+2$, and so on. This is okay because as long as the replica
has executed the command in log entry $i$, it is safe to execute $x$.
\iftoggle{techreportenabled}{}{%
  See our technical report~\cite{whittaker2020scaling} for a proof that this
  protocol correctly implements linearizable reads.
}

{
\tikzstyle{proc}=[draw, circle, thick, inner sep=2pt]
\tikzstyle{newproc}=[draw=flatred]
\tikzstyle{client}=[proc, fill=clientcolor!25]
\tikzstyle{proposer}=[proc, fill=proposercolor!25]
\tikzstyle{proxyleader}=[proc, fill=proxyleadercolor!25]
\tikzstyle{acceptor}=[proc, fill=acceptorcolor!25]
\tikzstyle{replica}=[proc, fill=replicacolor!25]

\tikzstyle{proclabel}=[inner sep=0pt, align=center, font=\footnotesize]

\tikzstyle{component}=[draw, thick, flatgray, rounded corners]

\tikzstyle{comm}=[-latex, thick]
\tikzstyle{readcomm}=[red, comm, dash pattern=on 2pt off 1pt]
\tikzstyle{writecomm}=[blue, comm]
\tikzstyle{commnum}=[fill=white, text=black, inner sep=0pt]
\tikzstyle{newcomm}=[comm, flatred]
\tikzstyle{newcommnum}=[commnum, flatred]
\tikzstyle{oldcomm}=[comm, black!50]
\tikzstyle{oldcommnum}=[commnum, black!50]

\begin{figure}[ht]
  \centering
  \begin{tikzpicture}[xscale=1.5]
    \node[client] (c1) at (0, 2) {$c_1$};
    \node[client] (c2) at (0, 1) {$c_2$};
    \node[client] (c3) at (0, 0) {$c_3$};
    \node[proposer] (p1) at (1, 1.5) {$p_1$};
    \node[proposer] (p2) at (1, 0.5) {$p_2$};
    \node[proxyleader] (pl1) at (2, 2) {$l_1$};
    \node[proxyleader] (pl2) at (2, 1) {$l_2$};
    \node[proxyleader] (pl3) at (2, 0) {$l_3$};
    \node[acceptor] (a1) at (3, 2) {$a_1$};
    \node[acceptor] (a2) at (4, 2) {$a_2$};
    \node[acceptor] (a3) at (3, 0) {$a_3$};
    \node[acceptor] (a4) at (4, 0) {$a_4$};
    \node[replica] (r1) at (5, 2) {$r_1$};
    \node[replica] (r2) at (5, 1) {$r_2$};
    \node[replica] (r3) at (5, 0) {$r_3$};

    \crown{(p1.north)++(0,-0.15)}{0.333}{0.25}
    \node[proclabel] (clients) at (0, 3) {Clients};
    \node[proclabel] (proposers) at (1, 3) {$f+1$\\Proposers};
    \node[proclabel] (proxyleaders) at (2, 3) {$\geq f+1$\\Proxy Leaders};
    \node[proclabel] (acceptors) at (3.5, 3) {$(\geq\!f\!+\!1) \times (\geq\!f\!+\!1)$\\Acceptors};
    \node[proclabel] (replicas) at (5, 3) {$\geq f+1$\\Replicas};
    \halffill{clients}{clientcolor!25}
    \quarterfill{proposers}{proposercolor!25}
    \quarterfill{proxyleaders}{proxyleadercolor!25}
    \quarterfill{acceptors}{acceptorcolor!25}
    \quarterfill{replicas}{replicacolor!25}


    \draw[readcomm] (c3) to node[commnum]{1} (a3);
    \draw[readcomm, near start, bend right=15] (c3) to node[commnum]{1} (a4);
    \draw[readcomm, near start, bend right=20] (a3) to node[commnum]{2} (c3);
    \draw[readcomm, near start, bend left=30] (a4) to node[commnum]{2} (c3);
    \draw[readcomm, near start, bend right=40] (c3) to node[commnum]{3} (r3);
    \draw[readcomm, near start, bend left=55] (r3) to node[commnum]{4} (c3);

    \draw[writecomm, near start] (c1) to node[commnum]{1} (p1);
    \draw[writecomm] (p1) to node[commnum]{2} (pl2);
    \draw[writecomm, near start] (pl2) to node[commnum]{3} (a1);
    \draw[writecomm, near start] (pl2) to node[commnum]{3} (a3);
    \draw[writecomm, near start, bend left=20] (a1) to node[commnum]{4} (pl2);
    \draw[writecomm, near start, bend right=20] (a3) to node[commnum]{4} (pl2);
    \draw[writecomm, bend right=10] (pl2) to node[commnum]{5} (r1);
    \draw[writecomm] (pl2) to node[commnum]{5} (r2);
    \draw[writecomm, bend left=10] (pl2) to node[commnum]{5} (r3);
    \draw[writecomm, bend right=15] (r1) to node[commnum]{6} (c1);
  \end{tikzpicture}
  \caption{%
    An example execution of Compartmentalized MultiPaxos' read and write path
    ($f=1$) with a $2 \times 2$ acceptor grid. The write path is shown using
    solid blue lines. The read path is shown using red dashed lines.
  }%
  \figlabel{LeaderlessReads}
\end{figure}
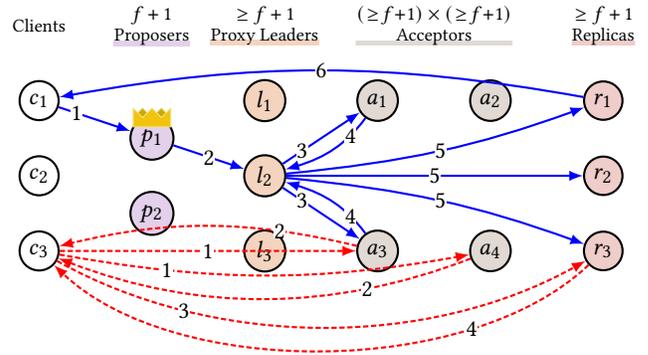}

\paragraph{Scale}
The decoupled read and write paths are shown in \figref{LeaderlessReads}. Reads
are sent to a row (read quorum) of acceptors, so we can increase the number of
rows to decrease the read load on every individual acceptor, eliminating the
acceptors as a read bottleneck. Reads are also sent to a single replica, so we
can increase the number of replicas to eliminate them as a read bottleneck as
well.

\paragraph{Discussion}
Note that read-heavy workloads are not a special case. Many workloads are
read-heavy~\cite{ghemawat2003google, nishtala2013scaling, atikoglu2012workload,
moraru2013there}. Chubby~\cite{burrows2006chubby} observes that fewer than 1\%
of operations are writes, and Spanner~\cite{corbett2013spanner} observes that
fewer than 0.3\% of operations are writes.

Also note that increasing the number of columns in an acceptor grid reduces the
write load on the acceptors, and increasing the number of rows in an acceptor
grid reduces the read load on the acceptors. There is no throughput trade-off
between the two. The number of rows and columns can be adjusted independently.
Increasing read throughput (by increasing the number of rows) does not decrease
write throughput, and vice versa. However, increasing the number of rows
does increase the \emph{size} (but not number) of columns, so increasing the
number of rows might increase the tail latency of writes, and vice versa.

\iftoggle{techreportenabled}{\subsection{Correctness}
We now define linearizability and prove that our protocol implements
linearizable reads.

Linearizability is a correctness condition for distributed
systems~\cite{herlihy1990linearizability}.  Intuitively, a linearizable
distributed system is indistinguishable from a system running on a single
machine that services all requests serially. This makes a linearizable system
easy to reason about. We first explain the intuition behind linearizability
and then formalize the intuition.

Consider a distributed system that implements a single register. Clients can
send requests to the distributed system to read or write the register. After a
client sends a read or write request, it waits to receive a response before
sending another request. As a result, a client can have at most one operation
pending at any point in time.

{\begin{figure*}[ht]
  \centering

  \begin{subfigure}[b]{0.3\textwidth}
    \centering
    \begin{tikzpicture}[xscale=0.8]
      \drawclients{2}{5.5};
      \operation{1}{0}{2}{$w(0)$}{OK};
      \operation{2}{1}{3}{$w(1)$}{OK};
      \operation{1}{4}{5}{$r()$}{$0$};
    \end{tikzpicture}
    \caption{An example execution}\figlabel{LinearizableExampleNoLin}
  \end{subfigure}\hspace{12pt}
  \begin{subfigure}[b]{0.3\textwidth}
    \centering
    \begin{tikzpicture}[xscale=0.8]
      \drawclients{2}{5.5};
      \operation{1}{0}{2}{$w(0)$}{OK};
      \operation{2}{1}{3}{$w(1)$}{OK};
      \operation{1}{4}{5}{$r()$}{$0$};
      \notlinpoint{1}{1}{p0};
      \notlinpoint{2}{2}{p1};
      \notlinpoint{1}{4.5}{p2};
      \draw[notlinline, -latex] (p0) to (p1);
      \draw[notlinline, -latex] (p1) to (p2);
    \end{tikzpicture}
    \caption{An incorrect linearization}\figlabel{LinearizableExampleBadLin}
  \end{subfigure}\hspace{12pt}
  \begin{subfigure}[b]{0.3\textwidth}
    \centering
    \begin{tikzpicture}[xscale=0.8]
      \drawclients{2}{5.5};
      \operation{1}{0}{2}{$w(0)$}{OK};
      \operation{2}{1}{3}{$w(1)$}{OK};
      \operation{1}{4}{5}{$r()$}{$0$};
      \linpoint{2}{1.25}{p0};
      \linpoint{1}{1.75}{p1};
      \linpoint{1}{4.5}{p2};
      \draw[linline, -latex] (p0) to (p1);
      \draw[linline, -latex, bend right] (p1) to (p2);
    \end{tikzpicture}
    \caption{A linearization}\figlabel{LinearizableExampleGoodLin}
  \end{subfigure}

  \caption{}\figlabel{LinearizableExample}
\end{figure*}}

As a simple example, consider the execution illustrated in
\figref{LinearizableExampleNoLin} where the $x$-axis represents the passage of
time (real time, not logical time~\cite{lamport2019time}). This execution
involves two clients, $c_1$ and $c_2$. Client $c_1$ sends a $w(0)$ request to
the system, requesting that the value $0$ be written to the register. Then,
client $c_2$ sends a $w(1)$ request, requesting that the value $1$ be written
to the register. The system then sends acknowledgments to $c_1$ and $c_2$
before $c_1$ sends a read request and receives the value $0$.

For every client request, let's associate the request with a point in time that
falls between when the client sent the request and when the client received the
corresponding response. Next, let us imagine that the system executes every
request instantaneously at the point in time associated with the request. This
hypothetical execution may or may not be consistent with the real execution.

For example, in \figref{LinearizableExampleNoLin}, we have associated every
request with a point halfway between its invocation and response. Thus, in this
hypothetical execution, the system executes $c_1$'s $w(0)$ request, then
$c_2$'s $w(1)$ request, and finally $c_1$'s $r()$ request. In other words, it
writes $0$ into the register, then $1$, and then reads the value $1$ (the
latest value written). This hypothetical execution is \emph{not} consistent
with the real execution because $c_1$ reads $1$ instead of $0$.

Now consider the hypothetical execution in \figref{LinearizableExampleGoodLin}
in which we execute $w(1)$, then $w(0)$, and then $r()$. This execution is
consistent with the real execution. Note that $c_1$ reads $0$ in both
executions. Such a hypothetical execution---one that is consistent with the
real execution---is called a \defword{linearization}. Note that from the
clients' perspective, the real execution is indistinguishable from its
linearization. Maybe the distributed register really is executing our requests
at exactly the points in time that we selected? There's no way for the clients
to prove otherwise.

If an execution has a linearization, we say the execution is
\defword{linearizable}. Similarly, if a system only allows linearizable
executions, we say the system is linearizable. Note that not every execution is
linearizable. The execution in \figref{NotLinearizableExample}, for example, is
not linearizable. Try to find a linearization. You'll see that it's impossible.

{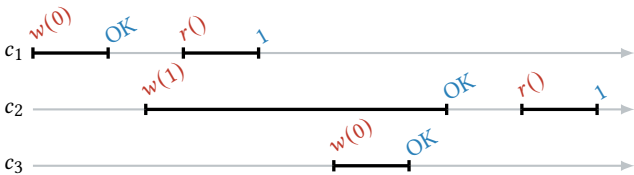
\begin{figure}[ht]
  \centering
  \begin{tikzpicture}[yscale=0.75]
    \drawclients{3}{8};
    \operation{1}{0}{1}{$w(0)$}{OK};
    \operation{1}{2}{3}{$r()$}{$1$};
    \operation{2}{1.5}{5.5}{$w(1)$}{OK};
    \operation{2}{6.5}{7.5}{$r()$}{$1$};
    \operation{3}{4}{5}{$w(0)$}{OK};
  \end{tikzpicture}
  \caption{An execution that is not linearizable}%
  \figlabel{NotLinearizableExample}
\end{figure}}

\newcommand{\invocation}[2]{#2.\textcolor{flatred}{#1}}
\newcommand{\response}[2]{#2.\textcolor{flatblue}{#1}}
\newcommand{\subhistory}[2]{#1\,|\,#2}

We now formalize our intuition on
linearizability~\cite{herlihy1990linearizability}. A \defword{history} is a
finite sequence of operation \defword{invocation} and \defword{response}
events. For example, the following history:
\[
  H_{wwr} =
  \invocation{w(0)}{c_1};\>
  \invocation{w(1)}{c_2};\>
  \response{\text{OK}}{c_1};\>
  \response{\text{OK}}{c_2};\>
  \invocation{r()}{c_1};\>
  \response{0}{c_1}
\]
is the history illustrated in \figref{LinearizableExampleNoLin}. We draw
invocation events in red, and response events in blue. We call an invocation
and matching response an \defword{operation}. In $H_{wwr}$, every invocation is
followed eventually by a corresponding response, but this is not always the
case. An invocation in a history is \defword{pending} if there does not exist a
corresponding response. For example, in the history $H_\text{pending}$ below,
$c_2$'s invocation is pending:
\[
  H_{\text{pending}} =
  \invocation{w(0)}{c_1};\>
  \invocation{w(1)}{c_2};\>
  \response{\text{OK}}{c_1};\>
  \invocation{r()}{c_1};\>
  \response{0}{c_1}
\]
$H_{\text{pending}}$ is illustrated in \figref{PendingLinearizableExample}.
complete$(H)$ is the subhistory of $H$ that only includes non-pending
operations. For example,
\[
  \text{complete}(H_{\text{pending}}) =
  \invocation{w(0)}{c_1};\>
  \response{\text{OK}}{c_1};\>
  \invocation{r()}{c_1};\>
  \response{0}{c_1}
\]

{\begin{figure}[ht]
  \centering
  \begin{tikzpicture}[yscale=0.75]
    \drawclients{2}{5.5};
    \operation{1}{0}{2}{$w(0)$}{OK};
    \pendingoperation{2}{1}{5}{$w(1)$};
    \operation{1}{4}{5}{$r()$}{$0$};
  \end{tikzpicture}
  \caption{A history, $H_\text{pending}$, with a pending invocation}%
  \figlabel{PendingLinearizableExample}
\end{figure}
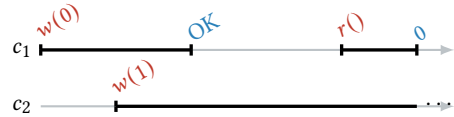
}

A \defword{client subhistory}, $\subhistory{H}{c_i}$, of a history $H$ is the
subsequence of all events in $H$ associated with client $c_i$. Referring again
to $H_{wwr}$ above, we have:
\begin{gather*}
  \subhistory{H_{wwr}}{c_1}
    = \invocation{w(0)}{c_1};\>
      \response{\text{OK}}{c_1};\>
      \invocation{r()}{c_1};\>
      \response{0}{c_1} \\
  \subhistory{H_{wwr}}{c_2}
    = \invocation{w(1)}{c_2};\>
      \response{\text{OK}}{c_2}
\end{gather*}
$\subhistory{H_{wwr}}{c_1}$ is illustrated in \figref{SubhistoryExample}.

{\begin{figure}[ht]
  \centering

  \begin{tikzpicture}
    \drawclients{1}{5.5};
    \operation{1}{0}{2}{$w(0)$}{OK};
    \operation{1}{4}{5}{$r()$}{$0$};
  \end{tikzpicture}

  \caption{$\subhistory{H_{wwr}}{c_1}$}\figlabel{SubhistoryExample}
\end{figure}
}

Two histories $H$ and $H'$ are \defword{equivalent} if for every client $c_i$,
$\subhistory{H}{c_i} = \subhistory{H'}{c_i}$. For example, consider the
following history:
\[
  H_{wrw} =
  \invocation{w(0)}{c_1};\>
  \response{\text{OK}}{c_1};\>
  \invocation{r()}{c_1};\>
  \invocation{w(1)}{c_2};\>
  \response{0}{c_1};\>
  \response{\text{OK}}{c_2}
\]
$H_{wrw}$ is illustrated in \figref{EquivalentExample}. $H_{wwr}$ is
equivalent to $H_{wrw}$ because
\begin{gather*}
  \subhistory{H_{wwr}}{c_1}
    = \invocation{w(0)}{c_1};\>
      \response{\text{OK}}{c_1};\>
      \invocation{r()}{c_1};\>
      \response{0}{c_1}
    = \subhistory{H_{wrw}}{c_1} \\
  \subhistory{H_{wwr}}{c_2}
    = \invocation{w(1)}{c_2};\>
      \response{\text{OK}}{c_2};\>
    = \subhistory{H_{wrw}}{c_2}
\end{gather*}

{\begin{figure}[ht]
  \centering
  \begin{tikzpicture}[yscale=0.75]
    \drawclients{2}{5.5};
    \operation{1}{0}{1}{$w(0)$}{OK};
    \operation{1}{2}{4}{$r()$}{$0$};
    \operation{2}{3}{5}{$w(1)$}{OK};
  \end{tikzpicture}
  \caption{$H_{wrw}$}\figlabel{EquivalentExample}
\end{figure}
}

A history $H$ induces an irreflexive partial order $<_H$ on operations where
$o_1 <_H o_2$ if the response of $o_1$ precedes the invocation of $o_2$ in $H$.
If $o_1 <_H o_2$, we say $o_1$ \defword{happens before} $o_2$. In $H_{wwr}$ for
example, $c_2$'s operation happens before $c_1$'s second operation. In
$H_{wrw}$, on the other hand, the two operations are not ordered by the happens
before relation. This shows that equivalent histories may not have the same
happens before relation.

Finally, a history $H$ is \defword{linearizable} if it can be extended (by
appending zero or more response events) to some history $H'$ such that (a)
complete($H'$) is equivalent to some sequential history $S$, and (b) $<_S$
respects $<_H$ (i.e.\ if two operations are ordered in $H$, they must also be
ordered in $S$). $S$ is called a \defword{linearization}. The history
$H_{wwr}$, for example, is linearizable with the linearization
\[
  S_{wwr} =
  \invocation{w(1)}{c_2};\>
  \response{\text{OK}}{c_2};\>
  \invocation{w(0)}{c_1};\>
  \response{\text{OK}}{c_1};\>
  \invocation{r()}{c_1};\>
  \response{0}{c_1}
\]
illustrated in \figref{LinearizationExample}

{\begin{figure}[ht]
  \centering
  \begin{tikzpicture}[yscale=0.75]
    \drawclients{2}{5.5};
    \operation{2}{0}{1}{$w(1)$}{OK};
    \operation{1}{2}{3}{$w(0)$}{OK};
    \operation{1}{4}{5}{$r()$}{$0$};
  \end{tikzpicture}
  \caption{$S_{wrw}$}\figlabel{LinearizationExample}
\end{figure}
}

We now prove that our protocol correctly implements linearizable reads.

\newcommand{\complete}{\text{complete}}
\begin{proof}
  Let $H$ be an arbitrary history permitted by our protocol. To prove that our
  protocol is linearizable, we must extend $H$ to a history $H'$ such that
  $\complete(H')$ is equivalent to a sequential history that respects $<_H$.

  Recall that extending $H$ to $H'$ is sometimes necessary because of
  situations like the one shown in \figref{WhyExtendHistory}. This example
  involves a single register with an initial value of 0. $c_1$ issues a request
  to write the value of 1, but has not yet received a response. $c_2$ issues a
  read request and receives the value 1. If we do not extend the history to
  include a response to $c_1$'s write, then there will not exist an equivalent
  sequential history.

  {\begin{figure}[ht]
  \centering
  \begin{tikzpicture}[yscale=0.75]
    \drawclients{2}{3.5};
    \pendingoperation{1}{0}{3}{$w(1)$}
    \operation{2}{1}{2}{$r()$}{$1$}
  \end{tikzpicture}
  \caption{A motivating example of history extension}\figlabel{WhyExtendHistory}
\end{figure}
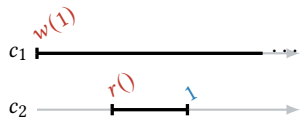
}

  So, which operations should we include in $H'$? Let $k$ be the largest log
  index written in or read from in $\complete(H)$. First note that for every
  index $0 \leq i \leq k$, there exists a (potentially pending) write in $H$
  that has been chosen in index $i$. Why?  Well, our protocol executes commands
  in log order, so a write at index $k$ can only complete after all writes with
  smaller indices have been chosen (and executed by some replica). Similarly,
  if a read operation reads from slot $k$, then the write in slot $k$ must have
  been executed, so again all writes with smaller indices have also been
  chosen.
  We extend $H$ to history $H'$ by including responses for all pending write
  invocations with indices $0 \leq i \leq k$. The responses are formed by
  executing the $k + 1$ commands in log order.

  For example, consider the history $G$ shown in \figref{ExtendExample}. $w_i$
  represents a write chosen in log index $i$, $r_i$ represents a read operation
  that reads from slot $i$, $w_?$ represents a pending write which has not been
  chosen in any particular log index, and $r_?$ represents a pending read.
  $\complete(G)$ includes $w_1$ and $r_2$, so here $k=2$ and we must include
  all writes in indices $0$, $1$, and $2$. That is, we extend $G$ to complete
  $w_0$ and $w_2$. $w_4$ is left pending, as is $w_?$ and $r_?$. Also note
  that we could not complete $w_4$ even if we wanted to because there is no
  $w_3$.

  {\begin{figure}[ht]
  \begin{tikzpicture}[yscale=0.75]
    \drawclients{5}{7.5}
    \pendingoperation{1}{0}{7}{$w_0$}
    \operation{2}{1}{2}{$w_1$}{}
    \pendingoperation{2}{3}{7}{$w_2$}
    \operation{3}{2.5}{3.5}{$r_2$}{}
    \pendingoperation{3}{4}{7}{$w_4$}
    \operation{4}{4}{5}{$r_2$}{}
    \pendingoperation{4}{6}{7}{$w_?$}
    \pendingoperation{5}{2}{7}{$r_?$}
  \end{tikzpicture}
  \caption{%
    An example history $G$. Responses are not shown, as they are not important
    for this example.
  }\figlabel{ExtendExample}
\end{figure}
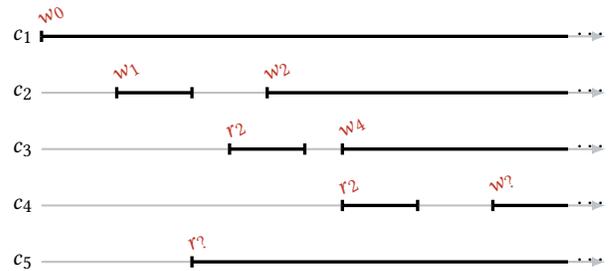}

  Now, we must prove that (1) $\complete(H')$ is equivalent to some legal
  sequential history $S$, and (2) $<_S$ respects $<_H$. We let $S$ be the
  sequential history formed from executing all writes in log order and from
  executing every read from index $i$ after the write in index $i$.
  If there are multiple reads from index $i$, the reads are ordered in an
  arbitrary way that respects $<_H$. For example, the history $G$ in
  \figref{ExtendExample} has the sequential history $S_G$ shown in
  \figref{ExampleS}. Note that $c_4$'s read comes after $c_3$'s read. This is
  essential because we must respect $<_G$. If the two reads were concurrent in
  $G$, they could be ordered arbitrarily in $S_G$.

  {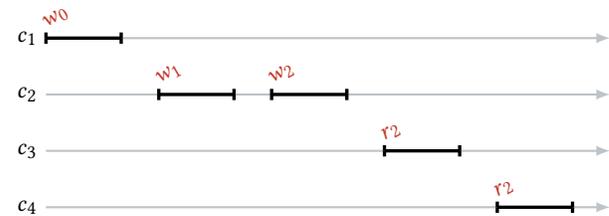
\begin{figure}[ht]
  \centering
  \begin{tikzpicture}[yscale=0.75]
    \drawclients{4}{7.5}
    \operation{1}{0}{1}{$w_0$}{}
    \operation{2}{1.5}{2.5}{$w_1$}{}
    \operation{2}{3}{4}{$w_2$}{}
    \operation{3}{4.5}{5.5}{$r_2$}{}
    \operation{4}{6}{7}{$r_2$}{}
  \end{tikzpicture}
  \caption{A linearization $S_G$ of the history in $G$ \figref{ExtendExample}}%
  \figlabel{ExampleS}
\end{figure}}

  To prove (1) and (2), we show that if two distinct operations $x$ and $y$
  that write to (or read from) log indices $i$ and $j$ are related in
  $H$---i.e. $x <_H y$, or $x$ finishes before $y$ begins---then $i \leq j$. We
  perform a case analysis on whether $x$ and $y$ are reads or writes.

  \begin{itemize}
    \item \textbf{$x$ and $y$ are both writes:}
      At the time $x$ completes in index $i$, all commands in indices less than
      $i$ have been chosen because our protocol executes commands in log order.
      Thus, when $y$ later begins, it cannot be chosen in a log entry less than
      $i$, since every log entry implements consensus. Thus, $i < j$.

    \item \textbf{$x$ and $y$ are both reads:}
      When $x$ completes, command $i$ has been chosen. Thus, some write quorum
      $w$ of acceptors must have voted for the command in log entry $i$. When
      $y$ begins, it sends \msgfont{PreRead} messages to some read quorum $r$
      of acceptors. $r$ and $w$ intersect, so the client executing $y$ will
      receive a \msg{PreReadAck}{w_i} message from some acceptor in $r$ with
      $w_i \geq i$. Therefore, $y$ is guaranteed to read from some $j \geq i$.

    \item \textbf{$x$ is a read and $y$ is a write:}
      When $x$ completes, all commands in indices $i$ and smaller have been
      chosen. By the first case above, $y$ must be chosen in some index $j >
      i$.

    \item \textbf{$x$ is a write and $y$ is a read:}
      When $x$ completes, command $i$ has been chosen. As with the second case
      above, when $y$ begins it will contact an acceptor group with a vote
      watermark at least as large as $i$ and will subsequently read from at
      least $i$.
  \end{itemize}

  From this, (1) is immediate since every client's operations are in the same
  order in $H'$ and in $S$. (2) holds because $S$ is ordered by log index with
  ties broken respecting $<_H$, so if $x <_H y$, then $i \leq j$ and $x <_S y$.
\end{proof}

}{}
\iftoggle{techreportenabled}{\subsection{Non-Linearizable Reads}\seclabel{NonLinearizableReads}
Our protocol implements linearizable reads, the strongest form of
non-transactional consistency. However, we can extend the protocol to support
reads with better performance but weaker consistency. Notably, we can
implement sequentially consistent~\cite{lamport1979make} and eventually
consistent reads. Writes are always linearizable. The decision of which
consistency level to choose depends on the application.

\textbf{Sequentially Consistent Reads.}
Sequential consistency is a lot like linearizability but without the real-time
ordering requirements. Specifically, a history $H$ is sequentially consistent
if we can extend it to some history $H'$ such that complete($H'$) is equivalent
to some sequential history $S$. Unlike with linearizability, we do not require
that $<_S$ respects $<_H$.

To implement sequentially consistent reads, every client needs to (a) keep
track of the largest log entry it has ever written to or read from, and (b)
make sure that all future operations write to or read from a log entry as least
as large. Concretely, we make the following changes:
\begin{itemize}
  \item
    Every client $c_i$ maintains an integer-valued watermark $w_i$, initially
    $-1$.

  \item
    When a replica executes a write $w$ in log entry $j$ and returns the result
    of executing $w$ to a client $c_i$, it also includes $j$. When $c_i$
    receives a write index $j$ from a replica, it updates $w_i$ to the max of
    $w_i$ and $j$.

  \item
    To execute a sequentially consistent read $r$, a client $c_i$ sends a
    \msg{Read}{r, w_i} message to any replica. The replica waits until it has
    executed the write in log entry $w_i$ and then executes $r$. It then
    replies to the client with the result of executing $r$ and the log entry
    $j$ from which $r$ reads. Here, $j \geq w_i$. When a client receives a read
    index $j$, it updates $w_i$ to the max of $w_i$ and $j$.
\end{itemize}

Note that a client can finish a sequentially consistent read after one
round-trip of communication (in the best case), whereas a linearizable read
requires at least two. Moreover, sequentially consistent reads do not involve
the acceptors. This means that we can increase read throughput by scaling up
the number of replicas without having to scale up the number of acceptors. Also
note that sequentially consistent reads are also causally consistent.

\textbf{Eventually Consistent Reads.}
Eventually consistent reads are trivial to implement. To execute an eventually
consistent read, a client simply sends the read request $r$ directly to any
replica. The replica executes the read immediately and returns the result back
to the client. Eventually consistent reads do not require any watermark
bookkeeping, do not involve acceptors, and never wait for writes. Moreover, the
reads are always executed against a consistent prefix of the log.
}{}
}
{\section{Batching}
All state machine replication protocols, including MultiPaxos, can take
advantage of batching to increase throughput. The standard way to implement
batching~\cite{santos2012tuning, santos2013optimizing} is to have clients send
their commands to the leader and to have the leader group the commands
together into batches, as shown in \figref{Batching}. The rest of the protocol
remains unchanged, with command batches replacing commands. The one notable
difference is that replicas now execute one batch of commands at a time, rather
than one command at a time. After executing a single command, a replica has to
send back a single result to a client, but after executing a batch of commands,
a replica has to send a result to every client with a command in the batch.

{
\tikzstyle{proc}=[draw, circle, thick, inner sep=2pt]
\tikzstyle{newproc}=[draw=flatgreen]
\tikzstyle{client}=[proc, fill=clientcolor!25]
\tikzstyle{proposer}=[proc, fill=proposercolor!25]
\tikzstyle{proxyleader}=[proc, fill=proxyleadercolor!25]
\tikzstyle{acceptor}=[proc, fill=acceptorcolor!25]
\tikzstyle{replica}=[proc, fill=replicacolor!25]

\tikzstyle{proclabel}=[inner sep=0pt, align=center, font=\footnotesize]

\tikzstyle{component}=[draw, thick, flatgray, rounded corners]

\tikzstyle{comm}=[-latex, thick]
\tikzstyle{batched}=[ultra thick]
\tikzstyle{commnum}=[fill=white, inner sep=0pt]
\tikzstyle{newcomm}=[comm, flatgreen]
\tikzstyle{newcommnum}=[commnum, flatgreen]
\tikzstyle{oldcomm}=[comm, black!50]
\tikzstyle{oldcommnum}=[commnum, black!50]

\begin{figure}[ht]
  \centering
  \begin{tikzpicture}[xscale=1.5]
    \node[client] (c1) at (0, 2) {$c_1$};
    \node[client] (c2) at (0, 1) {$c_2$};
    \node[client] (c3) at (0, 0) {$c_3$};
    \node[proposer] (p1) at (1, 1.5) {$p_1$};
    \node[proposer] (p2) at (1, 0.5) {$p_2$};
    \node[proxyleader] (pl1) at (2, 2) {$l_1$};
    \node[proxyleader] (pl2) at (2, 1) {$l_2$};
    \node[proxyleader] (pl3) at (2, 0) {$l_3$};
    \node[acceptor] (a1) at (3, 2) {$a_1$};
    \node[acceptor] (a2) at (4, 2) {$a_2$};
    \node[acceptor] (a3) at (3, 0) {$a_3$};
    \node[acceptor] (a4) at (4, 0) {$a_4$};
    \node[replica] (r1) at (5, 2) {$r_1$};
    \node[replica] (r2) at (5, 1) {$r_2$};
    \node[replica] (r3) at (5, 0) {$r_3$};

    \crown{(p1.north)++(0,-0.15)}{0.333}{0.25}
    \node[proclabel] (clients) at (0, 3) {Clients};
    \node[proclabel] (proposers) at (1, 3) {$f+1$\\Proposers};
    \node[proclabel] (proxyleaders) at (2, 3) {$\geq f+1$\\Proxy Leaders};
    \node[proclabel] (acceptors) at (3.5, 3) {$(\geq\!f\!+\!1) \times (\geq\!f\!+\!1)$\\Acceptors};
    \node[proclabel] (replicas) at (5, 3) {$\geq f+1$\\Replicas};
    \halffill{clients}{clientcolor!25}
    \quarterfill{proposers}{proposercolor!25}
    \quarterfill{proxyleaders}{proxyleadercolor!25}
    \quarterfill{acceptors}{acceptorcolor!25}
    \quarterfill{replicas}{replicacolor!25}

    \draw[newcomm] (c1) to node[commnum, near start]{1} (p1);
    \draw[newcomm] (c2) to node[commnum, near start]{1} (p1);
    \draw[newcomm] (c3) to node[commnum, near start]{1} (p1);
    \draw[oldcomm, batched] (p1) to node[commnum]{2} (pl2);
    \draw[oldcomm, batched, near start] (pl2) to node[commnum]{3} (a1);
    \draw[oldcomm, batched, near start] (pl2) to node[commnum]{3} (a3);
    \draw[oldcomm, batched, near start, bend right=20] (a1) to node[commnum]{4} (pl2);
    \draw[oldcomm, batched, near start, bend left=20] (a3) to node[commnum]{4} (pl2);
    \draw[oldcomm, batched, bend right=10] (pl2) to node[commnum]{5} (r1);
    \draw[oldcomm, batched] (pl2) to node[commnum]{5} (r2);
    \draw[oldcomm, batched, bend left=10] (pl2) to node[commnum]{5} (r3);
    \draw[newcomm, bend right=40] (r2) to node[commnum]{6} (c1);
    \draw[newcomm, bend left=20] (r2) to node[commnum]{6} (c2);
    \draw[newcomm, bend left=25] (r2) to node[commnum]{6} (c3);
  \end{tikzpicture}
  \caption{%
    An example execution of Compartmentalized MultiPaxos with batching ($f=1$).
    Messages that contain a batch of commands, rather than a single command,
    are drawn thicker. Note how replica $r_2$ has to send multiple messages
    after executing a batch of commands.
  }%
  \figlabel{Batching}
\end{figure}
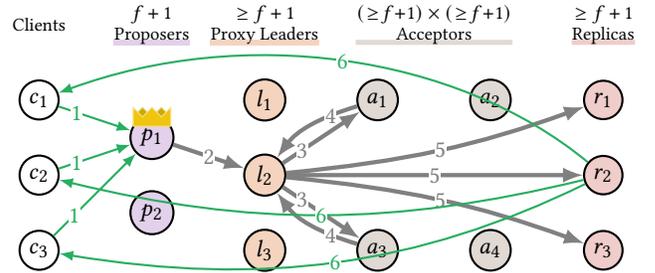}

\subsection{Compartmentalization 5: Batchers}\seclabel{Batchers}
\bds{leader}
    {batch formation and batch sequencing}
    {the number of batchers}

\paragraph{Bottleneck}
We first discuss write batching and discuss read batching momentarily.
Batching increases throughput by amortizing the communication and computation
cost of processing a command. Take the acceptors for example. Without batching,
an acceptor processes two messages \emph{per command}. With batching, however,
an acceptor only processes two messages \emph{per batch}. The acceptors process
fewer messages per command as the batch size increases. With batches of size
10, for example, an acceptor processes $10\times$ fewer messages per command
with batching than without.

Refer again to \figref{Batching}. The load on the proxy leaders and the
acceptors both decrease as the batch size increases, but this is not the case
for the leader or the replicas. We focus first on the leader. To process a
single batch of $n$ commands, the leader has to receive $n$ messages and send
one message. Unlike the proxy leaders and acceptors, the leader's communication
cost is linear in the number of commands rather than the number of batches.
This makes the leader a very likely throughput bottleneck.

\paragraph{Decouple}
The leader has two responsibilities. It forms batches, and it sequences
batches. We decouple the two responsibilities by introducing a set of at least
$f+1$ \defword{batchers}, as illustrated in \figref{Batchers}. The batchers are
responsible for forming batches, while the leader is responsible for sequencing
batches.

{
\tikzstyle{proc}=[draw, circle, thick, inner sep=1.5pt]
\tikzstyle{newproc}=[draw=flatgreen]
\tikzstyle{client}=[proc, fill=clientcolor!25]
\tikzstyle{batcher}=[proc, fill=batchercolor!25]
\tikzstyle{proposer}=[proc, fill=proposercolor!25]
\tikzstyle{proxyleader}=[proc, fill=proxyleadercolor!25]
\tikzstyle{acceptor}=[proc, fill=acceptorcolor!25]
\tikzstyle{replica}=[proc, fill=replicacolor!25]

\tikzstyle{proclabel}=[inner sep=0pt, align=center, font=\footnotesize]

\tikzstyle{component}=[draw, thick, flatgray, rounded corners]

\tikzstyle{comm}=[-latex, thick]
\tikzstyle{batched}=[ultra thick]
\tikzstyle{commnum}=[fill=white, inner sep=0pt]
\tikzstyle{newcomm}=[comm, flatgreen]
\tikzstyle{newcommnum}=[commnum, flatgreen]
\tikzstyle{oldcomm}=[comm, black!50]
\tikzstyle{oldcommnum}=[commnum, black!50]

\begin{figure}[ht]
  \centering
  \begin{tikzpicture}[xscale=1.25]
    \node[client] (c1) at (0, 2) {$c_1$};
    \node[client] (c2) at (0, 1) {$c_2$};
    \node[client] (c3) at (0, 0) {$c_3$};
    \node[batcher, newproc] (b1) at (1, 2) {$b_1$};
    \node[batcher, newproc] (b2) at (1, 1) {$b_2$};
    \node[batcher, newproc] (b3) at (1, 0) {$b_3$};
    \node[proposer] (p1) at (2, 1.5) {$p_1$};
    \node[proposer] (p2) at (2, 0.5) {$p_2$};
    \node[proxyleader] (pl1) at (3, 2) {$l_1$};
    \node[proxyleader] (pl2) at (3, 1) {$l_2$};
    \node[proxyleader] (pl3) at (3, 0) {$l_3$};
    \node[acceptor] (a1) at (4, 2) {$a_1$};
    \node[acceptor] (a2) at (5, 2) {$a_2$};
    \node[acceptor] (a3) at (4, 0) {$a_3$};
    \node[acceptor] (a4) at (5, 0) {$a_4$};
    \node[replica] (r1) at (6, 2) {$r_1$};
    \node[replica] (r2) at (6, 1) {$r_2$};
    \node[replica] (r3) at (6, 0) {$r_3$};

    \crown{(p1.north)++(0,-0.15)}{0.4}{0.25}
    \node[proclabel] (clients) at (0, 3) {Clients};
    \node[proclabel] (batchers) at (1, 3) {$\geq f+1$\\Batchers};
    \node[proclabel] (proposers) at (2, 3) {$f+1$\\Proposers};
    \node[proclabel] (proxyleaders) at (3, 3) {$\geq f+1$\\Proxy\\Leaders};
    \node[proclabel] (acceptors) at (4.5, 3) {$(\geq\!f\!+\!1) \times (\geq\!f\!+\!1)$\\Acceptors};
    \node[proclabel] (replicas) at (6, 3) {$\geq f+1$\\Replicas};
    \halffill{clients}{clientcolor!25}
    \quarterfill{batchers}{batchercolor!25}
    \quarterfill{proposers}{proposercolor!25}
    \quarterfill{proxyleaders}{proxyleadercolor!25}
    \quarterfill{acceptors}{acceptorcolor!25}
    \quarterfill{replicas}{replicacolor!25}

    \draw[newcomm] (c1) to node[commnum, near start]{1} (b1);
    \draw[newcomm] (c2) to node[commnum, near start]{1} (b1);
    \draw[newcomm] (c3) to node[commnum, near start]{1} (b1);
    \draw[newcomm, batched] (b1) to node[commnum, near start]{2} (p1);
    \draw[oldcomm, batched] (p1) to node[commnum, near start]{3} (pl2);
    \draw[oldcomm, batched, near start] (pl2) to node[commnum]{4} (a1);
    \draw[oldcomm, batched, near start] (pl2) to node[commnum]{4} (a3);
    \draw[oldcomm, batched, near start, bend right=20] (a1) to node[commnum]{5} (pl2);
    \draw[oldcomm, batched, near start, bend left=20] (a3) to node[commnum]{5} (pl2);
    \draw[oldcomm, batched, bend right=10] (pl2) to node[commnum]{6} (r1);
    \draw[oldcomm, batched] (pl2) to node[commnum]{6} (r2);
    \draw[oldcomm, batched, bend left=10] (pl2) to node[commnum]{6} (r3);
    \draw[oldcomm, bend right=28] (r2) to node[commnum]{7} (c1);
    \draw[oldcomm, bend left=15] (r2) to node[commnum]{7} (c2);
    \draw[oldcomm, bend left=30] (r2) to node[commnum]{7} (c3);
  \end{tikzpicture}
  \caption{%
    An example execution of Compartmentalized MultiPaxos with batchers ($f=1$).
  }%
  \figlabel{Batchers}
\end{figure}
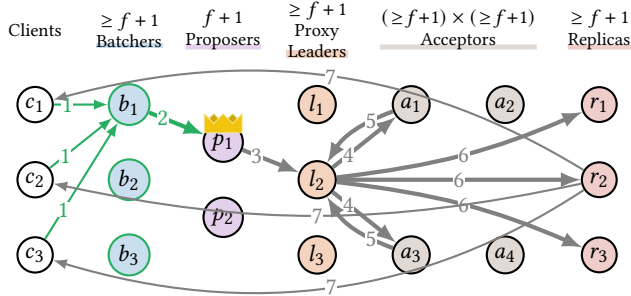}

More concretely, when a client wants to propose a state machine command, it
sends the command to a randomly selected batcher (1). After receiving
sufficiently many commands from the clients (or after a timeout expires), a
batcher places the commands in a batch and forwards it to the leader (2). When
the leader receives a batch of commands, it assigns it a log entry, forms a
\msgfont{Phase 2a} message, and sends the \msgfont{Phase2a} message to a proxy
leader (3). The rest of the protocol remains unchanged.

Without batchers, the leader has to receive $n$ messages per batch of $n$
commands. With batchers, the leader only has to receive one. This either
reduces the load on the bottleneck leader or eliminates it as a bottleneck
completely.

\paragraph{Scale}
The batchers are embarrassingly parallel. We can increase the number of
batchers until they're not a throughput bottleneck.

\paragraph{Discussion}
Read batching is very similar to write batching. Clients send reads to randomly
selected batchers, and batchers group reads together into batches. After a
batcher has formed a read batch $X$, it sends a \msg{PreRead}{} message to a
read quorum of acceptors, computes the resulting watermark $i$, and sends a
\msg{Read}{X, i} request to any one of the replicas.

\subsection{Compartmentalization 6: Unbatchers}\seclabel{Unbatchers}
\bds{replicas}
    {batch processing and batch replying}
    {the number of unbatchers}

\paragraph{Bottleneck}
After executing a batch of $n$ commands, a replica has to send $n$ messages
back to the $n$ clients. Thus, the replicas (like the leader without batchers)
suffer communication overheads linear in the number of commands rather than the
number of batches.

\paragraph{Decouple}
The replicas have two responsibilities. They execute batches of commands, and
they send replies to the clients. We decouple these two responsibilities by
introducing a set of at least $f+1$ \defword{unbatchers}, as illustrated in
\figref{Unbatchers}. The replicas are responsible for executing batches of
commands, while the unbatchers are responsible for sending the results of
executing the commands back to the clients. Concretely, after executing a batch
of commands, a replica forms a batch of results and sends the batch to a
randomly selected unbatcher (7). Upon receiving a result batch, an unbatcher
sends the results back to the clients (8). This decoupling reduces the load on
the replicas.

{
\tikzstyle{proc}=[draw, circle, thick, inner sep=1pt]
\tikzstyle{newproc}=[draw=flatgreen]
\tikzstyle{client}=[proc, fill=clientcolor!25]
\tikzstyle{batcher}=[proc, fill=batchercolor!25]
\tikzstyle{proposer}=[proc, fill=proposercolor!25]
\tikzstyle{proxyleader}=[proc, fill=proxyleadercolor!25]
\tikzstyle{acceptor}=[proc, fill=acceptorcolor!25]
\tikzstyle{replica}=[proc, fill=replicacolor!25]
\tikzstyle{proxyreplica}=[proc, fill=proxyreplicacolor!25]

\tikzstyle{proclabel}=[inner sep=0pt, align=center, font=\scriptsize]

\tikzstyle{component}=[draw, thick, flatgray, rounded corners]

\tikzstyle{comm}=[-latex, thick]
\tikzstyle{batched}=[ultra thick]
\tikzstyle{commnum}=[fill=white, inner sep=0pt]
\tikzstyle{newcomm}=[comm, flatgreen]
\tikzstyle{newcommnum}=[commnum, flatgreen]
\tikzstyle{oldcomm}=[comm, black!50]
\tikzstyle{oldcommnum}=[commnum, black!50]

\begin{figure}[ht]
  \centering
  \begin{tikzpicture}[xscale=1.1]
    \node[client] (c1) at (0, 2) {$c_1$};
    \node[client] (c2) at (0, 1) {$c_2$};
    \node[client] (c3) at (0, 0) {$c_3$};
    \node[batcher] (b1) at (1, 2) {$b_1$};
    \node[batcher] (b2) at (1, 1) {$b_2$};
    \node[batcher] (b3) at (1, 0) {$b_3$};
    \node[proposer] (p1) at (2, 1.5) {$p_1$};
    \node[proposer] (p2) at (2, 0.5) {$p_2$};
    \node[proxyleader] (pl1) at (3, 2) {$l_1$};
    \node[proxyleader] (pl2) at (3, 1) {$l_2$};
    \node[proxyleader] (pl3) at (3, 0) {$l_3$};
    \node[acceptor] (a1) at (4, 2) {$a_1$};
    \node[acceptor] (a2) at (5, 2) {$a_2$};
    \node[acceptor] (a3) at (4, 0) {$a_3$};
    \node[acceptor] (a4) at (5, 0) {$a_4$};
    \node[replica] (r1) at (6, 2) {$r_1$};
    \node[replica] (r2) at (6, 1) {$r_2$};
    \node[replica] (r3) at (6, 0) {$r_3$};
    \node[proxyreplica, newproc] (pr1) at (7, 2) {$d_1$};
    \node[proxyreplica, newproc] (pr2) at (7, 1) {$d_2$};
    \node[proxyreplica, newproc] (pr3) at (7, 0) {$d_3$};

    \crown{(p1.north)++(0,-0.1)}{0.4}{0.25}
    \node[proclabel] (clients) at (0, 3) {Clients};
    \node[proclabel] (batchers) at (1, 3) {$\geq f+1$\\Batchers};
    \node[proclabel] (proposers) at (2, 3) {$f+1$\\Proposers};
    \node[proclabel] (proxyleaders) at (3, 3) {$\geq f+1$\\Proxy\\Leaders};
    \node[proclabel] (acceptors) at (4.5, 3) {$(\geq\!f\!+\!1) \times (\geq\!f\!+\!1)$\\Acceptors};
    \node[proclabel] (replicas) at (6, 3) {$\geq f+1$\\Replicas};
    \node[proclabel] (proxyreplicas) at (7, 3) {$\geq f+1$\\Unbatchers};
    \halffill{clients}{clientcolor!25}
    \quarterfill{batchers}{batchercolor!25}
    \quarterfill{proposers}{proposercolor!25}
    \quarterfill{proxyleaders}{proxyleadercolor!25}
    \quarterfill{acceptors}{acceptorcolor!25}
    \quarterfill{replicas}{replicacolor!25}
    \quarterfill{proxyreplicas}{proxyreplicacolor!25}

    \draw[oldcomm] (c1) to node[commnum, near start]{1} (b1);
    \draw[oldcomm] (c2) to node[commnum, near start]{1} (b1);
    \draw[oldcomm] (c3) to node[commnum, near start]{1} (b1);
    \draw[oldcomm, batched] (b1) to node[commnum, near start]{2} (p1);
    \draw[oldcomm, batched] (p1) to node[commnum, near start]{3} (pl2);
    \draw[oldcomm, batched, near start] (pl2) to node[commnum]{4} (a1);
    \draw[oldcomm, batched, near start] (pl2) to node[commnum]{4} (a3);
    \draw[oldcomm, batched, near start, bend right=20] (a1) to node[commnum]{5} (pl2);
    \draw[oldcomm, batched, near start, bend left=20] (a3) to node[commnum]{5} (pl2);
    \draw[oldcomm, batched, bend right=10] (pl2) to node[commnum]{6} (r1);
    \draw[oldcomm, batched] (pl2) to node[commnum]{6} (r2);
    \draw[oldcomm, batched, bend left=10] (pl2) to node[commnum]{6} (r3);
    \draw[newcomm, batched] (r2) to node[commnum, near start]{7} (pr2);
    \draw[newcomm, bend right=26] (pr2) to node[commnum]{8} (c1);
    \draw[newcomm, bend left=30] (pr2) to node[commnum]{8} (c2);
    \draw[newcomm, bend left=30] (pr2) to node[commnum]{8} (c3);
  \end{tikzpicture}
  \caption{%
    An example execution of Compartmentalized MultiPaxos with unbatchers
    ($f=1$).
  }%
  \figlabel{Unbatchers}
\end{figure}
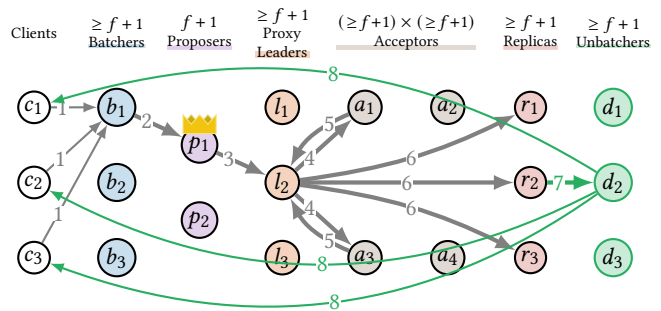}

\paragraph{Scale}
As with batchers, unbatchers are embarrassingly parallel, so we can increase
the number of unbatchers until they are not a throughput bottleneck.

\paragraph{Discussion}
Read unbatching is identical to write unbatching. After executing a batch of
reads, a replica forms the corresponding batch of results and sends it to a
randomly selected unbatcher.
}
{\section{Further Compartmentalization}
The six compartmentalizations that we've discussed are not exhaustive, and
MultiPaxos is not the only state machine replication protocol that can be
compartmentalized. Compartmentalization is a generally applicable technique.
There are many other compartmentalizations that can be applied to many other
protocols.
\iftoggle{techreportenabled}{%
  We now demonstrate this generality by compartmentalizing
  Mencius~\cite{mao2008mencius} and S-Paxos~\cite{biely2012s}. We are also
  currently working on compartmentalizing Raft~\cite{ongaro2014search} and
  EPaxos~\cite{moraru2013there}.
}{

  For example, Mencius~\cite{mao2008mencius} is a multi-leader MultiPaxos
  variant that partitions log entries between the leaders.
  S-Paxos~\cite{biely2012s} is a MultiPaxos variant in which every state
  machine command is given a unique id and persisted on a set of machines
  before MultiPaxos is used to order command ids rather than commands
  themselves. In our technical report~\cite{whittaker2020scaling}, we explain
  how to compartmentalize these two protocols. We compartmentalize Mencius very
  similarly to how we compartmentalized MultiPaxos. We compartmentalize S-Paxos
  by introducing new sets of nodes called \defword{disseminators} and
  \defword{stabilizers} which are analogous to proxy leaders and acceptors but
  are used to persist commands rather than order them. We also
  compartmentalized Scalog~\cite{ding2020scalog} and are currently working on
  compartmentalizing Raft~\cite{ongaro2014search} and
  EPaxos~\cite{moraru2013there}. Due to space constraints, we leave the details
  to our technical report~\cite{whittaker2020scaling}.
}
}
\iftoggle{techreportenabled}{%
  {\section{Mencius}\seclabel{Mencius}

\subsection{Background}
As discussed previously, the MultiPaxos leader is a throughput bottleneck
because all commands go through the leader and because the leader performs
disproportionately more work per command than the acceptors or replicas.
Mencius is a MultiPaxos variant that attempts to eliminate this bottleneck by
using more than one leader.

Rather than having a single leader sequence all commands in the log, Mencius
round-robin partitions the log among multiple leaders. For example, consider
the scenario with three leaders $l_1$, $l_2$, and $l_3$ illustrated in
\figref{MenciusLog}. Leader $l_1$ gets commands chosen in slots $0$, $3$, $6$,
etc.; leader $l_2$ gets commands chosen in slots $1$, $4$, $7$, etc.; and
leader $l_3$ gets commands chosen in slots $2$, $5$, $8$, etc.

{\newlength{\mlogentryinnersep}
\setlength{\mlogentryinnersep}{4pt}
\newlength{\mlogentrylinewidth}
\setlength{\mlogentrylinewidth}{1pt}
\newlength{\mlogentrywidth}
\setlength{\mlogentrywidth}{\widthof{$X$}+2\mlogentryinnersep}
\newcommand{\licolor}{flatred}
\newcommand{\liicolor}{flatgreen}
\newcommand{\liiicolor}{flatblue}

\tikzstyle{mlogentry}=[%
  draw,
  inner sep=\mlogentryinnersep,
  line width=\mlogentrylinewidth,
  minimum height=\mlogentrywidth,
  minimum width=\mlogentrywidth]
\tikzstyle{executed}=[fill=gray, opacity=0.2, draw opacity=1, text opacity=1]
\tikzstyle{logindex}=[black!75, font=\small]

\newcommand{\rightof}[1]{-\mlogentrylinewidth of #1}

\newcommand{\logentry}[4]{
  \node[mlogentry,
        label={[logindex]90:#1},
        label={[logindex]-90:#2},
        fill=#3,
        right=\rightof{#4}] (#1) {};
}

\begin{figure}[ht]
  \centering
  \begin{tikzpicture}
    \node[mlogentry,
          label={[logindex]90:0},
          label={[logindex]-90:$l_1$},
          fill=\licolor!15] (0) {};
    \logentry{1}{$l_2$}{\liicolor!15}{0}
    \logentry{2}{$l_3$}{\liiicolor!15}{1}
    \logentry{3}{$l_1$}{\licolor!15}{2}
    \logentry{4}{$l_2$}{\liicolor!15}{3}
    \logentry{5}{$l_3$}{\liiicolor!15}{4}
    \logentry{6}{$l_1$}{\licolor!15}{5}
    \node[right=\rightof{6}] {$\ldots$};

    \tikzstyle{proc}=[draw, circle, thick, inner sep=1.5pt]
    \node[proc, fill=\licolor!25] (l1) at (5, 1) {$l_1$};
    \node[proc, fill=\liicolor!25] (l2) at (5, 0) {$l_2$};
    \node[proc, fill=\liiicolor!25] (l3) at (5, -1) {$l_3$};

    \node[anchor=west, right=0 of l1] {$0, 3, 6, \ldots$};
    \node[anchor=west, right=0 of l2] {$1, 4, 7, \ldots$};
    \node[anchor=west, right=0 of l3] {$2, 5, 8, \ldots$};
  \end{tikzpicture}
  \caption{A Mencius log round robin partitioned among three leaders.}%
  \figlabel{MenciusLog}
\end{figure}
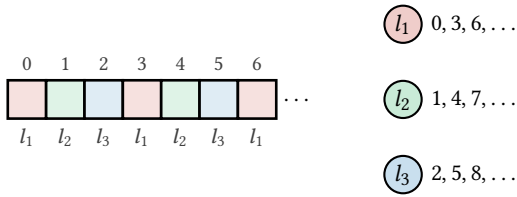
}

Having multiple leaders works well when all the leaders process commands at the
exact same rate. However, if one of the leaders is slower than the others, then
holes start appearing in the log entries owned by the slow leader. This is
illustrated in \figref{MenciusNoopsBefore}. \figref{MenciusNoopsBefore} depicts
a Mencius log partitioned across three leaders. Leaders $l_1$ and $l_2$ have
both gotten a few commands chosen (e.g., $a$ in slot 0, $b$ in slot 1, etc.),
but leader $l_3$ is lagging behind and has not gotten any commands chosen yet.
Replicas execute commands in log order, so they are unable to execute all of
the chosen commands until $l_3$ gets commands chosen in its vacant log entries.

{\newlength{\nlogentryinnersep}
\setlength{\nlogentryinnersep}{2pt}
\newlength{\nlogentrylinewidth}
\setlength{\nlogentrylinewidth}{1pt}
\newlength{\nlogentrywidth}
\setlength{\nlogentrywidth}{\widthof{$XX$}+2\nlogentryinnersep}
\newcommand{\licolor}{flatred}
\newcommand{\liicolor}{flatgreen}
\newcommand{\liiicolor}{flatblue}

\tikzstyle{nlogentry}=[%
  draw,
  align=center,
  inner sep=\nlogentryinnersep,
  line width=\nlogentrylinewidth,
  minimum height=\nlogentrywidth,
  minimum width=\nlogentrywidth]
\tikzstyle{logindex}=[black!75, font=\small]

\newcommand{\rightof}[1]{-\nlogentrylinewidth of #1}

\newcommand{\logentry}[5]{
  \node[nlogentry,
        label={[logindex]90:#1},
        label={[logindex]-90:#2},
        fill=#3,
        right=\rightof{#4}] (#1) {#5};
}

\begin{figure}[t]
  \centering
  \begin{subfigure}[b]{\columnwidth}
    \centering
    \begin{tikzpicture}
      \node[nlogentry,
            label={[logindex]90:0},
            label={[logindex]-90:$l_1$},
            fill=\licolor!35] (0) {$a$};
      \logentry{1}{$l_2$}{\liicolor!35}{0}{$b$}
      \logentry{2}{$l_3$}{\liiicolor!7}{1}{}
      \logentry{3}{$l_1$}{\licolor!35}{2}{$c$}
      \logentry{4}{$l_2$}{\liicolor!35}{3}{$d$}
      \logentry{5}{$l_3$}{\liiicolor!7}{4}{}
      \logentry{6}{$l_1$}{\licolor!35}{5}{$e$}
      \logentry{7}{$l_2$}{\liicolor!35}{6}{$f$}
      \logentry{8}{$l_3$}{\liiicolor!7}{7}{}
      \logentry{9}{$l_1$}{\licolor!35}{8}{$g$}
      \node[right=\rightof{9}] {$\ldots$};
    \end{tikzpicture}
    \caption{Before noops}\figlabel{MenciusNoopsBefore}
  \end{subfigure}

  \vspace{12pt}

  \begin{subfigure}[b]{\columnwidth}
    \centering
    \begin{tikzpicture}
      \node[nlogentry,
            label={[logindex]90:0},
            label={[logindex]-90:$l_1$},
            fill=\licolor!35] (0) {$a$};
      \logentry{1}{$l_2$}{\liicolor!35}{0}{$b$}
      \logentry{2}{$l_3$}{\liiicolor!35}{1}{no\\[-6pt]op}
      \logentry{3}{$l_1$}{\licolor!35}{2}{$c$}
      \logentry{4}{$l_2$}{\liicolor!35}{3}{$d$}
      \logentry{5}{$l_3$}{\liiicolor!35}{4}{no\\[-6pt]op}
      \logentry{6}{$l_1$}{\licolor!35}{5}{$e$}
      \logentry{7}{$l_2$}{\liicolor!35}{6}{$f$}
      \logentry{8}{$l_3$}{\liiicolor!35}{7}{no\\[-6pt]op}
      \logentry{9}{$l_1$}{\licolor!35}{8}{$g$}
      \node[right=\rightof{9}] {$\ldots$};
    \end{tikzpicture}
    \caption{After noops}\figlabel{MenciusNoopsAfter}
  \end{subfigure}
  \caption{%
    An example of using noops to deal with a slow leader. Leader $l_3$ is
    slower than leaders $l_1$ and $l_2$, so the log has holes in $l_3$'s slots.
    $l_3$ fills its holes with noops to allow commands in the log to be
    executed.
  }\figlabel{MenciusNoops}
\end{figure}
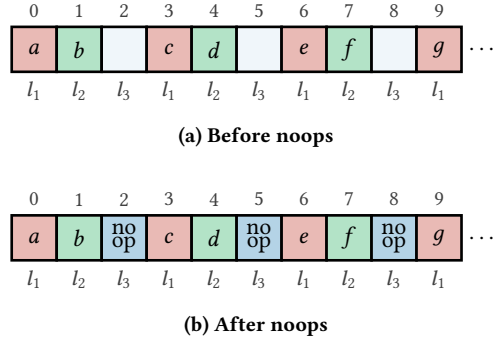
}

If a leader detects that it is lagging behind, then it fills its vacant log
entries with a sequence of noops. A \defword{noop} is a distinguished command
that does not affect the state of the replicated state machine. In
\figref{MenciusNoopsAfter}, we see that $l_3$ fills its vacant log entries with
noops. This allows the replicas to execute all of the chosen commands.

More concretely, a Mencius deployment that tolerates $f$ faults is implemented
with $2f+1$ \defword{servers}, as illustrated in
\figref{MenciusBackgroundDiagram}. Roughly speaking, every Mencius server plays
the role of a MultiPaxos leader, acceptor, and replica.

{
\tikzstyle{proc}=[draw, circle, thick, inner sep=2pt]
\tikzstyle{client}=[proc, fill=clientcolor!25]
\tikzstyle{proposer}=[proc, fill=proposercolor!25]
\tikzstyle{acceptor}=[proc, fill=acceptorcolor!25]
\tikzstyle{server}=[proc, fill=servercolor!25]
\tikzstyle{replica}=[proc, fill=replicacolor!25]

\tikzstyle{proclabel}=[inner sep=0pt, align=center]

\tikzstyle{component}=[draw, thick, flatgray, rounded corners]

\tikzstyle{comm}=[-latex, thick]
\tikzstyle{commnum}=[fill=white, inner sep=0pt]

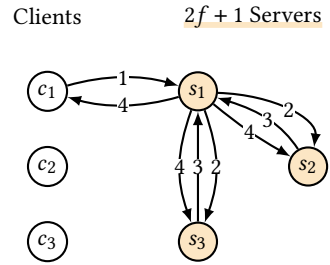
\begin{figure}[h]
  \centering
  \begin{tikzpicture}[xscale=2]
    \node[client] (c1) at (0, 2) {$c_1$};
    \node[client] (c2) at (0, 1) {$c_2$};
    \node[client] (c3) at (0, 0) {$c_3$};
    \node[server] (s1) at (1, 2) {$s_1$};
    \node[server] (s2) at (1.73205, 1) {$s_2$}; 
    \node[server] (s3) at (1, 0) {$s_3$};

    \node[proclabel] (clients) at (0, 3) {Clients};
    \node[proclabel] (servers) at (1.366025, 3) {$2f+1$ Servers};
    \halffill{clients}{clientcolor!25}
    \halffill{servers}{servercolor!25}

    \draw[comm, bend left] (c1) to node[commnum]{1} (s1);
    \draw[comm, bend left=45] (s1) to node[commnum]{2} (s2);
    \draw[comm, bend right=20] (s2) to node[commnum]{3} (s1);
    \draw[comm, bend right=5] (s1) to node[commnum]{4} (s2);
    \draw[comm, bend left=10] (s1) to node[commnum]{2} (s3);
    \draw[comm] (s3) to node[commnum]{3} (s1);
    \draw[comm, bend right=10] (s1) to node[commnum]{4} (s3);
    \draw[comm, bend left] (s1) to node[commnum]{4} (c1);

  \end{tikzpicture}
  \caption{An example execution of Mencius.}%
  \figlabel{MenciusBackgroundDiagram}
\end{figure}}

When a client wants to propose a state machine command $x$, it sends $x$ to any
of the servers (1). Upon receiving command $x$, a server $s_l$ plays the role
of a leader. It assigns the command $x$ a slot $i$ and sends a Phase 2a message
that includes $x$ and $i$ to the other servers (2). Upon receiving a Phase 2a
message, a server $s_a$ plays the role of an acceptor and replies with a Phase
2b message (3).

In addition, $s_a$ uses $i$ to determine if it is lagging behind $s_l$. If it
is, then it sends a \textsc{skip} message along with the Phase 2b message. The
\textsc{skip} message informs the other servers to choose a noop in every slot
owned by $s_a$ up to slot $i$. For example, if a server $s_a$'s next available
slot is slot $10$ and it receives a Phase 2a message for slot $100$, then it
broadcasts a \textsc{skip} message informing the other servers to place noops
in all of the slots between slots $10$ and $100$ that are owned by server
$s_a$. Mencius leverages a protocol called Coordinated Paxos to ensure noops
are chosen correctly. We refer to the reader to~\cite{mao2008mencius} for
details.

Upon receiving Phase 2b messages for command $x$ from a majority of the
servers, server $s_l$ deems the command $x$ chosen. It informs the other
servers that the command has been chosen and also sends the result of executing
$x$ back to the client.

\subsection{Compartmentalization}
Mencius uses multiple leaders to avoid being bottlenecked by a single leader.
However, despite this, Mencius still does not achieve optimal throughput. Part
of the problem is that every Mencius server plays three roles, that of a
leader, an acceptor, and a replica. Because of this, a server has to send and
receive a total of roughly $3f+5$ messages for every command that it leads
\emph{and also} has to send and receive messages acking other servers as they
simultaneously choose commands.

We can solve this problem by decoupling the servers. Instead of deploying a set
of heavily loaded servers, we instead view Mencius as a MultiPaxos variant and
deploy it as a set of proposers, a set of acceptors, and set of replicas. This
is illustrated in \figref{MenciusDecoupledDiagram}.

{
\tikzstyle{proc}=[draw, circle, thick, inner sep=2pt]
\tikzstyle{client}=[proc, fill=clientcolor!25]
\tikzstyle{proposer}=[proc, fill=proposercolor!25]
\tikzstyle{acceptor}=[proc, fill=acceptorcolor!25]
\tikzstyle{replica}=[proc, fill=replicacolor!25]

\tikzstyle{proclabel}=[inner sep=0pt, align=center]

\tikzstyle{component}=[draw, thick, flatgray, rounded corners]

\tikzstyle{comm}=[-latex, thick]
\tikzstyle{commnum}=[fill=white, inner sep=0pt]

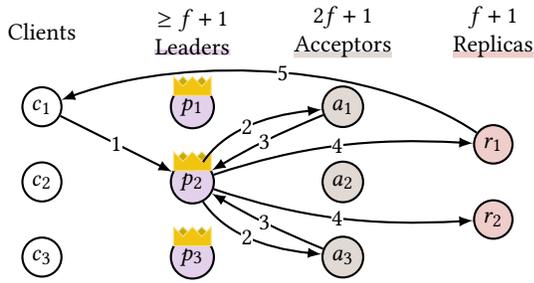
\begin{figure}[h]
  \centering
  \begin{tikzpicture}[xscale=2]
    \node[client] (c1) at (0, 2) {$c_1$};
    \node[client] (c2) at (0, 1) {$c_2$};
    \node[client] (c3) at (0, 0) {$c_3$};
    \node[proposer] (p1) at (1, 2) {$p_1$};
    \node[proposer] (p2) at (1, 1) {$p_2$};
    \node[proposer] (p3) at (1, 0) {$p_3$};
    \node[acceptor] (a1) at (2, 2) {$a_1$};
    \node[acceptor] (a2) at (2, 1) {$a_2$};
    \node[acceptor] (a3) at (2, 0) {$a_3$};
    \node[replica] (r1) at (3, 1.5) {$r_1$};
    \node[replica] (r2) at (3, 0.5) {$r_2$};

    \crown{(p1.north)++(0,-0.15)}{0.25}{0.25}
    \crown{(p2.north)++(0,-0.15)}{0.25}{0.25}
    \crown{(p3.north)++(0,-0.15)}{0.25}{0.25}
    \node[proclabel] (clients) at (0, 3) {Clients};
    \node[proclabel] (proposers) at (1, 3) {$\geq f+1$\\Leaders};
    \node[proclabel] (acceptors) at (2, 3) {$2f+1$\\Acceptors};
    \node[proclabel] (replicas) at (3, 3) {$f+1$\\Replicas};
    \halffill{clients}{clientcolor!25}
    \quarterfill{proposers}{proposercolor!25}
    \quarterfill{acceptors}{acceptorcolor!25}
    \quarterfill{replicas}{replicacolor!25}

    \draw[comm] (c1) to node[commnum]{1} (p2);
    \draw[comm, bend left=30] (p2) to node[commnum]{2} (a1);
    \draw[comm, bend right=30] (p2) to node[commnum]{2} (a3);
    \draw[comm, bend right=5] (a1) to node[commnum]{3} (p2);
    \draw[comm, bend left=5] (a3) to node[commnum]{3} (p2);
    \draw[comm, bend left=20] (p2) to node[commnum]{4} (r1);
    \draw[comm, bend right=20] (p2) to node[commnum]{4} (r2);
    \draw[comm, bend right=45] (r1) to node[commnum]{5} (c1);
  \end{tikzpicture}
  \caption{%
    An example execution of decoupled Mencius. Note that every proposer is a
    leader.
  }%
  \figlabel{MenciusDecoupledDiagram}
\end{figure}}

Now, Mencius is equivalent to MultiPaxos with the following key differences.
First, every proposer is a leader, with the log round-robin partitioned among
all the proposers. If a client wants to propose a command, it can send it to
any of the proposers. Second, the proposers periodically broadcast their next
available slots to one another. Every server uses this information to gauge
whether it is lagging behind. If it is, it chooses noops in its vacant
slots, as described above.

This decoupled Mencius is a step in the right direction, but it shares many of
the problems that MultiPaxos faced. The proposers are responsible for both
sequencing commands and for coordinating with acceptors; we have a single
unscalable group of acceptors; and we are deploying too few replicas.
Thankfully, we can compartmentalize Mencius in exactly the same way as
MultiPaxos by leveraging proxy leaders, acceptor grids, and more replicas. This
is illustrated in \figref{MenciusFullOptimizationDiagram}.

{
\tikzstyle{proc}=[draw, circle, thick, inner sep=2pt]
\tikzstyle{client}=[proc, fill=clientcolor!25]
\tikzstyle{proposer}=[proc, fill=proposercolor!25]
\tikzstyle{proxyleader}=[proc, fill=proxyleadercolor!25]
\tikzstyle{acceptor}=[proc, fill=acceptorcolor!25]
\tikzstyle{replica}=[proc, fill=replicacolor!25]

\tikzstyle{proclabel}=[inner sep=0pt, align=center, font=\footnotesize]

\tikzstyle{component}=[draw, thick, flatgray, rounded corners]

\tikzstyle{comm}=[-latex, thick]
\tikzstyle{commnum}=[fill=white, inner sep=0pt]

\begin{figure}[ht]
  \centering
  \begin{tikzpicture}[xscale=1.5]
    \node[client] (c1) at (0, 2) {$c_1$};
    \node[client] (c2) at (0, 1) {$c_2$};
    \node[client] (c3) at (0, 0) {$c_3$};
    \node[proposer] (p1) at (1, 2) {$p_1$};
    \node[proposer] (p2) at (1, 1) {$p_2$};
    \node[proposer] (p3) at (1, 0) {$p_3$};
    \node[proxyleader] (pl1) at (2, 2) {$l_1$};
    \node[proxyleader] (pl2) at (2, 1) {$l_2$};
    \node[proxyleader] (pl3) at (2, 0) {$l_3$};
    \node[acceptor] (a1) at (3, 2) {$a_1$};
    \node[acceptor] (a2) at (4, 2) {$a_2$};
    \node[acceptor] (a3) at (3, 0) {$a_3$};
    \node[acceptor] (a4) at (4, 0) {$a_4$};
    \node[replica] (r1) at (5, 2) {$r_1$};
    \node[replica] (r2) at (5, 1) {$r_2$};
    \node[replica] (r3) at (5, 0) {$r_3$};

    \crown{(p1.north)++(0,-0.15)}{0.333}{0.25}
    \crown{(p2.north)++(0,-0.15)}{0.333}{0.25}
    \crown{(p3.north)++(0,-0.15)}{0.333}{0.25}
    \node[proclabel] (clients) at (0, 3) {Clients};
    \node[proclabel] (proposers) at (1, 3) {$\geq f+1$\\Leaders};
    \node[proclabel] (proxyleaders) at (2, 3) {$\geq f+1$\\Proxy Leaders};
    \node[proclabel] (acceptors) at (3.5, 3) {$(\geq\!f\!+\!1) \times (\geq\!f\!+\!1)$\\Acceptors};
    \node[proclabel] (replicas) at (5, 3) {$\geq f+1$\\Replicas};
    \halffill{clients}{clientcolor!25}
    \quarterfill{proposers}{proposercolor!25}
    \quarterfill{proxyleaders}{proxyleadercolor!25}
    \quarterfill{acceptors}{acceptorcolor!25}
    \quarterfill{replicas}{replicacolor!25}

    \draw[comm] (c1) to node[commnum]{1} (p2);
    \draw[comm] (p2) to node[commnum]{2} (pl2);
    \draw[comm, near start] (pl2) to node[commnum]{3} (a1);
    \draw[comm, near start] (pl2) to node[commnum]{3} (a3);
    \draw[comm, near start, bend left=20] (a1) to node[commnum]{4} (pl2);
    \draw[comm, near start, bend right=20] (a3) to node[commnum]{4} (pl2);
    \draw[comm] (pl2) to node[commnum]{5} (r1);
    \draw[comm] (pl2) to node[commnum]{5} (r2);
    \draw[comm] (pl2) to node[commnum]{5} (r3);
    \draw[comm, bend right=20] (r1) to node[commnum]{6} (c1);
  \end{tikzpicture}
  \caption{%
    An execution of Mencius with proxy leaders, acceptor grids, and an
    increased number of replicas.
  }%
  \figlabel{MenciusFullOptimizationDiagram}
\end{figure}
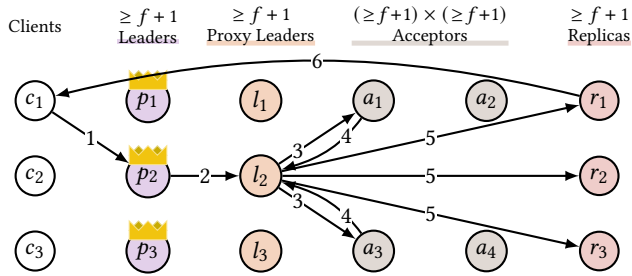}

This protocol shares all of the advantages of compartmentalized MultiPaxos.
Proxy leaders and acceptors both trivially scale so are not bottlenecks, while
leaders and replicas have been pared down to their essential responsibilities
of sequencing and executing commands respectively. Moreover, because Mencius
allows us to deploy multiple leaders, we can also increase the number of
leaders until they are no longer a bottleneck. We can also introduce batchers
and unbatchers like we did with MultiPaxos and can implement linearizable
leaderless reads.
}
  {\section{S-Paxos}\seclabel{SPaxos}

\subsection{Background}
S-Paxos~\cite{biely2012s} is a MultiPaxos variant that, like Mencius, aims to
avoid being bottlenecked by a single leader. Recall that when a MultiPaxos
leader receives a state machine command $x$ from a client, it broadcasts a
Phase 2a message to the acceptors that includes the command $x$. If the leader
receives a state machine command that is large (in terms of bytes) or receives
a large batch of modestly sized commands, the overheads of disseminating the
commands begin to dominate the cost of the protocol, exacerbating the fact that
command disseminating is performed solely by the leader.

S-Paxos avoids this by decoupling command dissemination from command
sequencing---separating control from from data flow---and distributing command
dissemination across all nodes.  More concretely, an S-Paxos deployment that
tolerates $f$ faults consists of $2f+1$ servers, as illustrated in
\figref{SPaxosBackgroundDiagram}. Every server plays the role of a MultiPaxos
proposer, acceptor, and replica. It also plays the role of a
\defword{disseminator} and \defword{stabilizer}, two roles that will become
clear momentarily.

{
\tikzstyle{proc}=[draw, circle, thick, inner sep=2pt]
\tikzstyle{client}=[proc, fill=clientcolor!25]
\tikzstyle{server}=[proc, fill=servercolor!25]

\tikzstyle{proclabel}=[inner sep=0pt, align=center]

\tikzstyle{comm}=[-latex, thick]
\tikzstyle{commnum}=[fill=white, inner sep=0pt]
\tikzstyle{bigcommand}=[ultra thick]

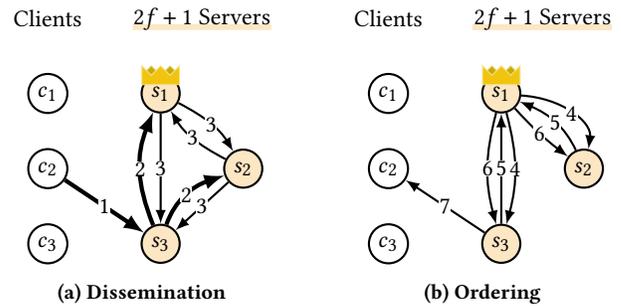
\begin{figure}[h]
  \centering
  \begin{subfigure}[b]{0.45\columnwidth}
    \centering
    \begin{tikzpicture}[xscale=1.5]
      \node[client] (c1) at (0, 2) {$c_1$};
      \node[client] (c2) at (0, 1) {$c_2$};
      \node[client] (c3) at (0, 0) {$c_3$};
      \node[server] (s1) at (1, 2) {$s_1$};
      \node[server] (s2) at (1.73205, 1) {$s_2$}; 
      \node[server] (s3) at (1, 0) {$s_3$};

      \crown{(s1.north)++(0,-0.15)}{0.333}{0.25}
      \node[proclabel] (clients) at (0, 3) {Clients};
      \node[proclabel] (servers) at (1.366025, 3) {$2f+1$ Servers};
      \halffill{clients}{clientcolor!25}
      \halffill{servers}{servercolor!25}

      \draw[comm, bigcommand] (c2) to node[commnum]{1} (s3);
      \draw[comm, bigcommand, bend left=15] (s3) to node[commnum]{2} (s1);
      \draw[comm, bigcommand, bend left=25] (s3) to node[commnum]{2} (s2);
      \draw[comm, bend left=15] (s1) to node[commnum]{3} (s2);
      \draw[comm] (s1) to node[commnum]{3} (s3);
      \draw[comm, bend left=15] (s2) to node[commnum]{3} (s1);
      \draw[comm] (s2) to node[commnum]{3} (s3);
    \end{tikzpicture}
    \caption{Dissemination}\figlabel{SPaxosBackgroundDiagramDissemination}
  \end{subfigure}\hspace{0.075\columnwidth}
  \begin{subfigure}[b]{0.45\columnwidth}
    \centering
    \begin{tikzpicture}[xscale=1.5]
      \node[client] (c1) at (0, 2) {$c_1$};
      \node[client] (c2) at (0, 1) {$c_2$};
      \node[client] (c3) at (0, 0) {$c_3$};
      \node[server] (s1) at (1, 2) {$s_1$};
      \node[server] (s2) at (1.73205, 1) {$s_2$}; 
      \node[server] (s3) at (1, 0) {$s_3$};

      \crown{(s1.north)++(0,-0.15)}{0.333}{0.25}
      \node[proclabel] (clients) at (0, 3) {Clients};
      \node[proclabel] (servers) at (1.366025, 3) {$2f+1$ Servers};
      \halffill{clients}{clientcolor!25}
      \halffill{servers}{servercolor!25}

      \draw[comm, bend left=45] (s1) to node[commnum]{4} (s2);
      \draw[comm, bend right=20] (s2) to node[commnum]{5} (s1);
      \draw[comm, bend right=5] (s1) to node[commnum]{6} (s2);
      \draw[comm, bend left=10] (s1) to node[commnum]{4} (s3);
      \draw[comm] (s3) to node[commnum]{5} (s1);
      \draw[comm, bend right=10] (s1) to node[commnum]{6} (s3);
      \draw[comm] (s3) to node[commnum]{7} (c2);
    \end{tikzpicture}
    \caption{Ordering}\figlabel{SPaxosBackgroundDiagramOrdering}
  \end{subfigure}
  \caption{%
    An example execution of S-Paxos. Messages that include client commands (as
    opposed to ids) are bolded.
  }%
  \figlabel{SPaxosBackgroundDiagram}
\end{figure}}

When a client wants to propose a state machine command $x$, it sends $x$ to any
of the servers. Upon receiving a command from a client, a server plays the part
of a disseminator. It assigns the command a globally unique id $\text{id}_x$
and begins a \defword{dissemination phase} with the goal of persisting the
command and its id on at least a majority of the servers. This is shown in
\figref{SPaxosBackgroundDiagramDissemination}. The server broadcasts $x$ and
$\text{id}_x$ to the other servers. Upon receiving $x$ and $\text{id}_x$, a
server plays the role of a stabilizer and stores the pair in memory. It then
broadcasts an acknowledgement to all servers. The acknowledgement contains
$\text{id}_x$ but not $x$.

One of the servers is the MultiPaxos leader. Upon receiving acknowledgements
for $\text{id}_x$ from a majority of the servers, the leader knows the command
is stable. It then uses the id $\text{id}_x$ as a proxy for the corresponding
command $x$ and runs the MultiPaxos protocol as usual (i.e.\ broadcasting Phase
2a messages, receiving Phase 2b messages, and notifying the other servers when
a command id has been chosen) as shown in
\figref{SPaxosBackgroundDiagramOrdering}. Thus, while MultiPaxos agrees on a
log of \emph{commands}, S-Paxos agrees on a log of \emph{command ids}.

The S-Paxos leader, like the MultiPaxos leader, is responsible for ordering
command ids and getting them chosen. But, the responsibility of disseminating
commands is shared by all the servers.

\subsection{Compartmentalization}
We compartmentalize S-Paxos similar to how we compartmentalize MultiPaxos and
Mencius. First, we decouple servers into a set of at least $f+1$ disseminators,
a set of $2f+1$ stabilizers, a set of proposers, a set of acceptors, and a set
of replicas. This is illustrated in \figref{SPaxosDecoupledDiagram}. To propose
a command $x$, a client sends it to any of the disseminators. Upon receiving
$x$, a disseminator persists the command and its id $\text{id}_x$ on at least a
majority of (and typically all of) the stabilizers. It then forwards the id to
the leader. The leader gets the id chosen in a particular log entry and informs
one of the stabilizers. Upon receiving $\text{id}_x$ from the leader, the
stabilizer fetches $x$ from the other stabilizers if it has not previously
received it. The stabilizer then informs the replicas that $x$ has been chosen.
Replicas execute commands in prefix order and reply to clients as usual.

{
\tikzstyle{proc}=[draw, circle, thick, inner sep=1.5pt]
\tikzstyle{client}=[proc, fill=clientcolor!25]
\tikzstyle{disseminator}=[proc, fill=disseminatorcolor!25]
\tikzstyle{stabilizer}=[proc, fill=stabilizercolor!25]
\tikzstyle{proposer}=[proc, fill=proposercolor!25]
\tikzstyle{acceptor}=[proc, fill=acceptorcolor!25]
\tikzstyle{replica}=[proc, fill=replicacolor!25]

\tikzstyle{proclabel}=[inner sep=0pt, align=center, font=\footnotesize]

\tikzstyle{comm}=[-latex, thick]
\tikzstyle{commnum}=[fill=white, inner sep=0pt]
\tikzstyle{bigcommand}=[ultra thick]

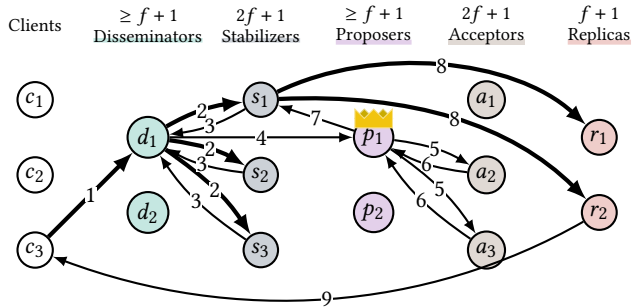
\begin{figure}[h]
  \centering
  \begin{tikzpicture}[xscale=1.5]
    \node[client] (c1) at (0, 2) {$c_1$};
    \node[client] (c2) at (0, 1) {$c_2$};
    \node[client] (c3) at (0, 0) {$c_3$};
    \node[disseminator] (d1) at (1, 1.5) {$d_1$};
    \node[disseminator] (d2) at (1, 0.5) {$d_2$};
    \node[stabilizer] (s1) at (2, 2) {$s_1$};
    \node[stabilizer] (s2) at (2, 1) {$s_2$};
    \node[stabilizer] (s3) at (2, 0) {$s_3$};
    \node[proposer] (p1) at (3, 1.5) {$p_1$};
    \node[proposer] (p2) at (3, 0.5) {$p_2$};
    \node[acceptor] (a1) at (4, 2) {$a_1$};
    \node[acceptor] (a2) at (4, 1) {$a_2$};
    \node[acceptor] (a3) at (4, 0) {$a_3$};
    \node[replica] (r1) at (5, 1.5) {$r_1$};
    \node[replica] (r2) at (5, 0.5) {$r_2$};

    \crown{(p1.north)++(0,-0.15)}{0.333}{0.25}
    \node[proclabel] (clients) at (0, 3) {Clients};
    \node[proclabel] (disseminators) at (1, 3) {$\geq f+1$\\Disseminators};
    \node[proclabel] (stabilizers) at (2, 3) {$2f+1$\\Stabilizers};
    \node[proclabel] (proposers) at (3, 3) {$\geq f+1$\\Proposers};
    \node[proclabel] (acceptors) at (4, 3) {$2f+1$\\Acceptors};
    \node[proclabel] (replicas) at (5, 3) {$f+1$\\Replicas};
    \halffill{clients}{clientcolor!25}
    \quarterfill{disseminators}{disseminatorcolor!25}
    \quarterfill{stabilizers}{stabilizercolor!25}
    \quarterfill{proposers}{proposercolor!25}
    \quarterfill{acceptors}{acceptorcolor!25}
    \quarterfill{replicas}{replicacolor!25}

    \draw[comm, bigcommand] (c3) to node[commnum]{1} (d1);
    \draw[comm, bigcommand, bend left=15] (d1) to node[commnum]{2} (s1);
    \draw[comm, bigcommand, bend left=15] (d1) to node[commnum]{2} (s2);
    \draw[comm, bigcommand, bend left=10] (d1) to node[commnum]{2} (s3);
    \draw[comm, bend left=15] (s1) to node[commnum]{3} (d1);
    \draw[comm, bend left=15] (s2) to node[commnum]{3} (d1);
    \draw[comm, bend left=10] (s3) to node[commnum]{3} (d1);
    \draw[comm] (d1) to node[commnum]{4} (p1);
    \draw[comm, bend left=15] (p1) to node[commnum]{5} (a2);
    \draw[comm, bend left=10] (p1) to node[commnum]{5} (a3);
    \draw[comm, bend left=15] (a2) to node[commnum]{6} (p1);
    \draw[comm, bend left=10] (a3) to node[commnum]{6} (p1);
    \draw[comm] (p1) to node[commnum]{7} (s1);
    \draw[comm, bigcommand, bend left=45] (s1) to node[commnum]{8} (r1);
    \draw[comm, bigcommand, bend left=30] (s1) to node[commnum]{8} (r2);
    \draw[comm, bend left=35] (r2) to node[commnum]{9} (c3);
  \end{tikzpicture}
  \caption{%
    An example execution of decoupled S-Paxos. Messages that include client
    commands (as opposed to ids) are bolded. Note that the MultiPaxos leader
    does not send or receive any messages that include a command, only messages
    that include command ids.
  }%
  \figlabel{SPaxosDecoupledDiagram}
\end{figure}}

Though S-Paxos relieves the MultiPaxos leader of its duty to broadcast commands,
the leader still has to broadcast command ids. In other words, the leader is no
longer a bottleneck on the data path but is still a bottleneck on the control
path. Moreover, disseminators and stabilizers are potential bottlenecks. We can
resolve these issues by compartmentalizing S-Paxos similar to how we
compartmentalized MultiPaxos. We introduce proxy leaders, acceptor
grids, and more replicas. Moreover, we can trivially scale up the number of
disseminators; we can deploy disseminator grids; and we can implement
linearizable leaderless reads. This is illustrated in
\figref{SPaxosFullOptimizationDiagram}. To support batching, we can again
introduce batchers and unbatchers.

{
\tikzstyle{proc}=[draw, circle, thick, inner sep=1pt]
\tikzstyle{client}=[proc, fill=clientcolor!25]
\tikzstyle{disseminator}=[proc, fill=disseminatorcolor!25]
\tikzstyle{stabilizer}=[proc, fill=stabilizercolor!25, font=\scriptsize]
\tikzstyle{proposer}=[proc, fill=proposercolor!25]
\tikzstyle{proxyleader}=[proc, fill=proxyleadercolor!25]
\tikzstyle{acceptor}=[proc, fill=acceptorcolor!25, font=\scriptsize]
\tikzstyle{replica}=[proc, fill=replicacolor!25]

\tikzstyle{proclabel}=[inner sep=0pt, align=center, font=\scriptsize]

\tikzstyle{comm}=[-latex, thick]
\tikzstyle{commnum}=[fill=white, inner sep=0pt]
\tikzstyle{bigcommand}=[ultra thick]

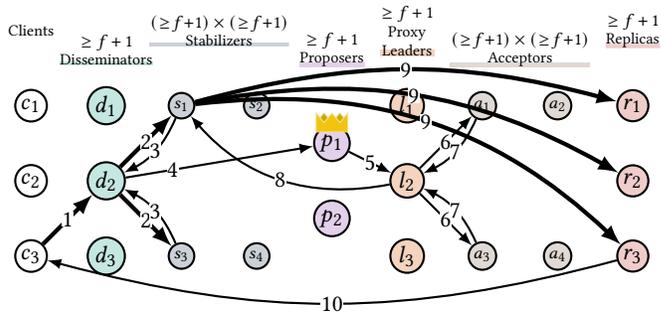
\begin{figure}[h]
  \centering
  \begin{tikzpicture}[xscale=1]
    \node[client] (c1) at (0, 2) {$c_1$};
    \node[client] (c2) at (0, 1) {$c_2$};
    \node[client] (c3) at (0, 0) {$c_3$};
    \node[disseminator] (d1) at (1, 2) {$d_1$};
    \node[disseminator] (d2) at (1, 1) {$d_2$};
    \node[disseminator] (d3) at (1, 0) {$d_3$};
    \node[stabilizer] (s1) at (2, 2) {$s_1$};
    \node[stabilizer] (s2) at (3, 2) {$s_2$};
    \node[stabilizer] (s3) at (2, 0) {$s_3$};
    \node[stabilizer] (s4) at (3, 0) {$s_4$};
    \node[proposer] (p1) at (4, 1.5) {$p_1$};
    \node[proposer] (p2) at (4, 0.5) {$p_2$};
    \node[proxyleader] (pl1) at (5, 2) {$l_1$};
    \node[proxyleader] (pl2) at (5, 1) {$l_2$};
    \node[proxyleader] (pl3) at (5, 0) {$l_3$};
    \node[acceptor] (a1) at (6, 2) {$a_1$};
    \node[acceptor] (a2) at (7, 2) {$a_2$};
    \node[acceptor] (a3) at (6, 0) {$a_3$};
    \node[acceptor] (a4) at (7, 0) {$a_4$};
    \node[replica] (r1) at (8, 2) {$r_1$};
    \node[replica] (r2) at (8, 1) {$r_2$};
    \node[replica] (r3) at (8, 0) {$r_3$};

    \crown{(p1.north)++(0,-0.1)}{0.42}{0.25}
    \node[proclabel] (clients) at (0, 3) {Clients};
    \node[proclabel] (disseminators) at (1, 2.75) {$\geq f+1$\\Disseminators};

    \node[proclabel] (stabilizers) at (2.5, 3) {$(\geq\!f\!+\!1) \times (\geq\!f\!+\!1)$\\Stabilizers};
    \node[proclabel] (acceptors) at (6.5, 2.75) {$(\geq\!f\!+\!1) \times (\geq\!f\!+\!1)$\\Acceptors};

    \node[proclabel] (proposers) at (4, 2.75) {$\geq f+1$\\Proposers};
    \node[proclabel] (proxyleaders) at (5, 3) {$\geq f+1$\\Proxy\\Leaders};
    \node[proclabel] (replicas) at (8, 3) {$\geq f+1$\\Replicas};
    \halffill{clients}{clientcolor!25}
    \quarterfill{disseminators}{disseminatorcolor!25}
    \quarterfill{stabilizers}{stabilizercolor!25}
    \quarterfill{proposers}{proposercolor!25}
    \quarterfill{proxyleaders}{proxyleadercolor!25}
    \quarterfill{acceptors}{acceptorcolor!25}
    \quarterfill{replicas}{replicacolor!25}

    \draw[comm, bigcommand] (c3) to node[commnum]{1} (d2);
    \draw[comm, bigcommand] (d2) to node[commnum]{2} (s1);
    \draw[comm, bigcommand] (d2) to node[commnum]{2} (s3);
    \draw[comm, bend left=20] (s1) to node[commnum]{3} (d2);
    \draw[comm, bend right=20] (s3) to node[commnum]{3} (d2);
    \draw[comm, near start] (d2) to node[commnum]{4} (p1);
    \draw[comm] (p1) to node[commnum]{5} (pl2);

    \draw[comm] (pl2) to node[commnum]{6} (a1);
    \draw[comm] (pl2) to node[commnum]{6} (a3);
    \draw[comm, bend left=20] (a1) to node[commnum]{7} (pl2);
    \draw[comm, bend right=20] (a3) to node[commnum]{7} (pl2);

    \draw[comm, bend left] (pl2) to node[commnum]{8} (s1);
    \draw[comm, bigcommand, bend left=15] (s1) to node[commnum]{9} (r1);
    \draw[comm, bigcommand, bend left=20] (s1) to node[commnum]{9} (r2);
    \draw[comm, bigcommand, bend left=25] (s1) to node[commnum]{9} (r3);
    \draw[comm, bend left=15] (r3) to node[commnum]{10} (c3);
  \end{tikzpicture}
  \caption{%
    An example execution of S-Paxos with stabilizer grids, proxy leaders,
    acceptor grids, and an increased number of replicas.  Messages that include
    client commands (as opposed to ids) are bolded.
  }%
  \figlabel{SPaxosFullOptimizationDiagram}
\end{figure}}
}
}{}
{
%
%
%

\section{Evaluation}\seclabel{Evaluation}

\subsection{Latency-Throughput}\seclabel{Eval/LatencyThroughput}

\begin{figure*}[ht]
  \centering

  \begin{subfigure}[c]{0.4\textwidth}
    \centering
    \includegraphics[width=\textwidth]{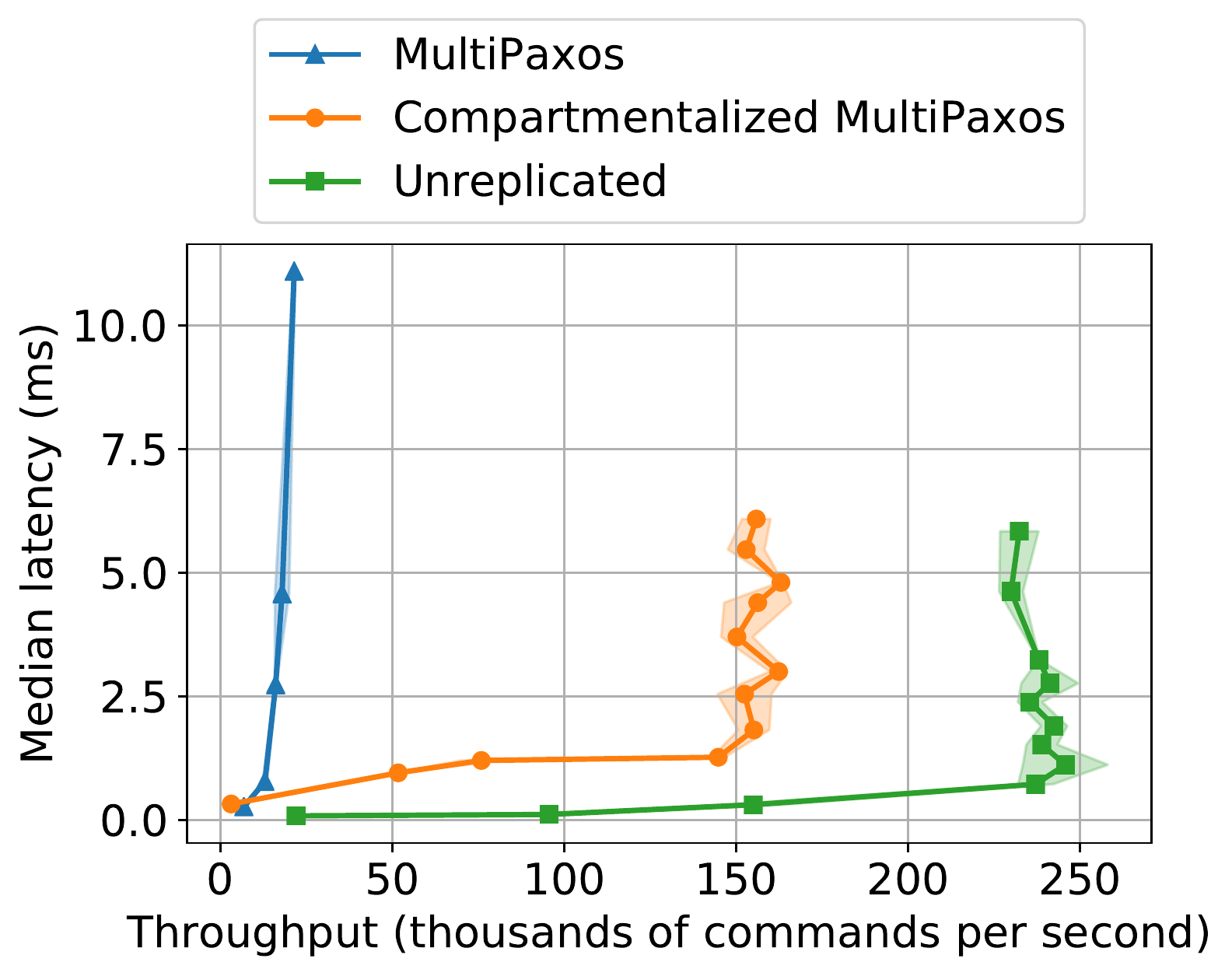}
    \caption{Without batching}%
    \figlabel{EvalMultiPaxosLtUnbatched}
  \end{subfigure}\hspace{0.1\textwidth}%
  \begin{subfigure}[c]{0.4\textwidth}
    \centering
    \includegraphics[width=\textwidth]{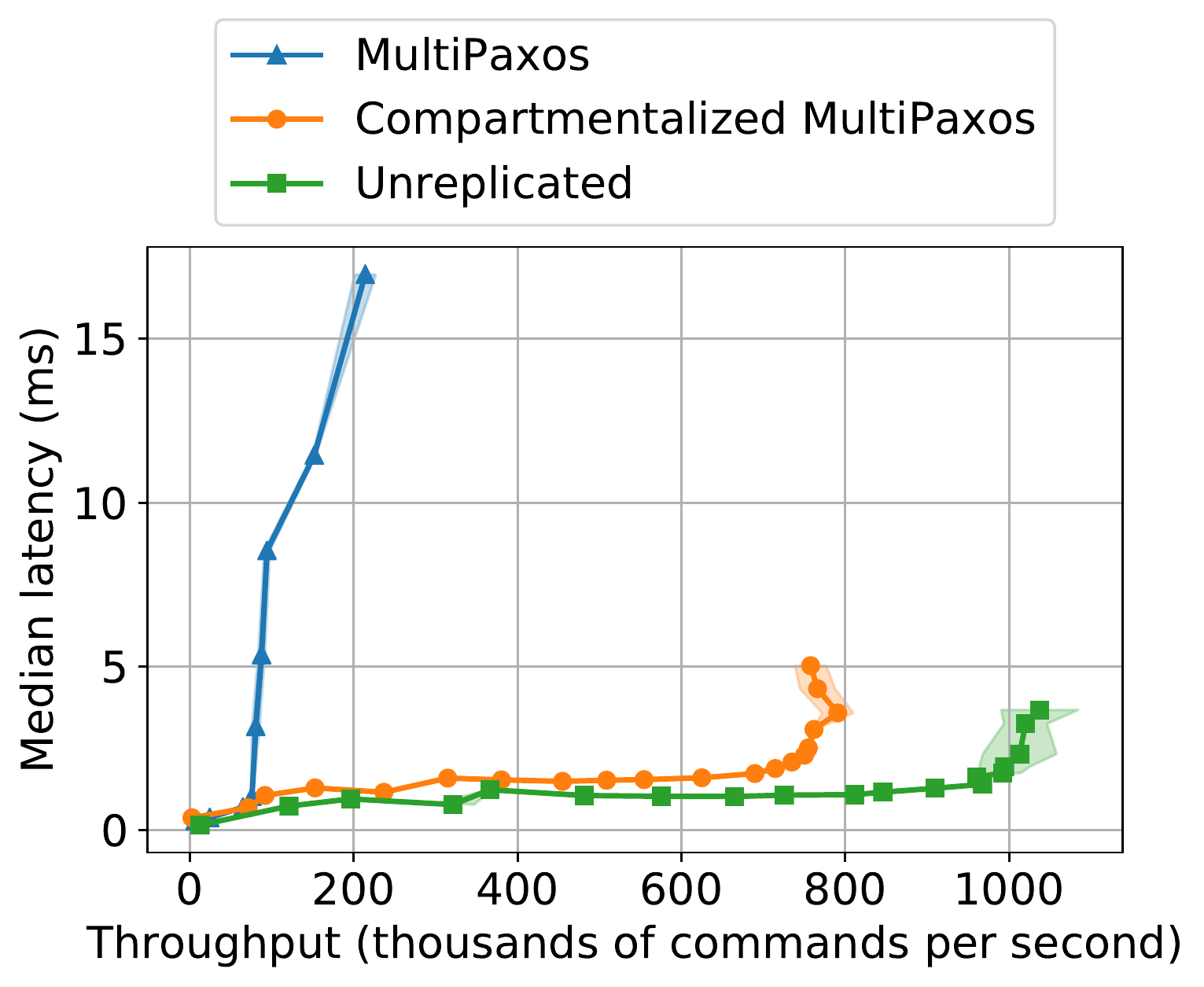}
    \caption{With batching}
    \figlabel{EvalMultiPaxosLtBatched}
  \end{subfigure}

  \caption{
    The latency and throughput of MultiPaxos, Compartmentalized MultiPaxos, and
    an unreplicated state machine.
  }%
  \figlabel{EvalMultiPaxosLt}
\end{figure*}

\paragraph{Experiment Description}
We call MultiPaxos with the six compartmentalizations described in this paper
\defword{Compartmentalized MultiPaxos}. We implemented MultiPaxos,
Compartmentalized MultiPaxos, and an unreplicated state machine in Scala using
the Netty networking library (see
\href{https://github.com/mwhittaker/frankenpaxos/}{github.com/mwhittaker/frankenpaxos}).
MultiPaxos employs $2f+1$ machines with each machine playing the role of a
MultiPaxos proposer, acceptor, and replica.
The unreplicated state machine is implemented as a single process on a single
server. Clients send commands directly to the state machine. Upon receiving a
command, the state machine executes the command and immediately sends back the
result. Note that unlike MultiPaxos and Compartmentalized MultiPaxos, the
unreplicated state machine is \emph{not} fault tolerant. If the single server
fails, all state is lost and no commands can be executed. Thus, the
unreplicated state machine should not be viewed as an apples-to-apples
comparison with the other two protocols. Instead, the unreplicated state
machine sets an upper bound on attainable performance.

We measure the throughput and median latency of the three protocols under
workloads with a varying numbers of clients. Each client issues state machine
commands in a closed loop. It waits to receive the result of executing its most
recently proposed command before it issues another. All three protocols
replicate a key-value store state machine where the keys are integers and the
values are 16 byte strings. In this benchmark, all state machine commands are
writes. There are no reads.

We deploy the protocols with and without batching for $f=1$. Without batching,
we deploy Compartmentalized MultiPaxos with two proposers, ten proxy leaders, a
two by two grid of acceptors, and four replicas. With batching, we deploy two
batchers, two proposers, three proxy replicas, a simple majority quorum system
of three acceptors, two replicas, and three unbatchers.
\begin{revisions}
  For simplicity, every node is deployed on its own machine, but in practice,
  nodes can be physically co-located. In particular, any two logical roles can
  be placed on the same machine without violating fault tolerance constraints,
  so long as the two roles are not the same.
\end{revisions}

We deploy the three protocols on AWS using a set of m5.xlarge machines within a
single availability zone.
\markrevisions{%
  Every m5.xlarge instance has 4 vCPUs and 16 GiB of memory. Everything is done
  in memory, and nothing is written to disk (because everything is replicated,
  data is persistent even without writing it to disk). In our experiments, the
  network is never a bottleneck.
}
All numbers presented are the average of three executions of the
benchmark. As is standard, we implement MultiPaxos and Compartmentalized
MultiPaxos with thriftiness enabled~\cite{moraru2013there}. For a given number
of clients, the batch size is set empirically to optimize throughput. For a
fair comparison, we deploy the unreplicated state machine with a set of
batchers and unbatchers when batching is enabled.

\paragraph{Results}
The results of the experiment are shown in \figref{EvalMultiPaxosLt}.
The standard deviation of throughput measurements are shown as a shaded region.
Without batching, MultiPaxos has a peak throughput of roughly 25,000 commands
per second, while Compartmentalized MultiPaxos has a peak throughput of roughly
150,000 commands per second, a $6\times$ increase. The unreplicated state
machine outperforms both protocols. It achieves a peak throughput of roughly
250,000 commands per second. Compartmentalized MultiPaxos underperforms the
unreplicated state machine because---despite decoupling the leader as much as
possible---the single leader remains a throughput bottleneck.
\markrevisions{%
  Note that after fully compartmentalizing MultiPaxos, either the leader or the
  replicas are guaranteed to be the throughput bottleneck because all other
  components (e.g., proxy leaders, acceptors, batchers, unbatchers) can be
  scaled arbitrarily. Implementation and deployment details (e.g., what state
  machine is being replicated) determine which component is the ultimate
  throughput bottleneck.
}
All three protocols have millisecond latencies at peak throughput. With
batching, MultiPaxos, Compartmentalized MultiPaxos, and the unreplicated state
machine have peak throughputs of roughly 200,000, 800,000 and 1,000,000
commands per second respectively.

Compartmentalized MultiPaxos uses $6.66\times$ more machines than MultiPaxos.
On the surface, this seems like a weakness, but in reality it is a strength.
MultiPaxos does not scale, so it is unable to take advantage of more machines.
Compartmentalized MultiPaxos, on the other hand, achieves a $6 \times$ increase
in throughput using $6.66\times$ the number of resources.
\markrevisions{Thus, we achieve 90\% of perfect linear scalability.}
In fact, with the mixed read-write workloads below, we are able to scale
throughput superlinearly with the number of resources. This is because
compartmentalization eliminates throughput bottlenecks. With throughput
bottlenecks, non-bottlenecked components are underutilized. When we eliminate
the bottlenecks, we eliminate underutilization and can increase performance
without increasing the number of resources.  Moreover, a protocol does not have
to be \emph{fully} compartmentalized. We can selectively compartmentalize some
but not all throughput bottlenecks to reduce the number of resources needed. In
other words, MultiPaxos and Compartmentalized MultiPaxos are not two
alternatives, but rather two extremes in a trade-off between throughput and
resource usage.

\iftoggle{techreportenabled}{
  \marktrrevisions{%
    We also compared the unbatched performance of Compartmentalized MultiPaxos
    and the unreplicated state machine with values being 100 bytes and 1000
    bytes. The results are shown in \figref{BiggerValuesLt}. Expectedly, the
    protocols' peak throughput decreases as we increase the value size.
  }
}{}

\iftoggle{techreportenabled}{%
  \begin{figure}[ht]
    \centering
    \includegraphics[width=\columnwidth]{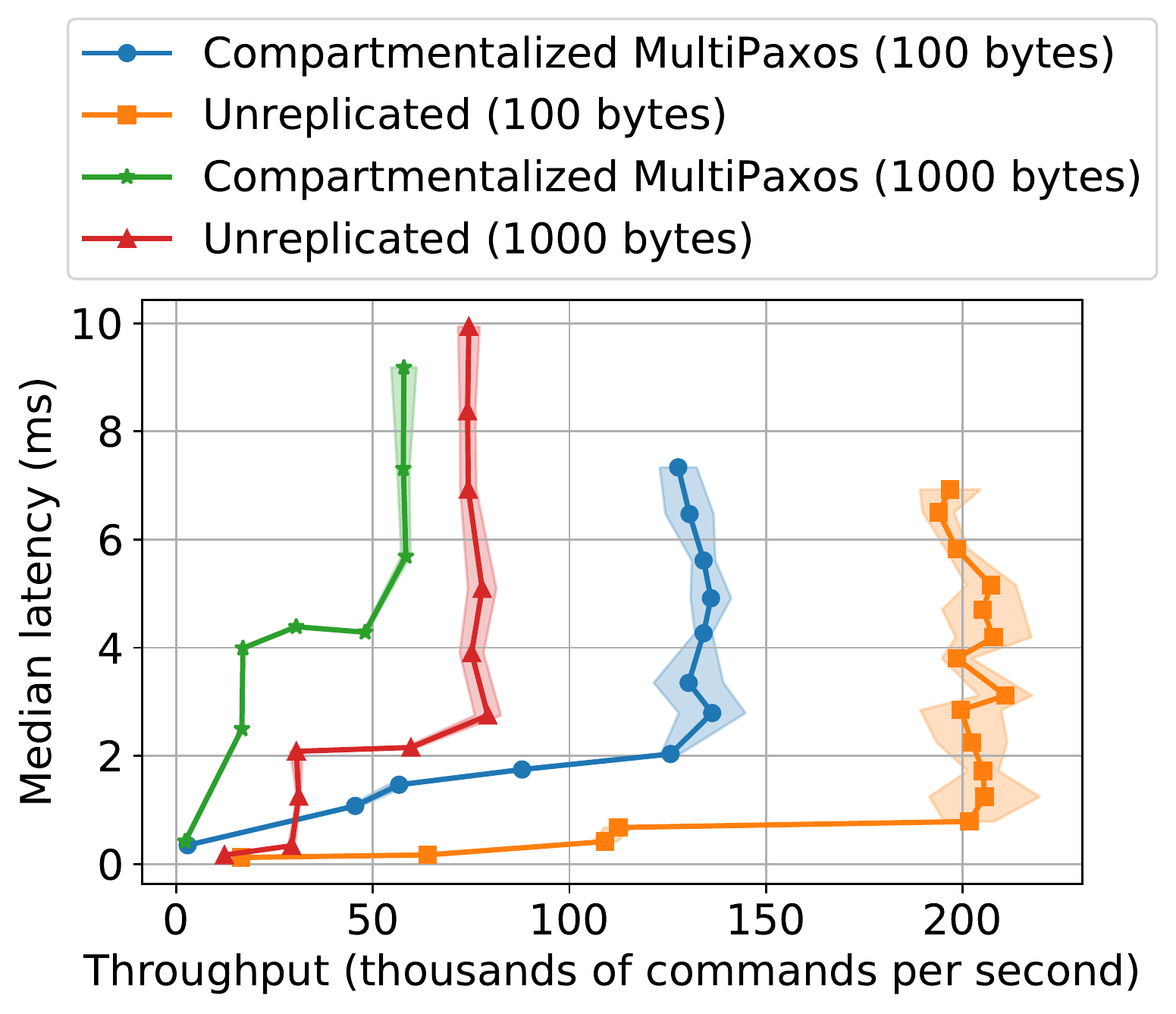}
    \caption{
      \marktrrevisions{%
        The latency and throughput of Compartmentalized MultiPaxos and an
        unreplicated state machine without batching and with larger value sizes.
      }
    }%
    \figlabel{BiggerValuesLt}
  \end{figure}
}{}

\iftoggle{techreportenabled}{
\subsection{\marktrrevisions{Mencius Latency-Throughput}}

\begin{figure*}[ht]
  \centering

  \begin{subfigure}[c]{0.4\textwidth}
    \centering
    \includegraphics[width=\textwidth]{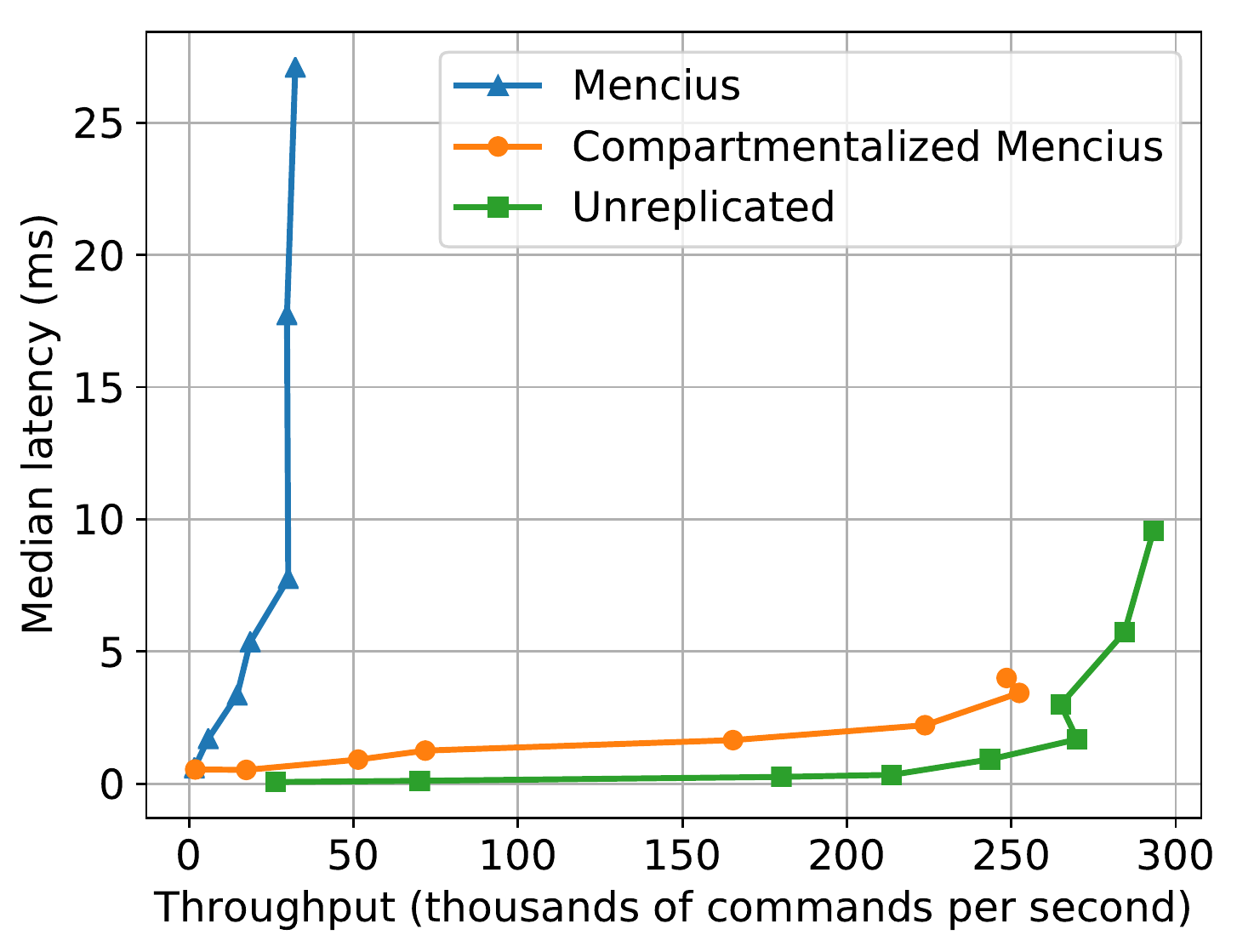}
    \caption{\marktrrevisions{Without batching}}%
    \figlabel{EvalMenciusLtUnbatched}
  \end{subfigure}\hspace{0.1\textwidth}%
  \begin{subfigure}[c]{0.4\textwidth}
    \centering
    \includegraphics[width=\textwidth]{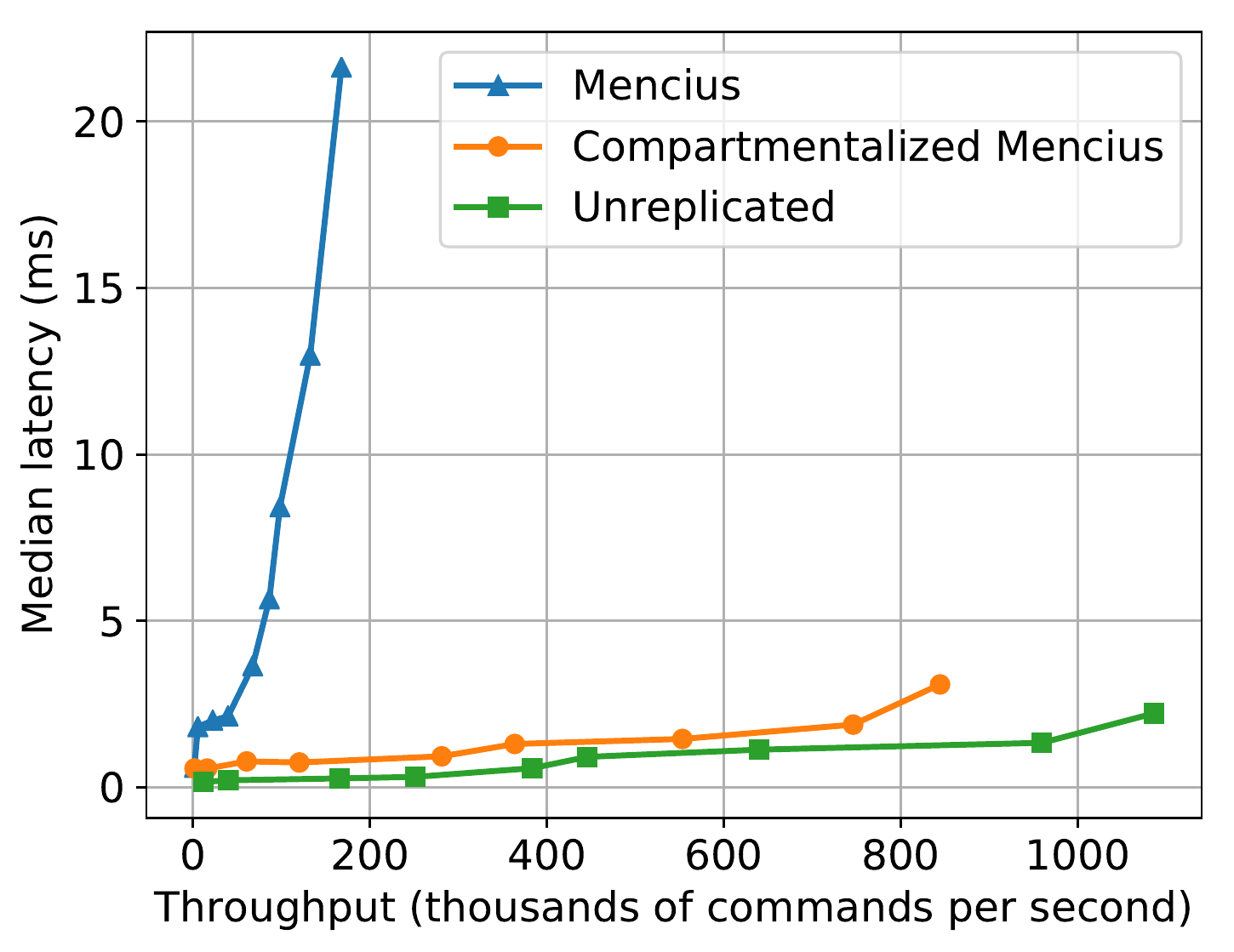}
    \caption{\marktrrevisions{With batching}}
    \figlabel{EvalMenciusLtBatched}
  \end{subfigure}

  \caption{
    \marktrrevisions{%
      The latency and throughput of Mencius, Compartmentalized Mencius, and an
      unreplicated state machine.
    }
  }%
  \figlabel{EvalMenciusLt}
\end{figure*}

\marktrrevisions{%
  We repeat the experiment above but with Mencius and Compartmentalized
  Mencius.  The results are shown in \figref{EvalMenciusLt}. Without batching,
  Mencius can process roughly 30,000 commands per second. Compartmentalized
  Mencius can process roughly 250,000 commands per second, an $8.3\times$
  improvement.  With batching, Mencius and Compartmentalized Mencius achieve
  peak throughputs of nearly 200,000 and 850,000 commands per second
  respectively, a $4.25\times$ improvement.
}
}{}

\iftoggle{techreportenabled}{
\subsection{\marktrrevisions{S-Paxos Latency-Throughput}}

\marktrrevisions{%
  We repeat the experiments above with S-Paxos and Compartmentalized S-Paxos.
  Without batching, Compartmentalized S-Paxos achieves a peak throughput of
  180,000 commands per second compared to S-Paxos' throughput of 22,000 (an
  $8.2\times$ improvement). With batching, Compartmentalized S-Paxos achieves a
  peak throughput of 750,000 commands per second compared to S-Paxos'
  throughput of 180,000 (a $4.16\times$ improvement). Note that our
  implementation of S-Paxos is not as optimized as our other two
  implementations, so its absolute performance is lower.
}
}{}

\subsection{Ablation Study}\seclabel{Eval/Ablation}
{\begin{figure}[ht]
  \centering

  \begin{subfigure}[b]{\columnwidth}
    \centering
    \includegraphics[width=0.85\textwidth]{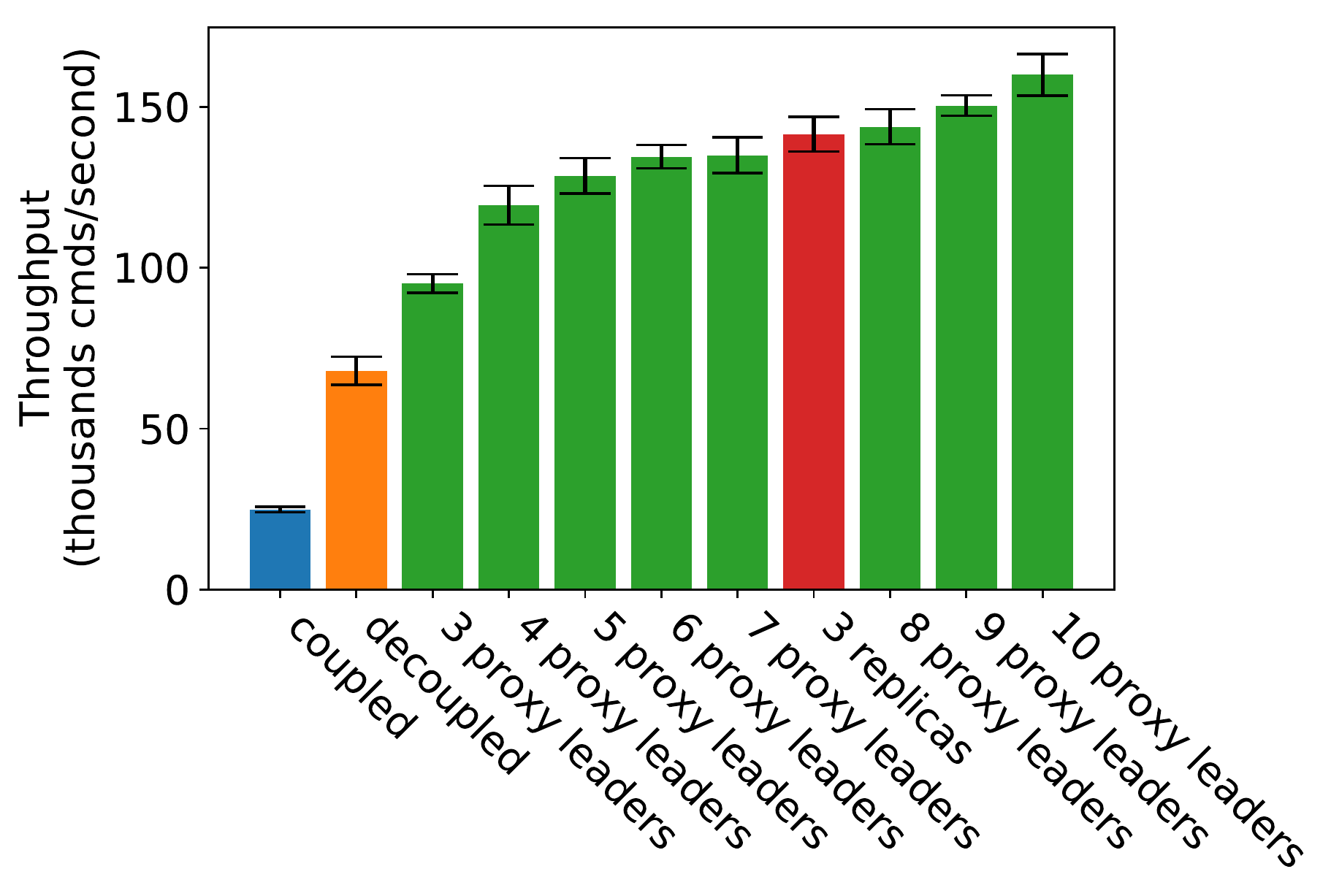}
    \vspace{-9pt}
    \caption{Without batching}%
    \figlabel{EvalMultiPaxosAblationUnbatched}
  \end{subfigure}

  \vspace{12pt}

  \begin{subfigure}[b]{\columnwidth}
    \centering
    \includegraphics[width=0.75\textwidth]{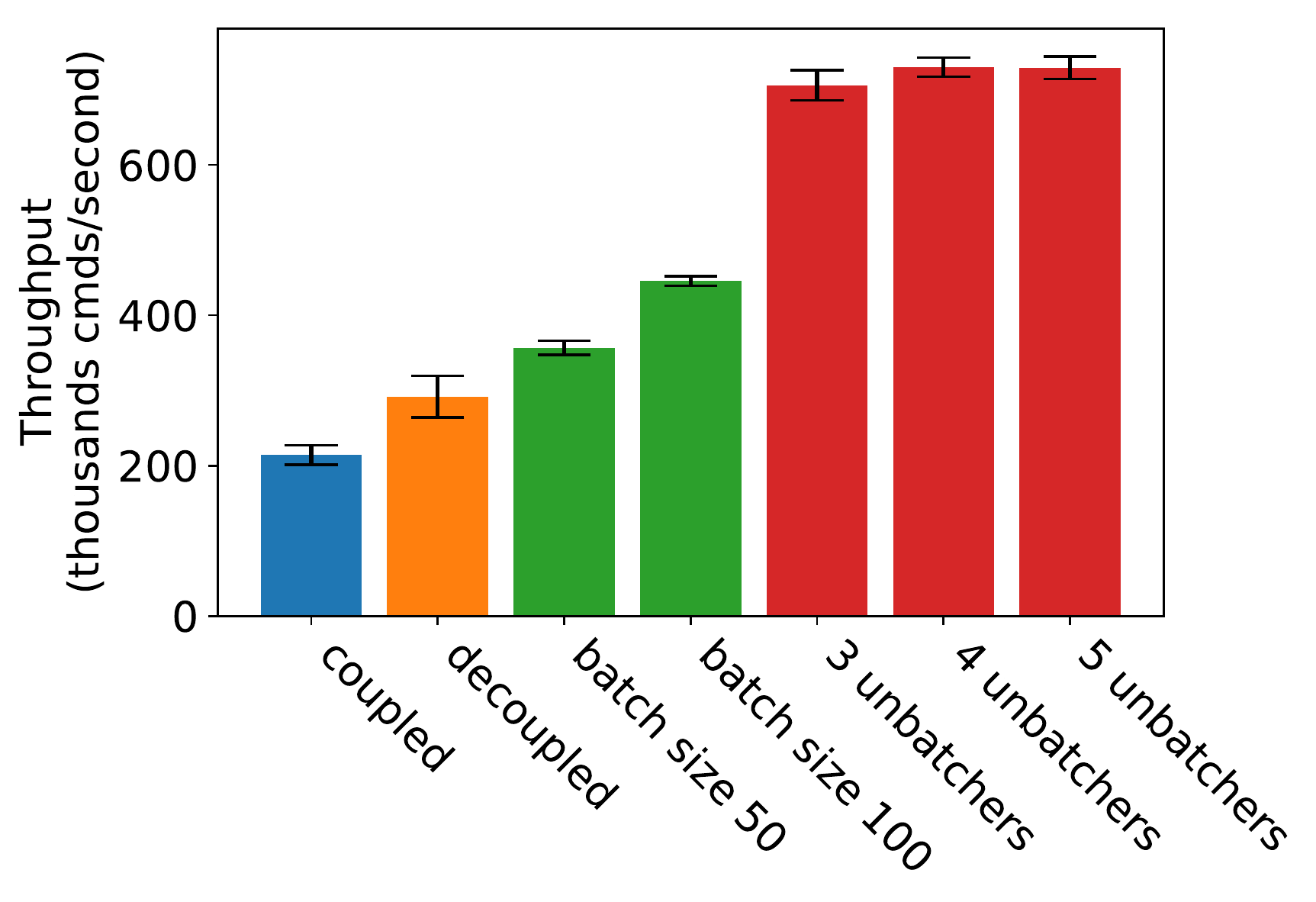}
    \vspace{-6pt}
    \caption{With batching}
    \figlabel{EvalMultiPaxosAblationBatched}
  \end{subfigure}

  \caption{%
    An ablation study. Standard deviations are shown using error bars.
  }%
  \figlabel{EvalMultiPaxosAblation}
\end{figure}
}

\paragraph{Experiment Description}
We now perform an ablation study to measure the effect of each
compartmentalization. In particular, we begin with MultiPaxos and then decouple
and scale the protocol according to the six compartmentalizations, measuring
peak throughput along the way. Note that we cannot measure the effect of each
individual compartmentalization in isolation because decoupling and scaling a
component only improves performance if that component is a bottleneck. Thus, to
measure the effect of each compartmentalization, we have to apply them all, and
we have to apply them in an order that is consistent with the order in which
bottlenecks appear. All the details of this experiment are the same as the
previous experiment unless otherwise noted.

\paragraph{Results}
The unbatched ablation study results are shown in
\figref{EvalMultiPaxosAblationUnbatched}. MultiPaxos has a throughput of
roughly 25,000 commands per second. When we decouple the protocol and introduce
proxy leaders (\secref{MultiPaxos/ProxyLeaders}), we increase the throughput to
roughly 70,000 commands per second. This decoupled MultiPaxos uses the bare
minimum number of proposers (2), proxy leaders (2), acceptors (3), and replicas
(2). We then scale up the number of proxy leaders from 2 to 7. The proxy
leaders are the throughput bottleneck, so as we scale them up, the throughput
of the protocol increases until it plateaus at roughly 135,000 commands per
second. At this point, the proxy leaders are no longer the throughput
bottleneck; the replicas are. We introduce an additional replica
(\secref{MultiPaxos/MoreReplicas}), though the throughput does not increase.
This is because proxy leaders broadcast commands to all replicas, so
introducing a new replica increases the load on the proxy leaders making them
the bottleneck again. We then increase the number of proxy leaders to 10 to
increase the throughput to roughly 150,000 commands per second. At this point,
we determined empirically that the leader was the bottleneck. In this
experiment, the acceptors are never the throughput bottleneck, so increasing
the number of acceptors does not increase the throughput
(\secref{MultiPaxos/AcceptorGrids}). However, this is particular to our
write-only workload. In the mixed read-write workloads discussed momentarily,
scaling up the number of acceptors is critical for high throughput.

The batched ablation study results are shown in
\figref{EvalMultiPaxosAblationBatched}. We decouple MultiPaxos and introduce
two batchers and two unbatchers with a batch size of 10 (\secref{Batchers},
\secref{Unbatchers}). This increases the throughput of the protocol from
200,000 commands per second to 300,000 commands per second. We then increase
the batch size to 50 and then to 100. This increases throughput to 500,000
commands per second. We then increase the number of unbatchers to 3 and reach a
peak throughput of roughly 800,000 commands per second. For this experiment,
two batchers and three unbatchers are sufficient to handle the clients' load.
With more clients and a larger load, more batchers would be needed to maximize
throughput.

\begin{revisions}
  Compartmentalization allows us to decouple and scale protocol components, but
  it doesn't automatically tell us the extent to which we should decouple and
  scale. Understanding this, through ablation studies like the one presented
  here, must currently be done by hand. As a line of future work, we are
  researching how to automatically deduce the optimal amount of decoupling and
  scaling.
\end{revisions}

\subsection{Read Scalability}\seclabel{Eval/ReadScalability}
\begin{figure*}[ht]
  \centering

  \begin{subfigure}[c]{0.45\textwidth}
    \centering
    \includegraphics[width=\textwidth]{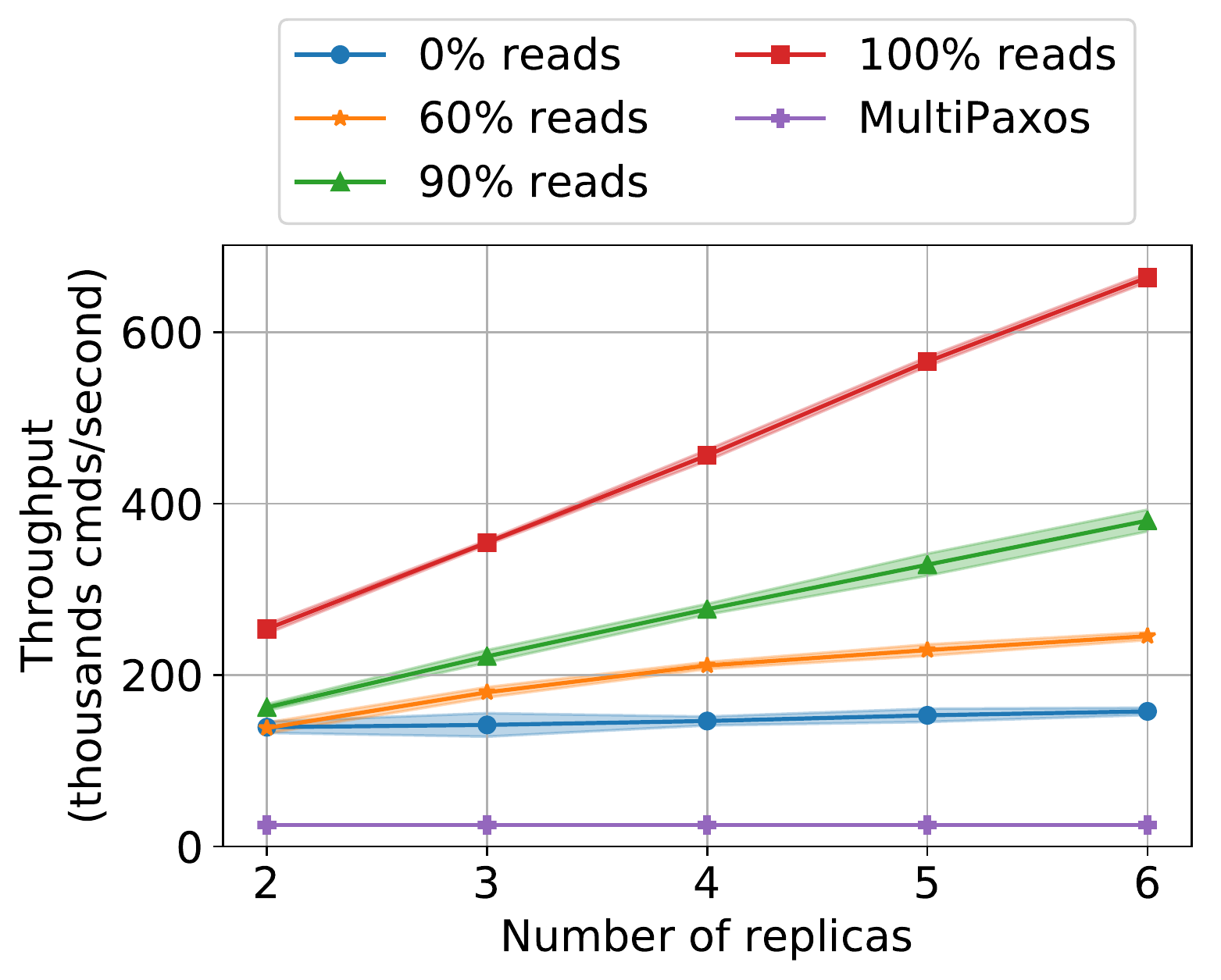}
    \caption{Unbatched linearizable reads}%
    \figlabel{ReadScaleUnbatched}
  \end{subfigure}\hspace{0.1\textwidth}%
  \begin{subfigure}[c]{0.45\textwidth}
    \centering
    \includegraphics[width=\textwidth]{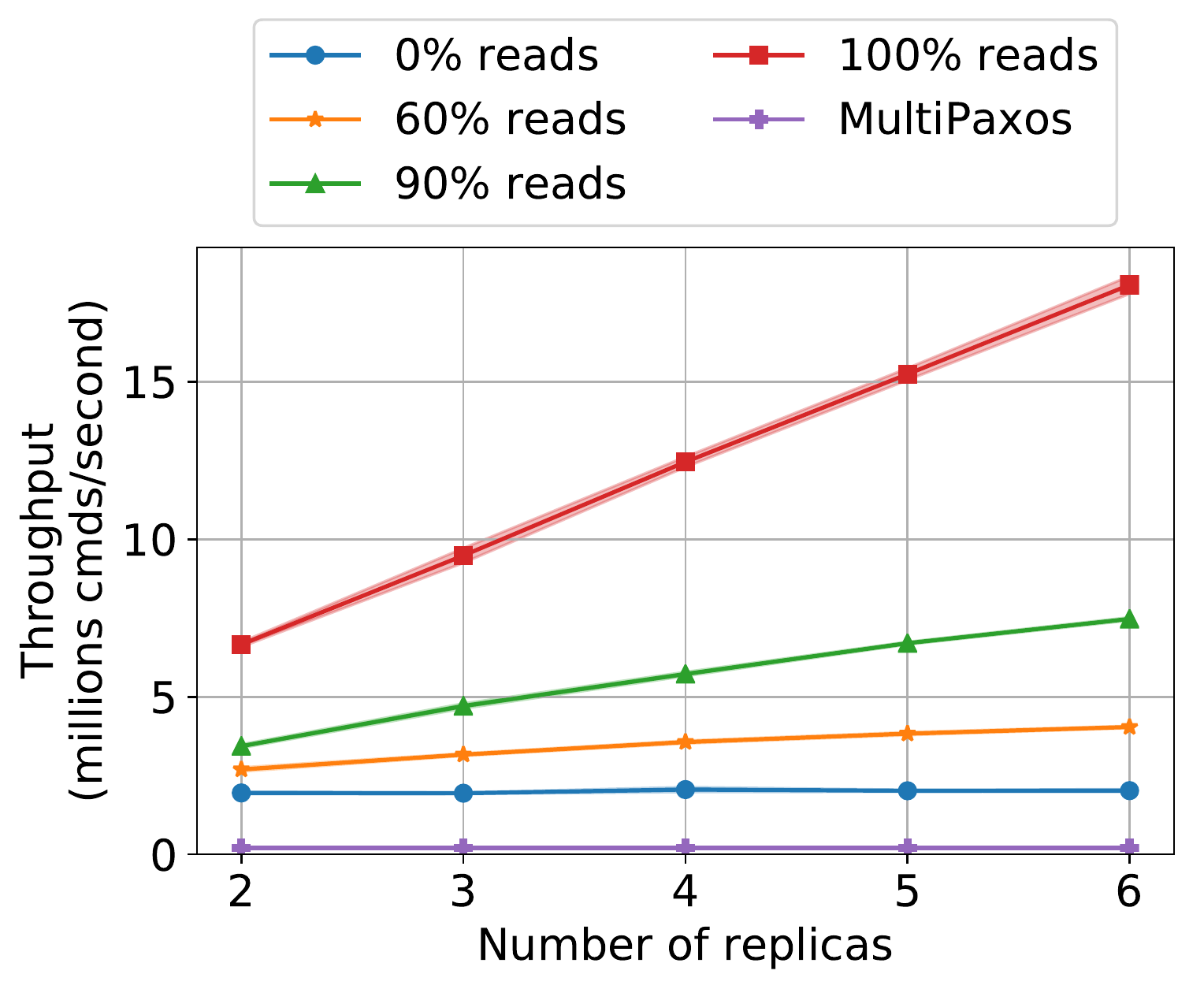}
    \caption{Batched linearizable reads}
    \figlabel{ReadScaleBatched}
  \end{subfigure}

  \caption{
    Peak throughput vs the number of replicas
  }%
  \figlabel{ReadScale}
\end{figure*}

\paragraph{Experiment Description}
Thus far, we have looked at write-only workloads. We now measure the throughput
of Compartmentalized MultiPaxos under a workload with reads \emph{and} writes.
In particular, we measure how the throughput of Compartmentalized MultiPaxos
scales as we increase the number of replicas. We deploy Compartmentalized
MultiPaxos with and without batching; with 2, 3, 4, 5, and 6 replicas; and with
workloads that have 0\%, 60\%, 90\%, and 100\% reads. For any given workload
and number of replicas, proxy leaders, and acceptors is chosen to maximize
throughput. The batch size is 50. In the batched experiments, we do \emph{not}
use batchers and unbatchers. Instead, clients form batches of commands
themselves. This has no effect on the throughput measurements. We did this only
to reduce the number of client machines that we needed to saturate the system.
This was not an issue with the write-only workloads because they had
significantly lower peak throughputs.

\paragraph{Results}
The unbatched results are shown in \figref{ReadScaleUnbatched}. We also show
MultiPaxos' throughput for comparison. MultiPaxos does not distinguish reads
and writes, so there is only a single line to compare against. With a 0\% read
workload, Compartmentalized MultiPaxos has a throughput of roughly 150,000
commands per second, and the protocol does not scale much with the number of
replicas. This is consistent with our previous experiments. For workloads with
reads and writes, our results confirm two expected trends. First, the higher
the fraction of reads, the higher the throughput. Second, the higher the
fraction of reads, the better the protocol scales with the number of replicas.
With a 100\% read workload, for example, Compartmentalized MultiPaxos scales
linearly up to a throughput of roughly 650,000 commands per second with 6
replicas. The batched results, shown in \figref{ReadScaleBatched}, are very
similar. With a 100\% read workload, Compartmentalized MultiPaxos scales
linearly up to a throughput of roughly 17.5 million commands per second.

Our results also show two \emph{counterintuitive} trends. First, a small
increase in the fraction of writes can lead to a disproportionately large
decrease in throughput. For example, the throughput of the 90\% read workload
is far less than 90\% of the throughput of the 100\% read workload. Second,
besides the 100\% read workload, throughput does \emph{not} scale linearly with
the number of replicas. We see that the throughput of the 0\%, 60\%, and 90\%
read workloads scale sublinearly with the number of replicas.
These results are not an artifact of our protocol; they are fundamental. Any
state machine replication protocol where writes are processed by every replica
and where reads are processed by a single replica~\cite{terrace2009object,
zhu2019harmonia, charapko2019linearizable} will exhibit these same two
performance anomalies.

We can explain this analytically. Assume that we have $n$ replicas; that every
replica can process at most $\alpha$ commands per second; and that we have a
workload with a $f_w$ fraction of writes and a $f_r = 1 - f_w$ fraction of
reads.
\iftoggle{techreportenabled}{%
  Let $T$ be peak throughput, measured in commands per second. Then, our
  protocol has a peak throughput of $f_w T$ writes per second and $f_r T$
  reads per second. Writes are processed by \emph{every} replica, so we impose
  a load of $nf_wT$ writes per second on the replicas. Reads are processed by
  \emph{a single} replica, so we impose a load of $f_rT$ reads per second on
  the replicas. The total aggregate throughput of the system is $n\alpha$, so
  we have $n\alpha = nf_wT + f_rT$. Solving for $T$, we find the peak
  throughput of our system is
  \[
    \frac{n \alpha}{nf_w + f_r}
  \]
}{%
  Because writes are processed by \emph{every} replica, and reads are processed
  by a \emph{single} replica, the peak throughput of our system is
  \[
    \frac{n \alpha}{nf_w + f_r}
  \]
}

This formula is plotted in \figref{TheoryTputVsReplicas} with $\alpha =
100,000$. The limit of our peak throughput as $n$ approaches infinity is
$\frac{\alpha}{f_w}$. This explains both of the performance anomalies described
above.
First, it shows that peak throughput has a $\frac{1}{f_w}$ relationship with
the fraction of writes, meaning that a small increase in $f_w$ can have a large
impact on peak throughput. For example, if we increase our write fraction from
1\% to 2\%, our throughput will half. A 1\% change in write fraction leads to a
50\% reduction in throughput.
Second, it shows that throughput does not scale linearly with the number of
replicas; it is upper bounded by $\frac{\alpha}{f_w}$. For example, a workload
with 50\% writes can never achieve more than twice the throughput of a 100\%
write workload, even with an infinite number of replicas.

\iftoggle{techreportenabled}{
  The results for sequentially consistent and eventually consistent reads are
  shown in \figref{WeakReadSCale}. The throughput of these weakly consistent
  reads are similar to that of linearizable reads, but they can be performed
  with far fewer acceptors.
}{}

\begin{figure}[ht]
  \centering
  \includegraphics[width=\columnwidth]{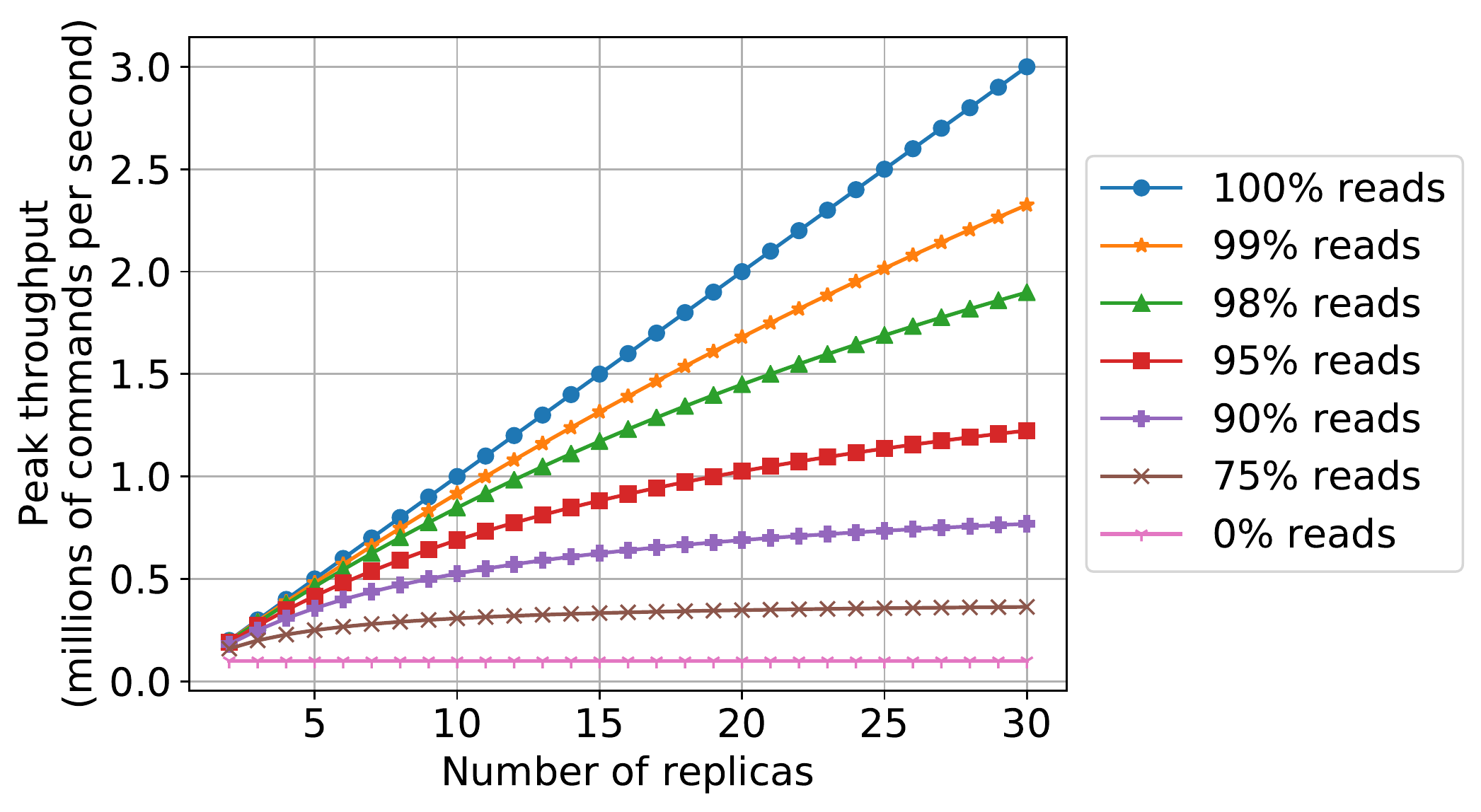}
  \caption{
    Analytical throughput vs the number of replicas.
  }%
  \figlabel{TheoryTputVsReplicas}
\end{figure}

\iftoggle{techreportenabled}{
  \begin{figure*}[ht]
    \centering

    \begin{subfigure}[c]{0.45\textwidth}
      \centering
      \includegraphics[width=\textwidth]{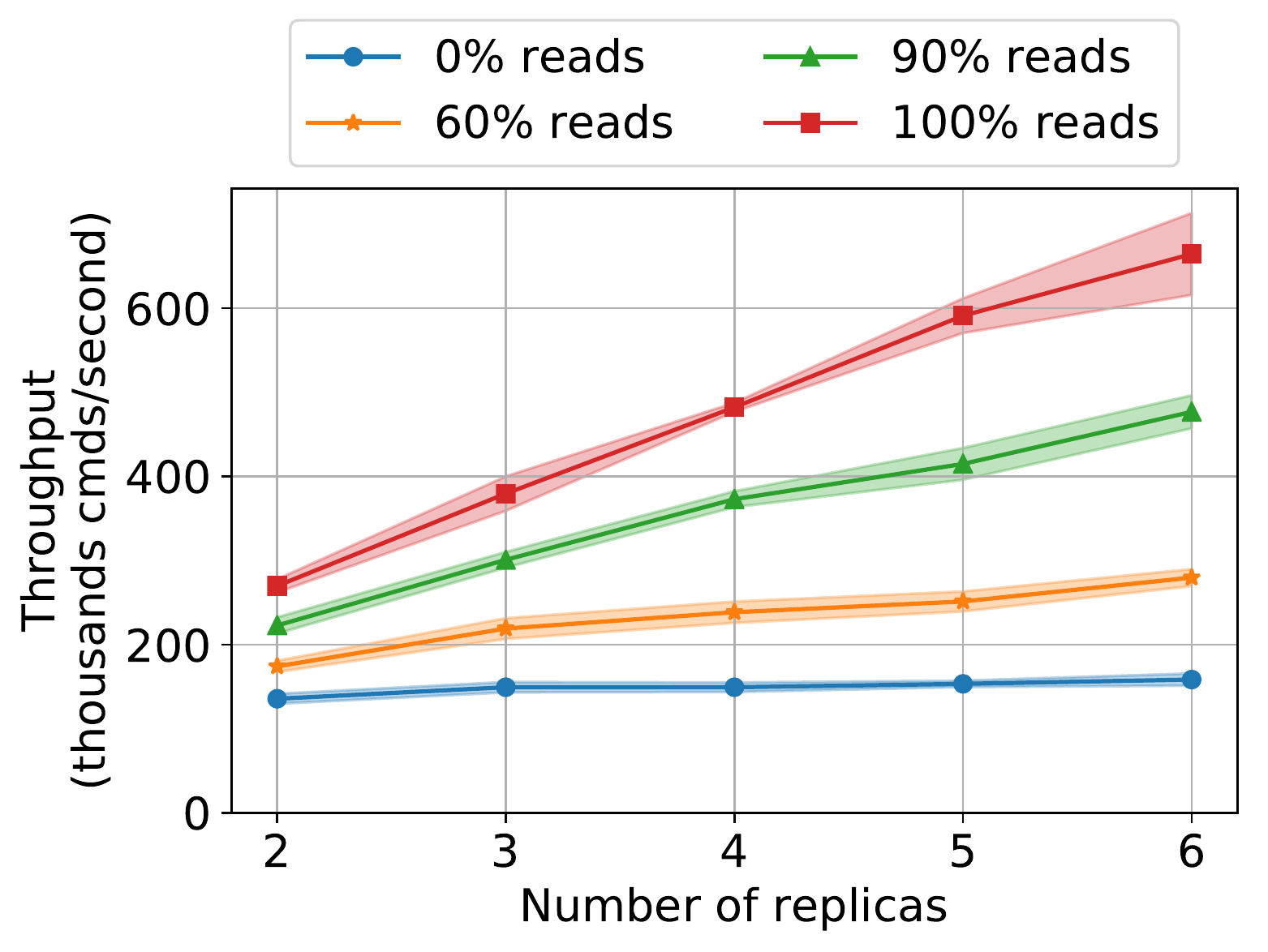}
      \caption{Unbatched eventually consistent reads}%
      \figlabel{TODO}
    \end{subfigure}\hspace{0.1\textwidth}%
    \begin{subfigure}[c]{0.45\textwidth}
      \centering
      \includegraphics[width=\textwidth]{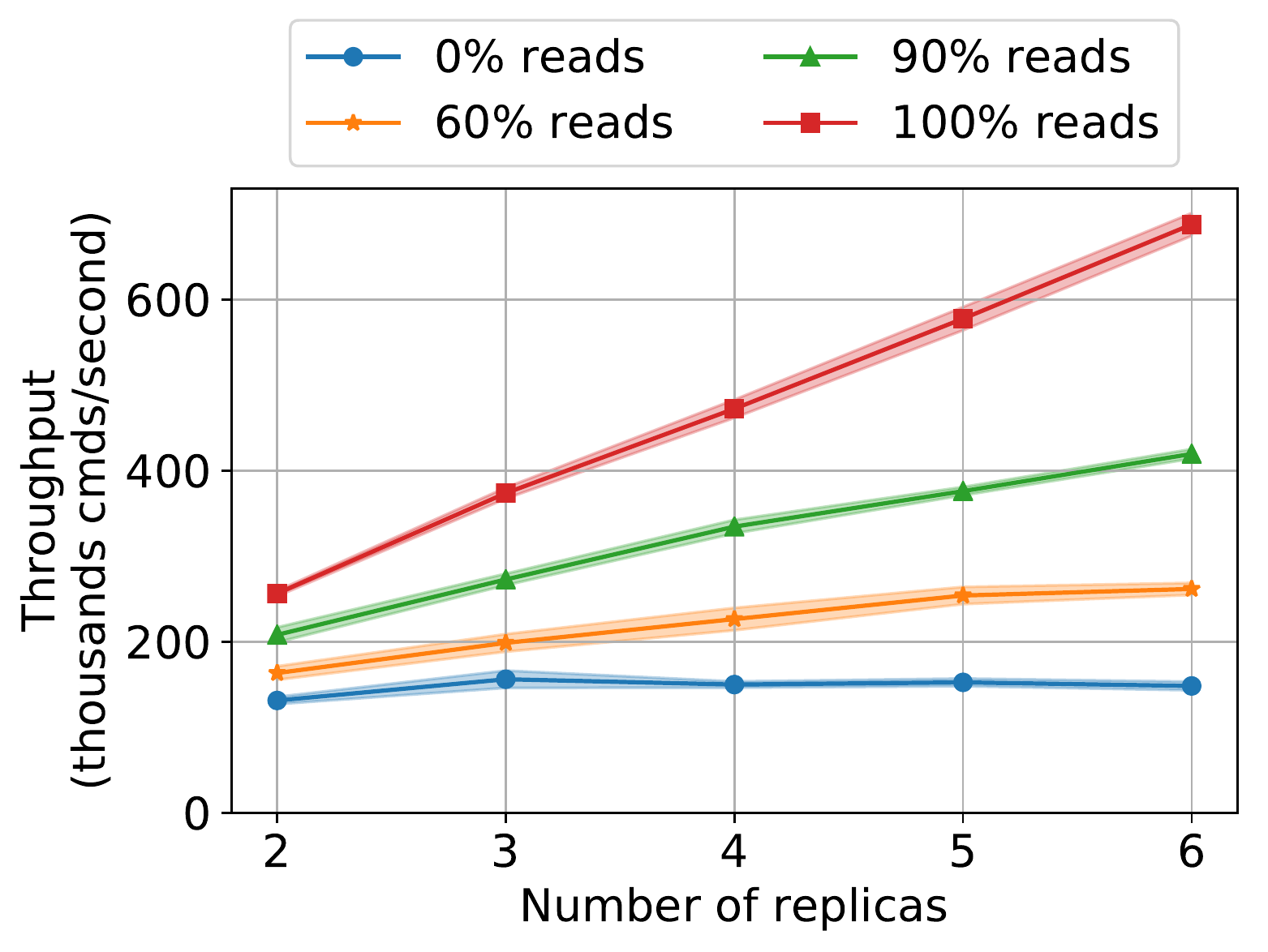}
      \caption{Unbatched sequentially consistent reads}
      \figlabel{TODO}
    \end{subfigure}

    \caption{
      Peak throughput vs the number of replicas
    }%
    \figlabel{WeakReadSCale}
  \end{figure*}
}{}

\subsection{Skew Tolerance}\seclabel{Eval/SkewTolerance}
\paragraph{Experiment Description}
CRAQ~\cite{terrace2009object} is a chain replication~\cite{van2004chain}
variant with scalable reads. A CRAQ deployment consists of at least $f+1$ nodes
arranged in a linked list, or chain. Writes are sent to the head of the chain
and propagated node-by-node down the chain from the head to the tail.  When the
tail receives the write, it sends a write acknowledgement to its predecessor,
and this ack is propagated node-by-node backwards through the chain until it
reaches the head. Reads are sent to any node. When a node receives a read of
key $k$, it checks to see if it has any unacknowledged write to that key. If it
doesn't, then it performs the read and replies to the client immediately. If it
does, then it forwards the read to the tail of the chain. When the tail
receives a read, it executes the read immediately and replies to the client.

We now compare Compartmentalized MultiPaxos with our implementation of CRAQ. In
particular, we show that CRAQ (and similar protocols like
Harmonia~\cite{zhu2019harmonia}) are sensitive to data skew, whereas
Compartmentalized MultiPaxos is not.
%
\markrevisions{%
  We deploy Compartmentalized MultiPaxos with two proposers, three proxy
  leaders, twelve acceptors, and six replicas, and we deploy CRAQ with six
  chain nodes. Though, our results hold for deployments with a different number
  of machines as well, as long as the number of Compartmentalized MultiPaxos
  replicas is equal to the number of CRAQ chain nodes.
}
Both protocols replicate a key-value store with 10,000 keys in the
range $1, \ldots, 10,000$. We subject both protocols to the following workload.
A client repeatedly flips a weighted coin, and with probability $p$ chooses to
read or write to key $0$. With probability $1-p$, it decides to read or write
to some other key $2, \ldots, 10,000$ chosen uniformly at random. The client
then decides to perform a read with 95\% probability and a write with 5\%
probability. As we vary the value of $p$, we vary the skew of the workload.
When $p=0$, the workload is completely uniform, and when $p=1$, the workload
consists of reads and writes to a single key. This artificial workload allows
to study the effect of skew in a simple way without having to understand more
complex skewed distributions.

\paragraph{Results}
The results are shown in \figref{Skew}, with $p$ on the $x$-axis. The
throughput of Compartmentalized MultiPaxos is constant; it is independent of
$p$. This is expected because Compartmentalized MultiPaxos is completely
agnostic to the state machine that it is replicating and is completely unaware
of the notion of keyed data. Its performance is only affected by the ratio of
reads to writes and is completely unaffected by what data is actually being
read or written. CRAQ, on the other hand, is susceptible to skew. As we
increase skew from $p=0$ to $p=1$, the throughput decreases from roughly
300,000 commands per second to roughly 100,000 commands per second. As we
increase $p$, we increase the fraction of reads which are forwarded to the
tail. In the extreme, all reads are forwarded to the tail, and the throughput
of the protocol is limited to that of a single node (i.e.\ the tail).

However, with low skew, CRAQ can perform reads in a single round trip to a
single chain node. This allows CRAQ to implement reads with lower latency and
with fewer nodes than Compartmentalized MultiPaxos. However, we also note that
Compartmentalized MultiPaxos outperforms CRAQ in our benchmark even with no
skew. This is because every chain node must process four messages per write,
whereas Compartmentalized MultiPaxos replicas only have to process two. CRAQ's
write latency also increases with the number of chain nodes, creating a hard
trade-off between read throughput and write latency. Ultimately, neither
protocol is strictly better than the other. For very read-heavy workloads with
low-skew, CRAQ will likely outperform Compartmentalized MultiPaxos
\markrevisions{using fewer machines},
and for workloads with more writes or more skew, Compartmentalized MultiPaxos
will likely outperform CRAQ.
\begin{revisions}
  For the 95\% read workload in our experiment, Compartmentalized MultiPaxos
  has strictly better throughput than CRAQ across all skews, but this is not
  true for workloads with a higher fraction of reads.
\end{revisions}

\begin{figure}[ht]
  \centering
  \includegraphics[width=\columnwidth]{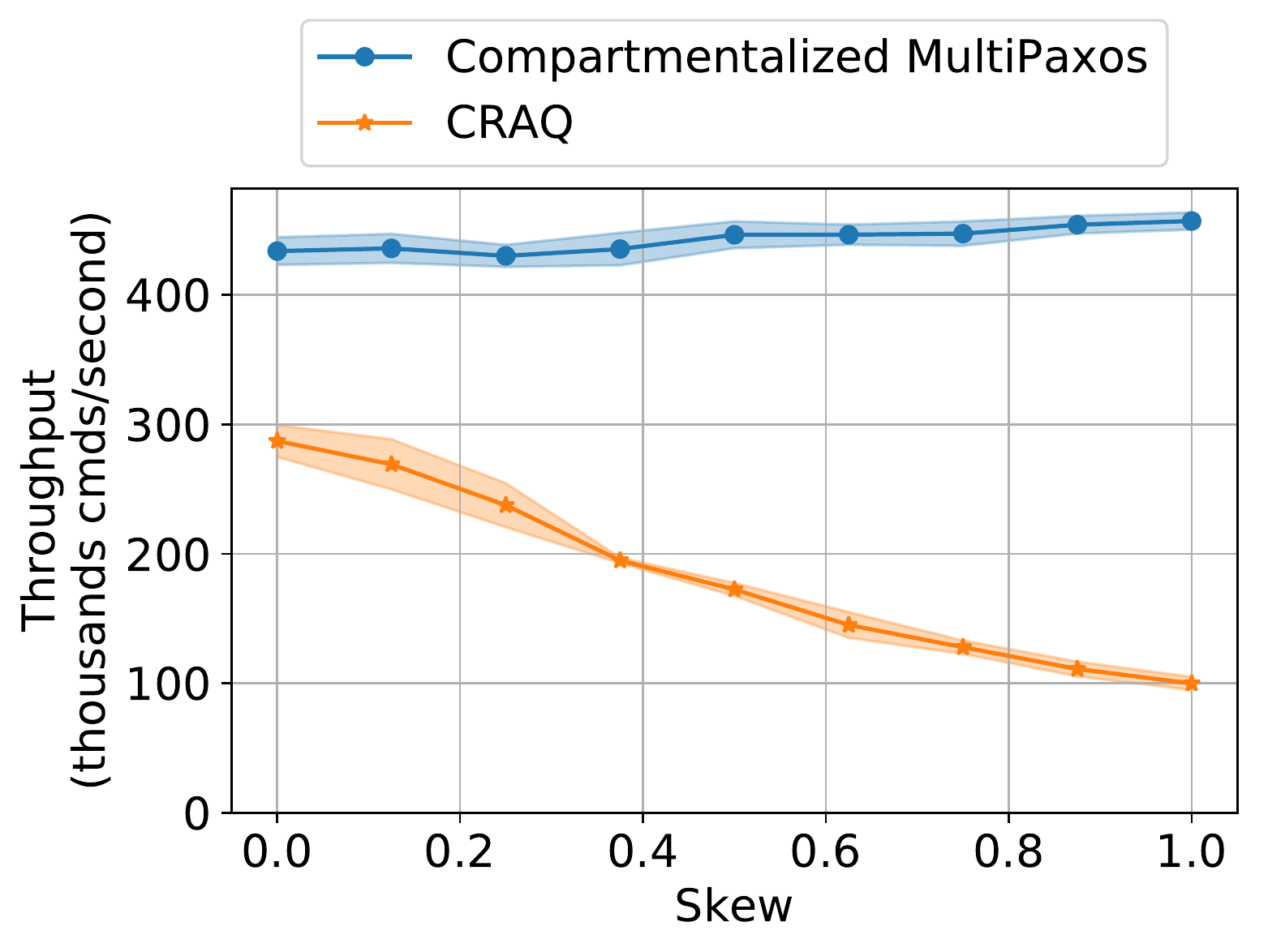}
  \caption{
    The effect of skew on Compartmentalized MultiPaxos and CRAQ.
  }%
  \figlabel{Skew}
\end{figure}

\iftoggle{techreportenabled}{
\subsection{\marktrrevisions{Comparison to Scalog}}\seclabel{Eval/Scalog}

\marktrrevisions{%
  \paragraph{Experiment Description}
  Scalog~\cite{ding2020scalog} is a replicated shared log protocol that
  achieves high throughput using an idea similar to Compartmentalized
  MultiPaxos' batchers and unbatchers. We implemented Scalog to compare against
  Compartmentalized MultiPaxos with batching, but the two protocols are not
  immediately comparable. Scalog is replicated shared log protocol, whereas
  Compartmentalized MultiPaxos is a state machine replication protocol.
  Practically, the main difference between the two is that Scalog doesn't have
  state machine replicas. To compare the two protocols fairly, we extended
  Scalog with replicas to implement state machine replication.
}

\marktrrevisions{%
  For simplicity, we implemented Scalog with a single aggregator, though we
  confirmed empirically that it was never the bottleneck. We deploy Scalog with
  three shards, two servers per shard, and five replicas. Servers send their
  local cuts to the aggregator after every 100 commands received. Re-running
  the same experiment from \secref{Eval/LatencyThroughput}, we measure the
  latency and throughput of the protocol on a write-only workload as we vary
  the number of clients.
}

\begin{figure}[ht]
  \centering
  \includegraphics[width=\columnwidth]{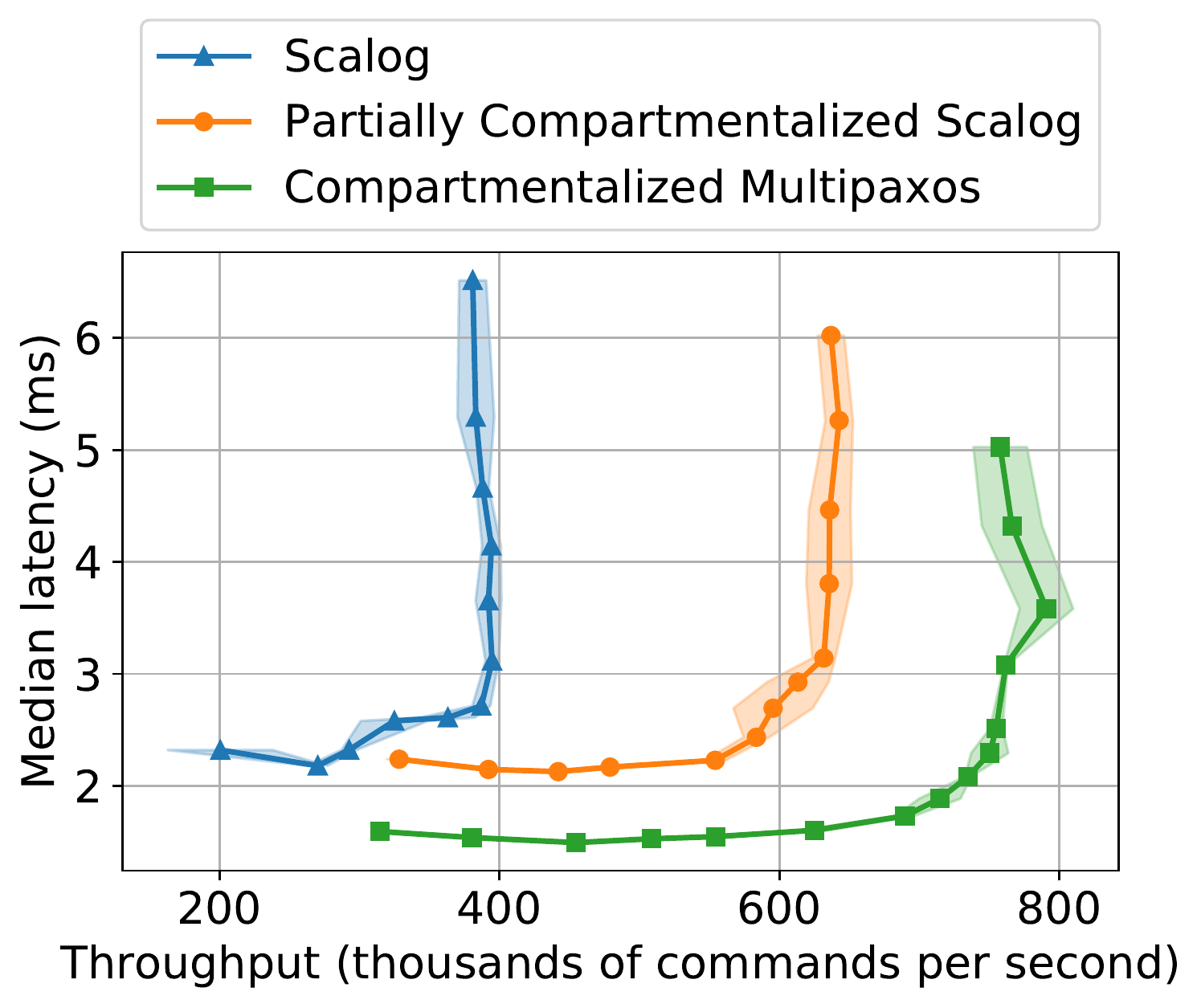}
  \caption{
    \marktrrevisions{%
      The latency and throughput of Scalog, Compartmentalized Scalog, and
      Compartmentalized MultiPaxos.
    }
  }%
  \figlabel{ScalogLt}
\end{figure}

\marktrrevisions{%
  \paragraph{Results}
  The results are shown in \figref{ScalogLt}. Scalog achieves a peak throughput
  of roughly 400,000 commands per second compared to Compartmentalized
  MultiPaxos' throughput of 800,000 commands per second. Scalog also uses 17
  machines, and Compartmentalized MultiPaxos uses 15. The Scalog servers
  implement batching very efficiently, however, batching is not the
  bottleneck in our experiment. The protocol is instead bottlenecked by the
  replicas. This demonstrates the importance of eliminating \emph{every}
  bottleneck rather than eliminating a single bottleneck. We compartmentalized
  the Scalog replicas by introducing proxy replicas. When deployed with two
  replicas and four proxy replicas, the protocol's throughput increased to
  roughly 650,000 requests per second. This is a $1.625\times$ increase in
  throughput using one additional machine (a $1.05\times$ increase in the
  number of machines).
}
}{}
}
{\section{Related Work}

\paragraph{MultiPaxos}
Unlike state machine replication protocols like Raft~\cite{ongaro2014search}
and Viewstamped Replication~\cite{liskov2012viewstamped},
MultiPaxos~\cite{lamport1998part, lamport2001paxos, van2015paxos} is designed
with the roles of proposer, acceptor, and replicas logically decoupled. This
decoupling alone is not sufficient for MultiPaxos to achieve the best possible
throughput, but the decoupling allows for the compartmentalizations described
in this paper.

\paragraph{PigPaxos}
PigPaxos~\cite{charapko2020pigpaxos} is a MultiPaxos variant that alters the
communication flow between the leader and the acceptors to improve scalability
and throughput. Similar to compartmentalization, PigPaxos realizes that the
leader is doing many different jobs and is a bottleneck in the system. In
particular, PigPaxos substitutes direct leader-to-acceptor communication with a
relay network. In PigPaxos the leader sends a message to one or more randomly
selected relay nodes, and each relay rebroadcasts the leader's message to the
peers in its relay-group and waits for some threshold of responses. Once each
relay receives enough responses from its peers, it aggregates them into a
single message to reply to the leader. The leader selects a new set of random
relays for each new message to prevent faulty relays from having a long-term
impact on the communication flow.  PigPaxos relays are comparable to our proxy
leaders, although the relays are simpler and only alter the communication flow.
As such, the relays cannot generally take over the other leader roles, such as
quorum counting or replying to the clients. Unlike PigPaxos, whose main goal is
to grow to larger clusters, compartmentalization is more general and improves
throughput under different conditions and situations.

\iftoggle{techreportenabled}{%
  \paragraph{Chain Replication}
  Chain Replication~\cite{van2004chain} is a state machine replication protocol
  in which the set of state machine replicas are arranged in a totally ordered
  chain. Writes are propagated through the chain from head to tail, and reads are
  serviced exclusively by the tail. Chain Replication has high throughput
  compared to MultiPaxos because load is more evenly distributed between the
  replicas, but every replica must process four messages per command, as opposed
  to two in Compartmentalized MultiPaxos. The tail is also a throughput
  bottleneck for read-heavy workloads. Finally, Chain Replication is not tolerant
  to network partitions and is therefore not appropriate in all situations.
}{}

\iftoggle{techreportenabled}{%
  \paragraph{Ring Paxos}
  Ring Paxos~\cite{marandi2010ring} is a MultiPaxos variant that decouples
  control flow from data flow (as in S-Paxos~\cite{biely2012s}) and that
  arranges nodes in a chain (as in Chain Replication). As a result, Ring Paxos
  has the same advantages as S-Paxos and Chain Replication. Like S-Paxos and
  Mencius, Ring Paxos eliminates some but not all throughput bottlenecks. It
  also does not optimize reads; reads are processed the same as writes.
}{}

\iftoggle{techreportenabled}{%
  \paragraph{NoPaxos}
  NoPaxos~\cite{li2016just} is a Viewstamped
  Replication~\cite{liskov2012viewstamped} variant that depends on an ordered
  unreliable multicast (OUM) layer. Each client sends commands to a centralized
  sequencer that is implemented on a network switch. The sequencer assigns
  increasing IDs to the commands and broadcasts them to a set of replicas. The
  replicas speculatively execute commands and reply to clients. In this paper, we
  describe how to use proxy leaders to avoid having a centralized leader.
  NoPaxos' on-switch sequencer is a hardware based alternative to avoid the
  bottleneck.
}{}

\paragraph{Scalog}
Scalog~\cite{ding2020scalog} is a replicated shared log protocol that achieves
high throughput using an idea similar to Compartmentalized MultiPaxos' batchers
and unbatchers. A client does not send values directly to a centralized leader
for sequencing in the log. Instead, the client sends its values to one of a
number of servers. Periodically, the servers' batches are sealed and assigned
an id. This id is then sent to a state machine replication protocol, like
MultiPaxos, for sequencing.
\iftoggle{techreportenabled}{}{%
  \markrevisions{%
    Compartmentalization and Scalog differ in many ways. The biggest difference
    is the fact that compartmentalization is a transferable technique, while
    Scalog is a specific protocol. Restricting our attention to
    Compartmentalized MultiPaxos, the two still differ. For exmaple, Scalog cannot perform
    fast linearizable reads like Compartmentalized MultiPaxos can (see
    \secref{MultiPaxos/LeaderlessReads}). Scalog aggregators must also be
    carefully managed.  If the root fails, for example, ``goodput'' in Scalog
    drops to zero.
  }
}

\iftoggle{techreportenabled}{%
\marktrrevisions{%
Compartmentalization and Scalog are fundamentally different because
compartmentalization is a generally applicable technique, while Scalog is a
protocol. Because compartmentalization is a technique, it can be applied to
many existing protocols. In this paper, we have shown how to compartmentalize
MultiPaxos, Mencius, and S-Paxos. In~\cite{whittaker2020bipartisan}, we show
how to compartmentalize EPaxos~\cite{moraru2013there}. In \secref{Eval/Scalog},
we even show how to partially compartmentalize Scalog. Scalog, on the other
hand, is a specific protocol.
}

\marktrrevisions{%
Restricting our attention to Compartmentalized MultiPaxos, Scalog and
Compartmentalized MultiPaxos are still fundamentally different. Scalog is a log
replication protocol, whereas Compartmentalized MultiPaxos is a complete state
machine replication protocol. In other words, Scalog doesn't have state machine
replicas, while Compartmentalized MultiPaxos does. These two types of protocols
are similar, but differ in a number of key ways. For example, to implement
state machine replication on top of Scalog, we have to implement the replicas
ourselves, and as discussed in \secref{Eval/Scalog}, these replicas become the
protocol's throughput bottleneck when they're not compartmentalized. As an
another example, Scalog cannot perform fast linearizable reads like
Compartmentalized MultiPaxos can, as described in
\secref{MultiPaxos/LeaderlessReads}.
}

\marktrrevisions{%
Ignoring this, the two protocols still have different focuses. Scalog focuses
on \emph{one} bottleneck. Specifically, the protocol implements an efficient
form of batching, with a focus on replicating commands before ordering them.
Compartmentalized MultiPaxos, on the other hand, provides a framework to
address \emph{every} bottleneck. Compartmentalized batchers solve the same
problem as Scalog servers, and the introduction of proxy leaders, flexible
quorums, unbatchers, decoupled read and write paths addresses the other
bottlenecks that we have seen arise. Scalog does not address these other
bottlenecks.  We show this empirically in \secref{Eval/Scalog}. This is a
contribution of our paper.
}

\marktrrevisions{%
There are other technical differences between the two protocols as well. For
example, there is complexity in managing the Scalog aggregators. If one fails,
we need to detect and replace it. If the root fails, for example, the short
term ``goodput'' drops to zero. Moreover, if a Compartmentalized batcher in
Compartmentalized MultiPaxos fails, we lose all of its unsent commands. If any
server in a Scalog shard fails, then \emph{any} unreplicated commands on
\emph{any} of the servers within the shard are lost.
}
}

\iftoggle{techreportenabled}{%
  \paragraph{Scalable Agreement}
  In~\cite{kapritsos2010scalable}, Kapritsos et al.\ present a protocol similar
  to Compartmentalized Mencius\iftoggle{techreportenabled}{.}{ (as described in
  our technical report~\cite{whittaker2020scaling}).} The protocol round-robin
  partitions log entries among a set of replica clusters co-located on a fixed
  set of machines.  Every cluster has $2f+1$ replicas, with every replica playing
  the role of a Paxos proposer and acceptor. Compartmentalized Mencius extends
  the protocol with the compartmentalizations described in this paper.
}{}

\iftoggle{techreportenabled}{%
  \paragraph{SEDA Architecture}
  The SEDA architecture~\cite{welsh2001seda} is a server architecture in which
  functionality is divided into a pipeline of multithreaded modules that
  communicate with one another using queues. This architecture introduces
  pipeline parallelism and allows individual components to be scaled up or down
  to avoid becoming the bottleneck. Our work on decoupling and scaling state
  machine replication protocols borrows these same ideas, except that we apply
  them at a fine grain to distributed protocols rather than a single server.
}{}

\iftoggle{techreportenabled}{%
  \paragraph{Multithreaded Replication}
  \cite{santos2013achieving} and \cite{behl2015consensus} both propose
  multithreaded state machine replication protocols.  The protocol
  in~\cite{santos2013achieving} is implemented using a combination of actors
  and the SEDA architecture~\cite{welsh2001seda}. A replica's functionality is
  decoupled into a number of modules, with each module run on its own thread.
  For example, a MultiPaxos leader has one module to receive messages, one to
  sequence them, and one to send them. \cite{behl2015consensus} argues for a
  Mencius-like approach in which each thread has complete functionality
  (receiving, sequencing, and sending), with slots round-robin partitioned
  across threads.  Multithreaded protocols like these are necessarily decoupled
  and scale within a single machine. This work is complementary to
  compartmentalization.  Compartmentalization works at the protocol level,
  while multithreading works on the process level. Both can be applied to a
  single protocol.
}{}

\iftoggle{techreportenabled}{%
  \paragraph{A Family of Leaderless Generalized Protocols}
  In~\cite{losa2016brief}, Losa et al.\ propose a template that can be used to
  implement state machine replication protocols that are both leaderless and
  generalized. The template involves a module to compute dependencies between
  commands and a module to choose and execute commands. The goal of this
  modularization is to unify existing protocols like
  EPaxos~\cite{moraru2013there}, and Caesar~\cite{arun2017speeding}. However, the
  modularity also introduces decoupling which can lead to performance gains. This
  is an example of compartmentalization.
}{}

\paragraph{Read Leases}
A common way to optimize reads in MultiPaxos is to grant a lease to the
leader~\cite{chandra2007paxos, corbett2013spanner, burrows2006chubby}. While
the leader holds the lease, no other node can become leader. As a result, the
leader can perform reads locally without contacting other nodes. Leases assume
some degree of clock synchrony, so they are not appropriate in all
circumstances. Moreover, the leader is still a read bottleneck.
Raft has a similar optimization that does not require
any form of clock synchrony, but the leader is still a read
bottleneck~\cite{ongaro2014search}. With Paxos Quorum
Leases~\cite{moraru2014paxos}, any set of nodes---not just the leader---can
hold a lease for a set of objects. These lease holders can read the objects
locally. Paxos Quorum Leases assume clock synchrony and are a special case of
Paxos Quorum Reads~\cite{charapko2019linearizable} in which read quorums
consist of any lease holding node and write quorums consist of any majority
that includes all the lease holding nodes. Compartmentalization MultiPaxos does
not assume clock synchrony and has no read bottlenecks.

\paragraph{Harmonia}
Harmonia~\cite{zhu2019harmonia} is a family of state machine replication
protocols that leverage specialized hardware---specifically, a specialized
network switch---to achieve high throughput and low latency. Like CRAQ,
Harmonia is sensitive to data skew. It performs extremely well under low
contention, but degrades in performance as contention grows. Harmonia also
assumes clock synchrony, whereas Compartmentalized MultiPaxos does not.
FLAIR~\cite{takruri2020flair} is replication protocol that also leverages
specialized hardware, similar to Harmonia.

\paragraph{Sharding}
In this paper, we have discussed state machine replication in its most general
form. We have not made any assumptions about the nature of the state machines
themselves. Because of this, we are not able to decouple the state machine
replicas. Every replica must execute every write. This creates a fundamental
throughput limit. However, if we are able to divide the state of the state
machine into independent shards, then we can further scale the protocols by
sharding the state across groups of replicas. For example,
in~\cite{bezerra2014scalable}, Bezerra et al.\ discuss how state machine
replication protocols can take advantage of sharding.

\begin{revisions}
  \paragraph{Low Latency Replication Protocols}
  While compartmentalization increases throughput, it also increases the number
  of network delays required to get a state machine command executed. For
  example, starting from a client, MultiPaxos can execute a state machine
  command and return a response to a client in four network delays, whereas
  Compartmentalized MultiPaxos requires six. Within a single data center, this
  translates to a small increase in latency, but when deployed on a wide area
  network, the latency is increased substantially. Thus, if your goal is to
  minimize latency, you should choose latency optimized protocols like
  CURP~\cite{park2019exploiting} or SpecPaxos~\cite{ports2015designing} over a
  compartmentalized protocol.
\end{revisions}
}
{\section{Conclusion}
In this paper, we analyzed the throughput bottlenecks in state machine
replication protocols and demonstrated how to eliminate them using a
combination of decoupling and scale, a technique we call compartmentalization.
Using compartmentalization, we establish a new baseline for MultiPaxos'
performance. We increase the protocol's throughput by a factor of $6 \times$ on
a write-only workload and $16 \times$ on a 90\% read workload, all without the
need for complex or specialized protocols.
}


\bibliographystyle{ACM-Reference-Format}
\bibliography{references}

\end{document}